\documentclass[aps,prl,onecolumn,showpacs,showkeys,preprintnumbers]{revtex4}
\usepackage{eurosym}
\usepackage{amsmath}
\usepackage{dcolumn}
\usepackage{bm}
\usepackage{subfigure}
\usepackage{amsfonts}
\usepackage{amssymb}
\usepackage{makeidx}
\usepackage{epsfig}
\usepackage{graphicx}

\begin{document}

\title{\textbf{Space-time second-quantization effects and the quantum origin
of cosmological constant in covariant quantum gravity}}
\author{Claudio Cremaschini}
\affiliation{Institute of Physics and Research Center for Theoretical Physics and
Astrophysics, Faculty of Philosophy and Science, Silesian University in
Opava, Bezru\v{c}ovo n\'{a}m.13, CZ-74601 Opava, Czech Republic}
\author{Massimo Tessarotto}
\affiliation{Department of Mathematics and Geosciences, University of Trieste, Via
Valerio 12, 34127 Trieste, Italy}
\affiliation{Institute of Physics, Faculty of Philosophy and Science, Silesian University
in Opava, Bezru\v{c}ovo n\'{a}m.13, CZ-74601 Opava, Czech Republic}
\date{\today }

\begin{abstract}
Space-time quantum contributions to the classical Einstein equations of
General Relativity are determined. The theoretical background is provided by
the non-perturbative theory of manifestly-covariant quantum gravity and the
trajectory-based representation of the related quantum wave equation in
terms of the Generalized Lagrangian path formalism. To reach the target an
extended functional setting is introduced, permitting the treatment of a
non-stationary background metric tensor allowed to depend on both space-time
coordinates and a suitably-defined invariant proper-time parameter. Based on
the Hamiltonian representation of the corresponding quantum hydrodynamic
equations occurring in such a context, the quantum-modified Einstein field
equations are obtained. As an application, the quantum origin of the
cosmological constant is investigated. This is shown to be ascribed to the
non-linear Bohm quantum interaction of the gravitational field with itself
in vacuum and to depend generally also on the realization of the quantum
probability density for the quantum gravitational field tensor. The emerging
physical picture predicts a generally non-stationary quantum cosmological
constant which originates from fluctuations (i.e., gradients) of vacuum
quantum gravitational energy density and is consistent with the existence of
quantum massive gravitons.
\end{abstract}

\maketitle

\subsection{1 - Introduction}

The theory of manifestly-covariant quantum gravity (\textit{CQG-theory})
recently proposed in a series of papers (see Refs.\cite{cqg-1,
cqg-2,cqg-3,cqg-4,cqg-5,cqg-6}) provides a possible new self-consistent
route to Quantum Gravity and the cosmological interpretation of quantum
vacuum. This refers specifically to the quantum prescription of the
cosmological constant and the long-standing question whether or not it can
be ascribed exclusively to suitable vacuum fluctuations arising at the
quantum level.

The crucial feature, that we intend to display in this paper, is in fact
that CQG-theory generates\ self-consistently quantum corrections to the
Einstein field equations,\ \textit{i.e.}, obtained without introducing the
semiclassical limit. More specifically, our claim is that CQG-theory
actually gives rise to a well-defined quantum prescription of the
cosmological constant, its physical interpretation being ascribed to the
action of the non-linear quantum vacuum interaction of the gravitational
field with itself. Remarkably, the new result is based purely\textbf{\ }on
the self-consistent (in the sense indicated above) prescription of quantum
vacuum density $\rho _{A}$\textbf{\ }obtained in such a framework. In
addition it is reached without introducing "ad hoc" phenomenological
prescriptions of quantum vacuum, nor possible modifications of the classical
Lagrangian formulation of GR based on higher-order classical curvature terms
\cite{eli0,eli1}.

While still not claiming its uniqueness, CQG-theory represents nevertheless
a possible new pathway for the establishment of a quantum theory for the
standard formulation of General Relativity (GR) and at the same time
provides a promising mathematical-physics framework for the investigation of
gravitational quantum vacuum effects.

Indeed it is generally agreed that a theory of this type should be, at the
same time, in agreement with the fundamental principles of quantum mechanics
and quantum field theory \cite{Messiah} as well as the classical Einstein
theory of GR \cite{ein1,LL,gravi,noi4}. The same principles - it should be
stressed - notably include the principles of covariance and manifest
covariance \cite{Jordi-Narciso}.

CQG-theory realizes, as such, a first-quantization picture of space-time
which embodies simultaneously all the required fundamental principles (for
an extended related discussion we refer again to Refs.\cite{cqg-1,
cqg-2,cqg-3,cqg-4,cqg-5,cqg-6}).\textbf{\ }One of its notable features is
realized, first of all, by the distinction between a continuum classical
background metric tensor $\widehat{g}\equiv \left\{ \widehat{g}_{\mu \nu
}\right\} $, yielding the geometric properties of the space-time, and the
quantum gravitational field $g_{\mu \nu }$ which dynamically evolves over $%
\widehat{g}$ according to a defined quantum wave equation (\textit{CQG-wave
equation}) \cite{cqg-4}. The latter is based on the identification of the
Hamiltonian structure associated with the classical space-time, with the
prescription of the corresponding manifestly covariant Hamilton equations
\cite{cqg-2} and the related Hamilton--Jacobi theory \cite{cqg-3} obtained
in the framework of a synchronous variational principle \cite{cqg-1}. This
leads to the realization in $4-$scalar form of the quantum Hamiltonian
operator and the CQG-wave equation for the corresponding CQG-state and wave
function, whose dynamics is parametrized in terms of an invariant
proper-time parameter.

Nevertheless, it must be remarked that relevant results established so far
in the framework of CQG-theory pertain both first- and second-quantization
effects. The first category includes:

\begin{itemize}
\item The establishment of the Schroedinger-like CQG-wave equation in
manifest covariant form, which is realized by a first-order PDE with respect
to the invariant proper-time \cite{cqg-4}.

\item The statistical interpretation of the CQG-wave equation in terms of
corresponding quantum hydrodynamic equations \cite{cqg-5}.

\item The fulfillment of generalized Heisenberg inequalities relating the
statistical measurement errors of quantum observables, represented in terms
of the standard deviations of the quantum gravitational tensor $g_{\mu \nu }$%
\ and its quantum conjugate momentum operator \cite{cqg-5}.

\item The formulation of a trajectory-based representation of CQG-theory
achieved in terms of a covariant Generalized Lagrangian-Path (GLP) approach
relying on a suitable statistical representation of Bohmian Lagrangian
trajectories \cite{cqg-6,FoP-2016}.

\item The construction of generally non-stationary analytical solutions for
the CQG-wave equation with non-vanishing cosmological constant and
exhibiting Gaussian-like probability densities that are non-dispersive in
proper-time \cite{cqg-6}.

\item The proof of the existence of an emergent gravity phenomenon occurring
in the context of CQG-theory (previously referred to as "\emph{second-type
emergent-gravity paradigm}" (see Ref.\cite{cqg-6})) according to which it
can/must be possible to represent the mean-field background space-time
metric tensor $\widehat{g}$ in terms of a suitable ensemble average. More
precisely, as shown in the same reference, this is identified in terms of a
statistical average with respect to stochastic fluctuations of the quantum
gravitational field $g_{\mu \nu }$, whose quantum-wave dynamics is actually
described by means of GLP trajectories \cite{cqg-6}.
\end{itemize}

Results belonging to second-quantization effects concern instead:

\begin{itemize}
\item The proof of existence of a discrete invariant-energy spectrum for
stationary solutions of the CQG-wave equation, obtained by implementing the
Dirac ladder method (\textit{i.e.}, a second-quantization method) for the
stationary wave equation with harmonic Hamiltonian potential \cite{cqg-4}.

\item The analytical estimate for the graviton mass and\ its quantum
discrete invariant energy spectrum, supporting the interpretation of the
graviton DeBroglie length as being associated with the quantum ground-state
related to the cosmological constant \cite{cqg-4}.\textbf{\ }It must be
stressed in this connection that the prediction of massive gravitons
represents an intrinsic property of CQG-theory which marks also an important
point of distinction with respect to past literature. In fact previous
perturbative treatments of quantum gravity based on linearized GR theories
typically exhibit - in analogy with the case of the electromagnetic field -
massless gravitons. Indeed in the framework of CQG-theory, as discovered in
Ref.\cite{cqg-4}, the existence of massive gravitons and their mass estimate
are found to be associated with a non-vanishing cosmological constant.
\end{itemize}

Based on these outcomes, the target of this paper is to show that in the
context of the CQG-theory the equation for the background metric tensor $%
\widehat{g}$\ can actually self-consistently be determined by the CQG-wave
equation itself, together with its relationship with the classical Einstein
equations. The consequences are of crucial importance since this feature of
CQG-theory makes possible the investigation of quantum
corrections/contributions to the classical GR equations themselves.\ In
particular, focus is given on the quantum origin of cosmological constant
and its quantum representation as predicted by CQG-theory. The consequent
second-quantization, \textit{i.e.}, non-linear quantum modifications of the
background space-time obtained in this way, represents the main subject of
investigation of the present work. As we intend to show, besides quantum
gravity theory itself, this is relevant in the context of theoretical
astrophysics and cosmology to reach\ a quantum-gravity
interpretation/explanation of selected physical evidences emerging from
large-scale phenomenology of the universe.

\subsection{1A - Physical evidence and open problems}

The current status of observations of the large-scale structure of the
universe is compatible with its identification in terms of a
coordinate-independent (\textit{i.e.}, frame independent) abstract setting
realized by a differential manifold $\left\{ \mathbf{Q}^{4},\widehat{g}%
\right\} $,\ being $\mathbf{Q}^{4}$\ a time-oriented $4-$dimensional Riemann
space-time with signature $\left\{ +,-,-,-\right\} $ and\textbf{\ }$\widehat{%
g}\equiv \left\{ \widehat{g}_{\mu \nu }\right\} $\textbf{\ }to be considered
a suitably-prescribed background metric tensor characterized by a number of
properties.

The first one (\emph{EVIDENCE \#1})\textbf{\ }is about the flatness of the
universe \cite{flat4,flat3,flat2,flat1}, \textit{i.e.}, the fact that the $%
4- $dimensional space-time curvature is very small and compatible with the
existence of a cosmological constant $\Lambda $\ which in magnitude is $%
\left\vert \Lambda \right\vert \ll 1$ \cite{Winberg2000,Carroll2004}.
However, open questions remain in this regard. These concern, in particular,
both the precise physical origin of the cosmological constant \cite%
{EV1-a,EV1-aa,EV1-aaa,EV1-aaaa} as well the actual meaning and possible
realization of a dynamical evolution of the universe to be realized either
in the context of classical or quantum gravity \cite%
{EV1-b,EV1-c,EV1-d,EV1-e,EV1-f,EV1-g,EV1-h,EV1-j}. The second physical
evidence (\emph{EVIDENCE \#2}) concerns the apparent lack of large-scale
correlations among distant regions of space in opposite directions (\textit{%
i.e.}, having typical light-ray separation $\gtrsim $\ $10^{9\text{ }}$light
years) and the consequent occurrence of the phenomenon of (large-scale)
homogeneity of the universe, whereby that latter appears to be the same in
all directions (isotropic property) \cite%
{EV2-a,EV2-b,EV2-c,EV2-d,EV2-e,EV2-f,EV2-g,EV2-h,EV2-i,EV2-j}. Also in this
case the physical origin of the phenomenon remains to be fully understood.
In fact, it is unclear how distant regions of the universe can undergo or
have undergone significant interactions with each other. The third evidence (%
\emph{EVIDENCE \#3}) concerns the discovery of an isotropic accelerating
expansion of the universe at large distances \cite%
{EV3-b,EV3-bb,EV3-bbb,EV3-a}. This implies, in turn, the fundamental
consequence that the same cosmological constant $\Lambda $\ must be slightly
positive in value \cite{Susskind2005}. The fourth evidence\ (\emph{EVIDENCE
\#4}) is about the validity of a "Big Bang hypothesis" \cite{EV4-a,EV4-b},
according to which the initial dynamical behavior of the universe should
have been characterized by an explosive, \textit{i.e.}, extremely fast,
expansion/acceleration of space-time starting from the initial condition
which in classical GR is understood as an initially-stationary primordial
black hole singularity \cite{EV4-d,EV4-dd,EV4-c}.\textbf{\ }Although several
different theories/models have been advanced, the precise physical nature
and explanation of the involved phenomena remain still unexplained. Finally,
the fifth evidence (\emph{EVIDENCE \#5}) refers to the conjecture of an
inflationary transient phase of the early universe \cite%
{inflation-5,inflation-4,inflation-3,inflation-2,inflation-1}. In the
original version of the inflationary theory \cite{inflation} the theoretical
inflation model was based on the action of a dynamically-varying scalar
field (to be distinguished from the gravitational field)\ in a local minimum
of its potential energy function and rolling the inflation in the primordial
era of the universe.

\subsection{1B - Issues about the cosmological constant}

Thanks to the evidence provided by astrophysical observations \cite{Planck},
the inclusion of a generic cosmological-constant term $\Lambda $\ in the
Einstein field equations \cite{Einstein1917} has nowadays become a
well-established part of GR theory.\textbf{\ }Nevertheless, the cosmological
constant $\Lambda $\ still emerges, for its possible conceptual
implications, as an unsolved issue of outmost importance. According to the
literature this can be cast in terms of a decomposition of the form%
\begin{equation}
\Lambda =\Lambda _{bare}+\Lambda _{QM},  \label{lambda-1}
\end{equation}%
where respectively $\Lambda _{bare}>0$ denotes a possible classical
contribution and $\Lambda _{QM}>0$ identifies a quantum contribution. In the
present case $\Lambda _{QM}$ will be identified with $\Lambda _{CQG}$,
namely the contribution arising specifically in the context of CQG-theory.

Regarding the possible realizations of Eq.(\ref{lambda-1}), disparate
theoretical models have been proposed in the past. These refer both to
classical and/or quantum derivations either for $\Lambda _{bare}$ or $%
\Lambda _{QM}$. In particular, concerning existing models for $\Lambda
_{bare}$ a typical common aspect concerns the adoption of modified classical
GR theories. These include for example the so-called Einstein-Cartan gravity
theory based on the introduction of torsion effects in the energy-momentum
tensor \cite{ivanov2015,ivanov2016}, the adoption of higher-dimensional
space-times (see for example Ref.\cite{azri2012}), Brans-Dicke theories
involving the introduction of coordinate-time dependent cosmological
constants \cite{Lu-2012}, coordinate-space or coordinate-time varying models
\cite{dym-1,dym-2,dym-3,maje2003,azri2017}, etc. Regarding, instead,
previous theoretical predictions/estimates of $\Lambda _{QM}$\ the list of
possible candidates is numerous. A historically famous one inspired by
quantum field theory is that $\Lambda _{QM}$ might be interpreted as due to
the quantum vacuum. This involves the conjecture that $\Lambda _{QM}$ should
actually be identified with the total quantum-vacuum energy density arising
from all possible quantum fields.\textbf{\ }In previous literature,
estimates of $\Lambda _{QM}$ based on such a conjecture, i.e., with the
inclusion of all the particles corresponding to the standard model and
typically a large number of bosonic field components, yield estimates which
exceed the experimentally-observed value of $\Lambda $ by nearly 120 orders
of magnitude \cite{vacuum}. Therefore, in the literature such a route is
usually regarded as to lead to unphysical predictions. It is worth
mentioning in this regard that recent numerical calculations including 28
bosonic fields and based on the estimate of the\ stochastic fluctuations of
the associated total quantum-vacuum energy density, rather than the energy
density itself, are claimed to provide lower estimates and a resulting
acceleration of the universe comparable to the observed one \cite{Wang0,Wang}%
. This type of studies should be regarded as complementary to the present
quantum theory, although they differ from it for the following main reasons:
1)\ they are based on numerical calculations, and therefore are subject to
the accuracy of the numerical codes actually implemented, while the theory
proposed here is analytical; 2) they assume that the source of the universe
expansion is the vacuum populated by bosonic fields, while in the present
model the cosmological constant is shown to arise purely from quantum
gravitational field with its quantum dynamics being predicted by CQG-theory,
without needing to invoke any additional field; 3) they realize
non-manifestly covariant solutions in which the coordinate time is singled
out with respect to space coordinates, while the investigation based on
CQG-theory preserves manifest covariance.

Nevertheless, in part independent of the reasons indicated above, several
alternative models have been developed in the literature to explain the
expansion/acceleration of the universe as well as the cosmological constant
itself (for a review see Ref.\cite{vilenkin2001}). These include, among
others: a) Scalar-field theories based on the introduction of scalar quantum
fields and, possibly, related Lagrangian functionals for the variational
derivation of the corresponding dynamics (\cite{Critte,Barrow} or the
so-called quintessence model \cite{Cicciarella2017,Asenjo2017}). By
comparison, as will be shown below, the present approach differs from these
ones in that there is no need to assume\textbf{\ "}\textit{a priori}\textbf{%
" }existence of external quantum fields other than the gravitational one. In
other words, the theory proposed here provides a representation for the
cosmological constant entering the Einstein equations which is purely
generated by quantum interaction of the vacuum gravitational field with
itself, independent of the possible additional action of external quantum
fields. b) Perturbative calculations in the framework of loop quantum
gravity (see for example Ref.\cite{Rovelli2011}). Here the main difference
is provided by the mathematical setting where calculations are performed. In
fact, models based on loop quantum gravity approach are intrinsically
non-manifestly covariant, in contrast to the present approach which
satisfies manifest covariance both at classical and quantum levels (see Refs.%
\cite{cqg-3,cqg-4}). In addition, as shown below, the calculation performed
in the present framework is not perturbative, but rather it realizes an
exact analytical result obtained adopting the trajectory-based
representation of CQG-wave equation given in Ref.\cite{cqg-6}. c)
Non-commutative approach to the Wheeler-DeWitt equation \cite{Nicolini}.
Even in this case there are no analogies with the present approach, both
because of the non-commutative framework and for the adoption of the
Wheeler-DeWitt equation, which is intrinsically non-manifestly covariant in
comparison to the CQG-wave equation and also makes use of a Hamiltonian
operator derived from preliminary space-time foliation. d) Theoretical
models obtained by introducing quantum corrections in the Raychaudhuri
equation \cite{ray1,ray2}. Despite referring to Bohmian trajectories, this
type of model is phenomenological. In addition, the Raychaudhuri equation is
a kinematical equation, not a dynamical one, and therefore inadequate by
itself to predict quantum dynamics (see also discussion in Ref.\cite{ray3}).
As a consequence, similarities with the present approach remain excluded
also in this case. The CQG-theory in fact is variational, following from
preliminary establishment of Lagrangian, Hamiltonian and Hamilton-Jacobi
theories for General Relativity, and then implementing a canonical
quantization approach. e) Phenomenological models associated with dark
matter and/or corresponding dark energy \cite{EV3-bbb,Perlmutter,Peebles}.
Despite being very popular in contemporary literature, dark matter/energy
models still lack both experimental evidence and definite theoretical
support.

However, despite the huge number of papers appeared so far, no convincing
theory or clear physical evidence exists which can explain the physical
origin of either $\Lambda _{bare}$\ or\ $\Lambda _{QM}$. In particular, a
number of questions regarding the cosmological constant remain. They include
in particular:

\begin{itemize}
\item \emph{ISSUE\ \#1:}\textbf{\emph{\ }}the possible quantum origin and,
more precisely, the quantum self-generation of the cosmological constant $%
\Lambda $, \textit{i.e.}, in which the same one is produced merely by the
presence of gravitons, as well as its precise estimate in the context of
Quantum Gravity.

\item \emph{ISSUE\ \#2: }the possible dynamical behavior of $\Lambda $ and
the search of an admissible dynamical parametrization in terms of physical
observables, including the relationship with its constant representation.

\item \emph{ISSUE\ \#3:}\textbf{\emph{\ }}the corresponding eventual
implications for cosmology, in particular in reference with the large-scale
phenomenology of the universe.
\end{itemize}

\subsection{1C - Goals and structure of the paper}

The problems addressed in the paper are cast in the framework provided by
the trajectory-based approach to CQG-theory formulated in Ref.\cite{cqg-6}
and referred to as generalized Lagrangian-path (GLP) approach.\ The notable
aspect of this representation is that it permits the construction of
dynamically-consistent analytic solutions of the CQG-wave equation which
lays at the basis of CQG-theory. These include in particular vacuum quantum
solutions in cosmological scenarios characterized by Gaussian-like or
Gaussian quantum probability density functions (PDF). The GLP-approach
developed in Ref.\cite{cqg-6} refers to the case of stationary background
space-time $\left\{ \mathbf{Q}^{4},\widehat{g}\right\} $, \textit{i.e.}, in
which the background metric tensor $\widehat{g}$ is considered stationary,
namely of the form%
\begin{equation}
\widehat{g}=\widehat{g}(r).  \label{Eq-2b}
\end{equation}%
However, the quantum wave-function determined in the same reference exhibits
an explicit dependence in terms of the observer's proper time, i.e., a
physical observable (see related discussion in Section 3 below). For this
reason it is reasonable to conjecture (as shall be shown explicitly in
subsequent Sections 6-9) that via second quantization effects, also the same
background tensor field might be expected to include an analogous type of
dependence. Therefore, the preliminary goal to be pursued as a first task of
the paper consists in the appropriate generalization of the theory to a
generally non-stationary metric tensor $\widehat{g}$\ of the type%
\begin{equation}
\widehat{g}_{\mu \nu }=\widehat{g}_{\mu \nu }(r,s),  \label{Eq-2a}
\end{equation}%
where $s$ denotes a suitably-prescribed invariant proper-time parameter. The
two settings (\ref{Eq-2b}) and (\ref{Eq-2a}) will be referred to here
respectively as\textbf{\ }\emph{stationary }and\emph{\ non-stationary
backgrounds}. More precisely, for this purpose the role of proper-time and
its definition as physical observable are first discussed in the context of
both covariant classical gravity (CCG) and covariant quantum gravity (CQG)
theories. Then, the extension to the case of a non-stationary background
metric tensor is ascertained both for CCG and CQG theories as well as for
the GLP-approach presented in Ref.\cite{cqg-6}.\ As a consequence, the
non-stationary quantum solutions of the CQG-quantum wave equation determined
in Ref.\cite{cqg-6} are shown to hold also in such a case.

The second task of the paper concerns, instead, the investigation of the
possible validity of the so-called "\textit{first-type emergent-gravity
paradigm}" (see Ref.\cite{cqg-6}). Accordingly, the functional form of the
Einstein GR field equations should be preserved when quantum corrections
implied by CQG-theory and the GLP-approach are retained, consistent with
the\ so-called\ emergent gravity picture. In other words, the determination
of the Einstein equations with quantum contributions included should not
depend on the evaluation of\ semiclassical continuum limit\ (namely obtained
letting in particular $\hslash \rightarrow 0$; see for example Ref.\cite{Han}
where the derivation of the Einstein field equations was discussed in the
context of loop quantum gravity) nor on the prescription of suitable
stochastic/quantum expectation values, but rather should be implied by the
quantum-wave equation itself. This task should involve the determination of
the PDE for the background field tensor $\widehat{g}_{\mu \nu }$\ with
inclusion of second-quantization effects arising from the quantum
gravitational field itself, to be associated with the corresponding
covariant quantum gravity wave equation, \textit{i.e.}, the CQG-wave
equation for the quantum state $\psi (g,r,s)$\ pointed out in Refs.\cite%
{cqg-4,cqg-5}. According to this procedure the CQG-wave equation should
deliver the so-called \emph{quantum-modified Einstein field equations}.
These are expected to have the same functional form of the classical
equations (see Eq.(\ref{EINSTEIN FIELD EQS}) below) but to retain at the
same time also well-definite quantum expectation values for the relevant
continuum fields, and in particular a quantum expectation value of the
cosmological constant $\Lambda $. To carry out this task a number of steps
are needed. First, the CQG-wave equation must be shown to imply the validity
of a set of Hamilton equations holding for suitable quantum canonical tensor
fields and denoted as quantum Hamilton equations. For this purpose the
equivalent set of quantum hydrodynamic equations, represented respectively
by the continuity and quantum Hamilton-Jacobi equations, are first recalled.
Second, by suitably prescribing the initial conditions, the same quantum
Hamilton equations must be proved to imply the validity of the
quantum-modified Einstein field equations. Third, by explicitly taking into
account the quantum solutions determined via the GLP-approach, the analytic
expression of the quantum\ cosmological constant $\Lambda $ needs to be
evaluated, with particular reference to its possible explicit dependence in
terms of the proper-time $s$ and its consequent identification as a
dynamically-evolving cosmological scalar field. Fourth, the solution of the
quantum-modified Einstein field equations must be investigated and shown to
take the general form of a non-stationary background metric tensor of the
type given by (\ref{Eq-2a}), where explicit proper-time dependences are
clearly identified as arising from quantum gravitational contributions.
Finally, the asymptotic behavior of the cosmological constant needs to be
analyzed in order to understand the physical role of the quantum corrections
to the gravitational field cosmological dynamics.

The implementation of this work-plan permits the establishment of a relevant
theoretical result, which concerns the investigation of the quantum origin
of the cosmological constant. In fact it is proved that CQG-theory predicts
quantum-modified Einstein equations which contain a cosmological constant
term purely generated by quantum interaction. More precisely, the quantum
cosmological constant is shown to arise from the quantum interaction\ of the
gravitational field with itself in vacuum. From the point of view of
mathematical treatment, this type of interaction is expressed by the quantum
Bohm potential term that is contained in the quantum-wave equation and is
made explicit after adoption of the Madelung representation for the quantum
wave function and the representation of the same equation in terms of
quantum hydrodynamic equations. A characteristic feature of the Bohm
potential is that of carrying a non-linear interaction expressed by
quadratic first-order derivatives and second-order derivatives of the
quantum probability density of the quantum gravitational field tensor. For
this reason, the Bohm potential depends also on the explicit realization of
the same quantum probability density, a feature which requires the
simultaneous solution of both continuity and quantum Hamilton-Jacobi
equations equivalent to the quantum-wave equation. The emerging physical
picture predicts a generally non-stationary quantum cosmological constant
which originates from fluctuations (i.e., gradients) of vacuum quantum
gravitational energy density and is consistent with the existence of quantum
massive gravitons (see also related discussion in Section 9).

Given these premises, the structure of the paper is as follows. In Section
2, the extension of the functional setting of CQG-theory is\ presented as
appropriate to the treatment of a non-stationary background metric tensor.
In Section 3 the role of proper-time in covariant classical/quantum gravity
is investigated. In Section 4 the formulation is presented of covariant
classical/quantum gravity in the framework of the extended functional
setting. In Section 5, the quantum Hamilton equations associated with the
corresponding set of quantum hydrodynamics equations are presented. These
lead in Section 6 to the construction of the corresponding quantum modified
Einstein field equations. In Section 7 the extension is considered of the
Generalized Lagrangian Path approach earlier formulated appropriate for the
treatment of the extended functional setting. In Section 8 the\ explicit
evaluation of the Bohm effective potential and corresponding source term,
together with the identification of the cosmological constant are presented.
In Section 9 the proper-time behavior of the quantum cosmological constant $%
\Lambda _{CQG}(s)$\ and related physical implications are discussed. Finally
in Section 10 concluding remarks are pointed out, while completing details
of algebraic calculations are reported in Appendices A-C.

\section{2 - Extended functional setting for CQG-theory}

The theoretical framework of the present paper is couched on the
manifestly-covariant Hamiltonian approach for massive gravitons recently
developed in Refs.\cite{cqg-1,cqg-2,cqg-3,cqg-4,cqg-5,cqg-6}. In its
classical formulation, referred to as covariant classical theory of gravity
(CCG-theory), this is based on a classical Hamiltonian representation for
the Einstein field equations of the gravitational field which permits to
recover the same Einstein equations as particular solutions of a suitable
set of manifestly-covariant continuum Hamilton equations.

The corresponding quantization approach for the space-time metric tensor,
denoted as $g-$quantization \cite{cqg-5}, realizes instead a
manifestly-covariant quantum gravity theory (CQG-theory).\ The
characteristic property of CQG-theory is that of yielding a non-perturbative
hyperbolic quantum wave equation, denoted as CQG-wave equation, advancing
the state of the quantum gravitational field with respect to an invariant
proper-time parameter $s$. In addition, both CCG-theory and CQG-theory are
manifestly covariant. In accordance, it follows that all the classical and
quantum Hamiltonian densities and operators as well as the corresponding
continuum coordinates and conjugate momenta, are all required to transform
as $4-$tensors, \textit{i.e.}, to fulfill as such well-definite covariance
tensor transformation laws%
\begin{equation}
r\equiv \left\{ r^{\mu }\right\} \rightarrow r^{\prime }\equiv \left\{
r^{\prime \mu }\right\} =r^{\prime }(r)  \label{LPT}
\end{equation}%
associated with local point transformations between reference systems (LPT
group \cite{noi4}), namely in which the rhs of the previous equation depends
on the local value of the initial and transformed $4-$positions $r\equiv
\left\{ r^{\mu }\right\} $\ and $r^{\prime }\equiv \left\{ r^{\prime \mu
}\right\} $\ respectively.

However, in order that the principle of manifest covariance can actually
apply, a background space-time picture must hold. This means that,
consistent with experimental evidence, the universe must be identified with
a suitable classical curved space-time\ $\left\{ \mathbf{Q}^{4},\widehat{g}%
\right\} $\ with the background metric tensor $\widehat{g}\equiv \left\{
\widehat{g}_{\mu \nu }\right\} $\ to be considered a classical tensor field.
In particular,\ this means that the LPT group must leave invariant the
differential manifold structure of a prescribed (but in principle arbitrary)
curved space-time\ $\left\{ \mathbf{Q}^{4},\widehat{g}\right\} $, to be
referred to as background space-time.\ Hence, no preferred GR reference
frames or coordinate systems are required. The latter occurrence follows for
example when decompositions or foliations of space-time (like the 3+1
representation) and the consequent adoption of non-tensor
Lagrangian/Hamiltonian variables are implemented. This typically involves
the singling out of the coordinate time\ to prescribe the dynamical
evolution of metric tensor hypersurfaces (see Refs.\cite%
{ADM,zzz2,alcu,Vaca5,Vaca6}). It must be stressed that although a
manifestly-covariant theory of this type needs not necessarily to be unique,
the involved notion of manifest covariance given above is certainly
unambiguously determined when the background space-time $\left\{ \mathbf{Q}%
^{4},\widehat{g}\right\} $ is prescribed.

A crucial aspect is therefore the prescription of its functional setting.

One notices in this regard that by assumption $\widehat{g}$\ determines the
geometric properties of the same space-time and is required to satisfy
suitable physical prescriptions. The first one is that $\widehat{g}$\ must
be considered as a deterministic, \textit{i.e.}, classical, tensor field. As
such, in the framework of CQG-theory this is assumed to realize a particular
solution of the Einstein field equations.\ In standard notation the latter
can be written
\begin{equation}
\widehat{R}_{\mu \nu }-\frac{1}{2}\left[ \widehat{R}-2\Lambda \right]
\widehat{g}_{\mu \nu }=\kappa \widehat{T}_{\mu \nu },
\label{EINSTEIN FIELD EQS}
\end{equation}%
with $\kappa $\ being the universal constant
\begin{equation}
\kappa \equiv \frac{8\pi G}{c^{4}},  \label{UNIVERSAL CONSTANT}
\end{equation}%
and where%
\begin{equation}
\widehat{G}_{\mu \nu }\equiv \widehat{R}_{\mu \nu }-\frac{1}{2}\widehat{R}%
\widehat{g}_{\mu \nu }  \label{Einstein field tensor}
\end{equation}%
is the Einstein field tensor. Moreover: 1) $\widehat{R}_{\mu \nu }$\ $\equiv
R_{\mu \nu }(\widehat{g}),$\ $\widehat{R}$\ $\equiv R(\widehat{g})\equiv
\widehat{g}^{\alpha \beta }\widehat{R}_{\alpha \beta }$\ and $\widehat{T}%
_{\mu \nu }=T_{\mu \nu }(\widehat{g})$\textbf{\ }identify respectively the
Ricci tensor, the Ricci $4-$scalar and stress-energy tensor\textbf{\ }(or
energy-momentum tensor of matter) all evaluated in terms of the background
metric tensor $\widehat{g}$; 2)\textbf{\ }$\Lambda $\ is the still to be
determined cosmological constant which can always be taken of the general
form (\ref{lambda-1}). Accordingly, the metric tensor $\widehat{g}_{\mu \nu
} $\ must raise and lower tensor indices of arbitrary tensor fields, such as
for example the second-order coordinate and momentum tensor fields $H_{\mu
\nu }=g_{\mu \nu },\pi _{\mu \nu }$, \textit{i.e.},%
\begin{equation}
H_{\mu \nu }=\widehat{g}_{\mu \alpha }\widehat{g}_{\nu \beta }H^{\alpha
\beta },  \label{Eq-1}
\end{equation}%
with $H_{\mu \nu }$\ and $H^{\alpha \beta }$\ denoting respectively
corresponding covariant and counter-variant components. The second
prescription is that\ $\widehat{g}$\ should determine the Riemann distance
on the space-time $\left\{ \mathbf{Q}^{4},\widehat{g}\right\} $ and
consequently the proper-time $s$ by means of the $4-$scalar equation%
\begin{equation}
ds^{2}=\widehat{g}_{\mu \nu }dr^{\mu }dr^{\nu }.  \label{Riemann distance}
\end{equation}%
Here $ds$\ is the so-called line element (arc length) and $dr^{\mu }$\ the
corresponding $4-$tensor displacement around a $4-$position $r\equiv \left\{
r^{\mu }\right\} $\ which belongs to the subset of $\left\{ \mathbf{Q}^{4},%
\widehat{g}\right\} $ where $\widehat{g}_{\mu \nu }dr^{\mu }dr^{\nu }\geq 0.$
As a consequence it follows by integration that
\begin{equation}
s-s_{1}=\int\limits_{r_{1}}^{r}\sqrt{\widehat{g}_{\mu \nu }dr^{\mu }dr^{\nu }%
},  \label{Integral Riemann distance}
\end{equation}%
where here $r\equiv r(s)$\ and $r_{1}\equiv r(s_{1})$\ denote two $4-$%
positions along an arbitrary curve (worldline) $r(s)$\ joining them (which
therefore belong to the same light cone), while $s$\ and $s_{1}$\ are the
corresponding proper-times. In particular, in accordance to Ref.\cite{cqg-4}%
, the worldlines on which the Riemann distance is evaluated can be
conveniently identified with appropriate non-null field geodetics. Hence,
for an arbitrary GR-frame endowed with a $4-$position $r^{\mu }$, such a
worldline can in principle be identified with one of the (infinite possible)
curves that cross the same position, \textit{i.e.}, an arbitrary observer's
geodetics $r(s)\equiv \left\{ r^{\mu }(s)\right\} $\ prescribed in such a
way that at proper-time $s$\ it coincides with the observer's position,
namely so that it satisfies the initial (crossing) condition
\begin{equation}
r^{\mu }=r^{\mu }(s).  \label{CROSSING CONDITION}
\end{equation}%
Here $s>0$\ denotes the arc length which is associated with the same
observer and therefore is referred to here as observer proper-time. As
discussed below (see following subsections 3A and 3B), under suitable
assumptions $s$\ can be interpreted as a classical $4-$scalar observable
which can be unambiguously associated with an arbitrary GR-frame. However,
the prescription of the proper-time $s$\ achieved in this way can also be
made unique for all observers thus yielding also a global observable. The
third requisite, in close analogy with the quantum wave-function determined
in Ref.\cite{cqg-6} and for consistency with the goals of the present
investigation, is that the background metric tensor $\widehat{g}\equiv
\left\{ \widehat{g}_{\mu \nu }\right\} $\ should be allowed for greater
generality to be non--stationary too. In the context of a
manifestly-covariant description, nevertheless,\ $\widehat{g}$ cannot depend
on a coordinate time but necessarily on an invariant time coordinate, to be
identified with the proper-time $s$. Therefore, the metric tensor $\widehat{g%
}$ should conveniently be allowed for greater generality\ to take the
non-stationary form (\ref{Eq-2a}).

Regarding classical GR the possibility of an extended functional setting of
this type has been already pointed in Ref.\cite{cqg-3} as being due either
to the action of suitable non-local point transformations acting on
GR-frames \cite{noi4} or to possible non-local source terms in the
stress-energy tensor of the Einstein equations. An example of the second
type (for the explicit proper-time dependence) arises in particular in the
case of electromagnetic radiation-reaction phenomena affecting the dynamics
of $N-$body systems of charged particles, with $N\geq 1$ \cite{EPJ1,EPJ2},
where the corresponding stress-energy tensor depends explicitly on the
proper-time of the particles subject to radiation-reaction.

In previous works dealing with CQG-theory the case of stationary background
metric tensor was actually treated, for which identically $\widehat{g}_{\mu
\nu }=\widehat{g}_{\mu \nu }(r)$. In the present context, however, requiring
validity of Eq.(\ref{Eq-2a}) poses two crucial questions. The first one is
whether a consistent generalization of the theory of covariant quantum
gravity and of the related GLP-theory developed in Ref.\cite{cqg-6} can
actually be achieved for a non-stationary background metric tensor of the
type (\ref{Eq-2a}). The issue concerns also the corresponding formulation of
CCG-theory and in particular how the extended and reduced-dimensional
variational Hamiltonian structures $\left\{ H,x\right\} $ determined in Refs.%
\cite{cqg-3,cqg-4} can be preserved under assumption (\ref{Eq-2a}). The
second question instead is about the possibility of predicting the $s$%
-dependence of $\widehat{g}_{\mu \nu }$ as arising specifically because of
second-quantization effects of the gravitational field, namely\ quantum
modifications of the background metric tensor $\widehat{g}_{\mu \nu }$ and
corresponding field equations due to non-linear dynamical interaction of its
quantum counterpart field $g_{\mu \nu }$. In fact, as shown in Ref.\cite%
{cqg-4} the occurrence of\ an explicit proper-time dependence in the quantum
wave-function is a characteristic feature of CQG-theory for the quantum
gravitational field, \textit{i.e.}, which arises in the actual construction
of particular solutions of the relevant quantum-wave equation of CQG-theory
based on GLP-parametrization (see also Ref.\cite{cqg-6}). As shown in this
work, the answer to these questions is deeply related with the investigation
of Issues \#1-\#3 posed above.

\section{3 - The role of proper-time in covariant classical/quantum gravity}

In the context of both CCG- and CQG-theories a crucial aspect concerns the
treatment of gravitons, \textit{i.e.}, the quanta of the gravitational field
and in particular the related prescription of the notion of proper-time ($s$%
). For this purpose one first notices that both in CCG- and CQG-theories the
background space-time $\widehat{g}_{\mu \nu }$ prescribing the coordinate
and the geometric properties of the reference system is not quantized. The
quantization pertains the fluctuations with respect to $\widehat{g}_{\mu \nu
}$ of the quantum gravitational field described by $g_{\mu \nu }$. The
implication is that gravitons still need to be treated as classical
particles, \textit{i.e.}, necessarily as point-like neutral, spin-$2$
collisionless particles, since in order to quantize them one should actually
perform a full quantization of the metric tensor defining the space-time,
and therefore the physical coordinates identified with position and
velocity. In addition, in view of the invariant discrete energy spectrum
discovered in Ref.\cite{cqg-4} gravitons must carry a non vanishing mass. As
a consequence their positions and velocities are considered as
deterministic. Therefore they are endowed with a purely geodesic motion
while their admissible (or\textbf{\ }\emph{virtual}) worldlines must be
identified with (deterministic) non-null subluminal geodetics which are
associated with the background metric tensor\textbf{\ }$\widehat{g}$.

Concerning the notion of proper-time, as recalled above, this can be
identified with the arc length of a non-null geodetics, \textit{i.e.}, the
virtual worldline of a graviton. Such geodetics are intrinsically
non-unique. In fact for an arbitrary observer (or GR-frame) defined by its $%
4-$position $r\equiv \left\{ r^{\mu }\right\} ,$ there are infinite geodesic
curves $r(s)\equiv \left\{ r^{\mu }(s)\right\} $\ fulfilling the crossing
condition (\ref{CROSSING CONDITION}) \textit{i.e.}, belonging to the same
observer. However, the notion of (observer) proper-time ($s$) makes sense
only if $s$ is an observable. Therefore there must exist a suitable way to
prescribe it. In this regard two choices are possible. According to the
first case, proper-time is an\textbf{\ }\emph{observer proper-time},\textbf{%
\emph{\ }}\textit{i.e.}, a \emph{local observable} which may have
nevertheless different realizations for each observer (\textit{i.e.},
GR-frames which are mutually connected via the LPT group). In this case the
proper-time $s$ is by construction the same one for all geodesic
trajectories which cross simultaneously the observer $4-$position (see Eq.(%
\ref{CROSSING CONDITION})).

The second possible realization is provided instead by the notion of \emph{%
global proper-time},\textbf{\emph{\ }}\textit{i.e.}, a \emph{global
observable }which is the same one also for a family of observers which are
properly "synchronized" with each other in such a way that the observer
proper-time $s$ indeed coincides for all of them. In this case the observer
proper-time $s$ takes therefore - by suitable construction - the same value
for all such observers. The two choices proposed here require in turn
well-definite prescriptions for the functional setting of the observers'
geodesic curves. We consider them below.

\subsection{3A - proper-time as a local or global observable}

In the first case one can show that a non-trivial definition of the observer
proper-time requires that:

1) For each observer, consistent with its identification with a graviton's
virtual worldline, the corresponding geodesic curves (\emph{observer
geodetics}) must be all non-vanishing and oriented (each one with its proper
orientation).

2) Because classical geodetics cannot cross event-horizons of arbitrary
black holes (just as classical particles with finite mass), curves
originating near them must have a origin point $r(s_{o})=r_{o}$ suitably
close to the same event horizons. It is understood that the indicated origin
point $r(s_{o})=r_{o}$ corresponds to a \emph{creation point }of a
graviton's virtual worldline, \textit{i.e.}, a point where a graviton may be
created. Hence it makes sense to assume that all the observer's geodesic
curves have proper origin points and hence are \emph{semi-infinite}. In
addition, the origin points of all observer geodesic curves cannot coincide
with event horizons but can be also arbitrarily close to them, so that the
limit of a suitable sequence of origin points actually may coincide with the
same event horizon. \

3) For all semi-infinite geodesic curves it makes sense to require that the
initial proper-time $s_{o}$ is positive or null. For the uniqueness of $s$
for a given observer - \textit{i.e.}, as a local observable - there must
exist among all the observer's geodesic curves a (possibly non-unique) \emph{%
observer's maximal geodetics, }\textit{i.e.}, a geodesic curve with origin
point $r^{\mu }(s_{o})$\ having the maximal arc length $s-s_{o}$ and subject
to the condition%
\begin{equation}
s_{o}=0,  \label{minimum proper time}
\end{equation}%
with $r^{\mu }(s_{o})$ coinciding (or being suitably close) to the \emph{Big
Bang event }$r^{\mu }(s_{o})\equiv r^{\mu }(s_{o}=0),$ with $s_{o}=0$\ to be
referred to as \emph{Big Bang proper-time}.

When interpreted in a cosmological scenario, such an assumption is
consistent with the Big Bang hypothesis (see \emph{EVIDENCE \#4}). Such
curves therefore should have originated suitably near the universe horizon
created during Big Bang, which is characterized by the lowest initial
proper-time. Thus, the root (\ref{minimum proper time}) identifies the
proper-time of a (possibly virtual) graviton generated in coincidence, or
immediately after, the Big Bang event.\textbf{\ }The remaining trajectories
which are associated with a given observer identify instead (again possibly
virtual) massive gravitons which are generated at later proper-times.

In order to be able to identify the proper-time $s$ also as a global
observable it is necessary to require, in addition, that:

4) For all observers which can be mutually connected by null geodetics (%
\textit{i.e.}, necessarily belong to the same light-cone) and for all
semi-infinite geodesic curves which are associated with them, the
corresponding initial proper-times $s_{o}$ are all positive or null.

5) Among them for all observers there is again for each one possibly a
non-unique "maximal length" geodetics with origin point $r^{\mu }(s_{o})$
such that the condition (\ref{minimum proper time}) holds.

\subsection{3B - Interpretation/meaning of proper-time}

An important issue about proper-time concerns its possible interpretation
and meaning. This concerns the customary interpretation occurring in the
context of General Relativity \cite{Wald,Rovelli}, \textit{i.e.}, in terms
of the Riemann distance on the space-time (\emph{geometric interpretation}).
Such an interpretation is based on equations (\ref{Riemann distance}) and (%
\ref{Integral Riemann distance}).\textbf{\ }However, it does not provide, by
itself, a unique prescription for $s$.\textbf{\ }In fact, once the reference
$4-$position $r=r(s)$\ (see Eq.(\ref{CROSSING CONDITION})) is prescribed,
the precise value of $s$\ depends both on the choice of the space-time curve
on which it is measured and that of the reference $4-$position $%
r_{1}=r(s_{1})$ on the same curve. As shown above, these indeterminacies can
be resolved if, for all observers belonging to the same light-cone,
proper-time is the arc length measured along an arbitrary observer geodetics
with origin point $r(s_{o})$ and in particular along an observer's maximal
geodetics having origin point $r(s_{o}=0)$ (\emph{CCG-theory geometric
interpretation}). Under the assumption of existence of massive gravitons,
proper-time acquires also the further interpretation according to which, for
all observers belonging to the same light-cone, it is the arc length of the
worldline of a graviton measured between its origin point $r(s_{o}=0)$\ and
the observer position\textbf{\ }$r(s)=r$ (\emph{dynamic interpretation}).
Finally, proper-time can also be interpreted as a global classical
observable realized by a $4-$scalar, which can be measured by an arbitrary
observer via an ideal measurement experiment or gedanken experiment (\emph{%
experimental interpretation}).

Some additional remarks are in order concerning the role of proper-time in
CQG-theory. First of all it must be stressed that the adoption of the
proper-time parametrization permits recovering the customary concepts and
formalism of standard quantum mechanics and relativistic quantum mechanics
also for CQG-theory, which are associated with the Hamiltonian and
Hamilton-Jacobi structures of the theory and the physical meaning of the
quantum wave function. From one side the proper-time is consistent with the
manifest covariance principle, since it is a $4-$scalar (contrary to the
coordinate time), while from the other side it indeed plays the role of
"time" dynamical variable in terms of which dynamical evolution of quantum
systems is parametrized, and therefore it represents also a convenient
choice for the quantum theory itself. It follows that CQG-theory is truly
founded on the notion of proper-time, which becomes necessary for the
representation of the fundamental equations of CQG-theory and its physical
interpretation. The role of the invariant proper-time is restored in
CQG-theory as dynamical parameter, in difference with the marginal role
played by coordinate time in loop quantum gravity\ (for a discussion of the
issue see Ref.\cite{Rovelli}).

\section{4 - Covariant classical/quantum gravity in the extended setting}

Let us now pose the problem of formulating the theory of covariant gravity
in the extended setting of the type (\ref{Eq-2a}), \textit{i.e.}, in the
general case of a non-stationary background metric tensor. This point, in
view of the goals set in the paper, is crucial. The expectation in fact is
that second-quantization effects arising due to non-linear quantum
corrections of the background metric tensor might give rise to a possible
explicit proper-time dependence of the same tensor field. For this purpose
the prerequisite is to ascertain whether respectively the classical and
quantum Hamiltonian structures determined in Refs.\cite{cqg-3} and \cite%
{cqg-4}, which are set at the basis of CCG and CQG theories, can actually be
preserved in such a case. This requires, more precisely, to uncover whether
and under which conditions the validity of the principle of manifest
covariance can be warranted. In the following subsections the issues are
discussed in detail.

\subsection{4A - The Classical Hamiltonian structure of GR}

Let us first consider the extension of the reduced continuum Hamiltonian
theory for GR and of the related classical Hamiltonian structure of GR
developed in Ref.\cite{cqg-3}. This is represented by a set $\left\{
x_{R},H_{R}\right\} ,$ formed by an appropriate $4-$tensor canonical state $%
x_{R}\equiv (g_{\mu \nu },\pi ^{\mu \nu })$ and an appropriate $4-$scalar
classical Hamiltonian density $H_{R}$. According to the same reference this
is identified with the function%
\begin{equation}
H_{R}\equiv T_{R}+V,  \label{classical Hamiltonian density}
\end{equation}%
where the effective kinetic and the normalized effective potential density $%
T_{R}$ and $V$ are reported for completeness in Appendix A. Then, adopting
for definiteness the Eulerian representation given in Ref.\cite{cqg-6} and
introducing the covariant $s-$derivative operator $\frac{d}{ds}$ (see Eq.(%
\ref{covariant s-derivative}) below), by assumption the same Hamiltonian
structure should generate the $4-$tensor (continuum) GR-Hamilton equations:%
\begin{equation}
\left\{
\begin{array}{c}
\frac{dg_{\mu \nu }}{ds}=\frac{\partial H_{R}}{\partial \pi ^{\mu \nu }}, \\
\frac{d\pi ^{\mu \nu }}{ds}=-\frac{\partial H_{R}}{\partial g_{\mu \nu }}.%
\end{array}%
\right.   \label{canonical evolution equations -2}
\end{equation}%
In terms of Eqs.(\ref{App-1}) and (\ref{App-2}) (see Appendix A) these
reduce to%
\begin{equation}
\left\{
\begin{array}{c}
\frac{dg_{\mu \nu }}{ds}=\frac{\pi _{\mu \nu }}{\alpha L}, \\
\frac{d\pi ^{\mu \nu }}{ds}=-\frac{\partial V}{\partial g_{\mu \nu }}.%
\end{array}%
\right.   \label{canonical eqs-2bis}
\end{equation}%
Omitting possible implicit dependences (\textit{i.e.}, with respect to the
tangent $4-$vector $t\equiv \left\{ t^{\alpha }\right\} $, see \textit{e.g.}
Eq.(\ref{TANGENT})$\ $below and Ref.\cite{cqg-4}) $H_{R}$ is assumed to be
of the form $H_{R}=H_{R}(x_{R},\widehat{g},r,s),$ where $\widehat{g}$ is
according to Eq.(\ref{Eq-2a}). Then, by introducing a proper-time
parametrization of the canonical state of the form%
\begin{equation}
x_{R}\equiv x_{R}(s)\equiv x_{R}(r(s),s),  \label{s-parametrization}
\end{equation}%
the same state is assumed to be subject to an initial condition of the type
\begin{equation}
\left\{
\begin{array}{c}
g_{\mu \nu }(s_{1})\equiv g_{\mu \nu }^{(o)}(r(s_{1}),s_{1}), \\
\pi ^{\mu \nu }(s_{1})\equiv \pi ^{(o)\mu \nu }(r(s_{1}),s_{1}),%
\end{array}%
\right.   \label{initial conditions-2}
\end{equation}%
being $s_{1}\geq s_{o}$\ and $r(s_{1})$\ respectively an initial proper-time
and a geodesic curve evaluated at the same proper-time. The mandatory
requisites in order to preserve the Hamiltonian structure indicated above,
\textit{i.e.}, for the validity of the canonical equations (\ref{canonical
evolution equations -2}), are that they should, at the same time: a) be
manifestly-covariant; b) by prescription of the initial conditions (\ref%
{initial conditions-2}), reduce identically for all $s\geq s_{o}$ to the
non-vacuum Einstein field equations (\ref{EINSTEIN FIELD EQS}); c) that a
corresponding classical Hamilton-Jacobi equation, equivalent to the
canonical equations (\ref{canonical evolution equations -2}), should hold.

As shown in Appendix B, the first requisite demands that the differential
operator $\frac{d}{ds}$ in Eqs.(\ref{canonical evolution equations -2}) and (%
\ref{canonical eqs-2bis}), when written in Eulerian form in analogy to Ref.%
\cite{cqg-6}, should take the form%
\begin{equation}
\frac{d}{ds}=\left. \frac{d}{ds}\right\vert _{s}+\left. \frac{d}{ds}%
\right\vert _{r}.  \label{covariant s-derivative}
\end{equation}%
Here the notation is as follows. First, $\left. \frac{d}{ds}\right\vert
_{s}\equiv t^{\alpha }\nabla _{\alpha }$\ identifies the\textbf{\ }\emph{%
directional covariant derivative}, with\textbf{\ }%
\begin{equation}
t^{\alpha }=\frac{dr^{\alpha }(s)}{ds}\equiv \left. \frac{d}{ds}\right\vert
_{s}r^{\alpha }(s)  \label{TANGENT}
\end{equation}%
being the tangent to the geodetic curve $r(s)\equiv \left\{ r^{\alpha
}(s)\right\} .$\ Second, $\left. \frac{d}{ds}\right\vert _{r}$\ denotes now
the \emph{covariant }$s-$\emph{partial derivative.} When it operates on a $%
4- $scalar this coincides with the ordinary partial derivative, so that
\begin{equation}
\left. \frac{d}{ds}\right\vert _{r}=\frac{\partial }{\partial s},
\label{cov-scalar-2}
\end{equation}%
and consequently in this case%
\begin{equation}
\frac{d}{ds}=\left. \frac{d}{ds}\right\vert _{s}+\frac{\partial }{\partial s}%
\equiv D_{s},  \label{cov-scalar-1}
\end{equation}%
with $D_{s}$ to be referred to as \emph{convective derivative}. However,
when acting on a second-order tensor it must be prescribed according to Eqs.(%
\ref{A-3}) (or equivalently Eqs.(\ref{A-3bis})) reported in the Appendix B
respectively for the countervariant and covariant components of a generic
second-order $4-$tensor. As a consequence one obtains respectively that the
operator $\frac{d}{ds}$ acts so that
\begin{equation}
\left\{
\begin{array}{c}
\frac{d}{ds}x^{\alpha \beta }=D_{s}x^{\alpha \beta }-\frac{1}{2}x^{pq}%
\widehat{g}_{\mu p}\widehat{g}_{\nu q}\frac{\partial }{\partial s}(\widehat{g%
}^{\alpha \mu }\widehat{g}^{\beta \nu }), \\
\frac{d}{ds}x_{\alpha \beta }=D_{s}x_{\alpha \beta }-\frac{1}{2}x_{pq}%
\widehat{g}^{\mu p}\widehat{g}^{\nu q}\frac{\partial }{\partial s}(\widehat{g%
}_{\alpha \mu }\widehat{g}_{\beta \nu }).%
\end{array}%
\right.  \label{COVARIANT S-OPERATOR-ii}
\end{equation}%
Thus, in particular, when $x^{\alpha \beta }\equiv \widehat{g}^{\alpha \beta
}(r,s)$\ or $x_{\alpha \beta }\equiv \widehat{g}_{\alpha \beta }(r,s),$\
namely the covariant and countervariant components of background metric
tensor are considered, it follows that the identities\textbf{\ }%
\begin{eqnarray}
\frac{d}{ds}\widehat{g}_{\mu \nu }(r,s) &\equiv &0,  \label{ID-1} \\
\frac{d}{ds}\widehat{g}^{\mu \nu }(r,s) &\equiv &0,  \label{ID-2}
\end{eqnarray}%
necessarily hold, where $r\equiv r(s)$\ denotes the (arbitrary) geodetics
indicated above.\textbf{\ }Regarding the second requisite, once Eq.(\ref%
{covariant s-derivative}) is set, then the same formally follows in a
straightforward way. In fact, introducing the initial conditions
\begin{equation}
\left\{
\begin{array}{c}
g_{\mu \nu }(s_{o})\equiv \widehat{g}_{\mu \nu }(r(s_{o}),s_{o}), \\
\pi ^{\mu \nu }(s_{o})\equiv \widehat{\pi }^{\mu \nu }(r(s_{o}),s_{o})=0,%
\end{array}%
\right.
\end{equation}%
and by requiring that the corresponding extremal fields are such that $%
\widehat{x}_{R}(s)\equiv (\widehat{g}_{\mu \nu }(s),\widehat{\pi }^{\mu \nu
}(s)\equiv 0)$, implies that thanks to the identities (\ref{ID-1}) and (\ref%
{ID-2}), Eqs.(\ref{canonical evolution equations -2}) become identically%
\begin{equation}
\left\{
\begin{array}{c}
\frac{d\widehat{g}_{\mu \nu }}{ds}\equiv 0, \\
-\left. \frac{\partial V}{\partial g_{\mu \nu }}\right\vert _{g_{\mu \nu
}(s)=\widehat{g}_{\mu \nu }(s)}=0.%
\end{array}%
\right.  \label{last}
\end{equation}%
Hence, the second equation coincides identically with the Einstein field
equations (\ref{EINSTEIN FIELD EQS}). The implication is therefore that the
same equations hold also in validity of non-stationary sources and
consequently in the case of a non-stationary background field tensor. These
conclusions overcome the conditions earlier stated in Ref.\cite{cqg-3} which
instead referred to the case of a stationary background field tensor.

Finally, let us consider the requirement of validity of the Hamilton-Jacobi
equation indicated above, originally first established in Ref.\cite{cqg-3}
for the case of stationary background field tensors (see THM.1 in the same
reference). The question arises whether also in the non-stationary case of
Eq.(\ref{Eq-2a}) the set of PDEs represented by the classical GR-Hamilton
equations (\ref{canonical evolution equations -2}) should be equivalent to a
single PDE to be referred to as GR-Hamilton-Jacobi equation, namely of the
type%
\begin{equation}
\frac{d\mathcal{S}(g,\widehat{g},r,s)}{ds}+H_{R}(g,\pi ,\widehat{g},r,s)=0,
\label{classical Hamilton-Jacobi}
\end{equation}%
which holds for a $4-$scalar function\ of the form $\mathcal{S}(g,\widehat{g}%
,r,s)$\textbf{\ }(Hamilton principal function), with $\widehat{g}\equiv
\widehat{g}(s)$\ to be understood everywhere in the following as a
non-stationary tensor of the type (\ref{Eq-2a}). In addition, due to the
arbitrariness in the definition of the same function $\mathcal{S}(g,\widehat{%
g},r,s)$, the latter should be prescribed so that: 1) first it results
identically%
\begin{equation}
\pi ^{\mu \nu }=\frac{\partial \mathcal{S}(g,\widehat{g},r,s)}{\partial
g_{\mu \nu }},  \label{AA-CON-1}
\end{equation}%
with $\pi ^{\mu \nu }$\ being the canonical momentum conjugate to $g_{\mu
\nu }$;\textbf{\ }2) second, denoting by $\left( G_{\mu \nu }\equiv g_{\mu
\nu }(s_{1}),P^{\mu \nu }\equiv \pi ^{\mu \nu }(s_{1})\right) $\ the initial
state prescribed according to the initial condition (\ref{initial
conditions-2}),\ the classical phase-function $\mathcal{S}(g,\widehat{g}%
,r,s) $\textbf{\ }should actually depend functionally on the initial
canonical state function $P$, to be identified either with the initial
coordinate $P\equiv \left\{ g_{\mu \nu }(s_{1})\right\} $, the conjugate
momentum $P\equiv \left\{ \pi ^{\mu \nu }(s_{1})\right\} $ or more generally
a function of both of them, \textit{i.e.}, to be of the form%
\begin{equation}
\mathcal{S}(g,\widehat{g},r,s)\equiv \mathcal{S}(g,\widehat{g},r,s;P).
\label{AA-CON-2}
\end{equation}%
In addition, in order that the classical Hamilton-Jacobi equation actually
warrants validity of Eqs.(\ref{canonical evolution equations -2}) the same
Hamilton principal function should satisfy identically also the
corresponding constraint equations%
\begin{equation}
\left\{
\begin{array}{c}
Q_{\mu \nu }=\frac{\partial \mathcal{S}(g,\widehat{g},r,s;P)}{\partial
P^{\mu \nu }}, \\
\left\vert \frac{\partial \mathcal{S}^{2}(g,\widehat{g},r,s;P)}{\partial
g_{\mu \nu }\partial P^{\mu \nu }}\right\vert \neq 0,%
\end{array}%
\right.  \label{AA-CON-3}
\end{equation}%
with $Q_{\mu \nu }$\ being a constant phase function, \textit{i.e.}, such
that $D_{s}Q_{\mu \nu }\equiv 0,$\ and $\left\vert \frac{\partial \mathcal{S}%
^{2}(g,\widehat{g},r,s)}{\partial g_{\mu \nu }\partial P^{\mu \nu }}%
\right\vert $\ being the determinant of the matrix\textbf{\ }$\left\{ \frac{%
\partial \mathcal{S}^{2}(g,\widehat{g},r,s;P)}{\partial g_{\mu \nu }\partial
P^{\mu \nu }}\right\} .$\ The latter, as usual in Hamilton-Jacobi theory, is
therefore required to be non-singular. To prove the validity of the
GR-Hamilton equations let us evaluate first the partial derivative of Eq.(%
\ref{classical Hamilton-Jacobi}) with respect to $g^{ik}$, keeping constant
both $\frac{\partial \mathcal{S}(g,\widehat{g},r,s;P)}{\partial g^{\iota \xi
}}$ and $\Pi \equiv \left\{ \Pi ^{\mu \nu }\right\} $. This gives%
\begin{eqnarray}
&&\left. \frac{\partial }{\partial g_{ik}}H_{R}\left( g^{\beta \gamma },%
\frac{\partial \mathcal{S}(g,\widehat{g},r,s;P)}{\partial g^{\iota \xi }},%
\widehat{g},r,s\right) +\right.  \notag \\
&&\left. \frac{\partial }{\partial g_{ik}}\frac{d}{ds}\mathcal{S}(g,\widehat{%
g},r,s;P)=0,\right.  \label{first-hjh}
\end{eqnarray}%
where the covariant $s-$derivative $\frac{d}{ds}\equiv D_{s}$ acting on the $%
4-$scalar $\mathcal{S}(g,\widehat{g},r,s;P)$ is performed keeping $g\equiv
\left\{ g^{\mu \nu }\right\} $ and $P\equiv \left\{ P^{\mu \nu }\right\} $
constant and therefore is necessarily prescribed according to Eqs.(\ref%
{covariant s-derivative})\ and (\ref{cov-scalar-2}). In addition, the
identities%
\begin{eqnarray}
\frac{\partial }{\partial g_{\mu \nu }}D_{s}\mathcal{S}\left( g,\widehat{g}%
,r,s;P\right) &=&\frac{d}{ds}\frac{\partial }{\partial g_{\mu \nu }}\mathcal{%
S}\left( g,\widehat{g},r,s;P\right) ,  \label{cov-1} \\
\frac{\partial }{\partial g^{_{\mu \nu }}}D_{s}\mathcal{S}\left( g,\widehat{g%
},r,s;P\right) &=&\frac{d}{ds}\frac{\partial }{\partial g^{_{\mu \nu }}}%
\mathcal{S}\left( g,\widehat{g},r,s;P\right) ,  \label{cov-2}
\end{eqnarray}%
hold respectively for the counter- and covariant components, where on the
lhs $D_{s}$ is identified with the operator (\ref{cov-scalar-1}). On the
other hand upon denoting $\frac{\partial }{\partial g_{\mu \nu }}\mathcal{S}%
(g,\widehat{g},r,s;P)\equiv \pi ^{\mu \nu }$ and $\frac{\partial }{\partial
g^{_{\mu \nu }}}\mathcal{S}(g,\widehat{g},r,s;P)\equiv \pi _{\mu \nu }$ (and
identifying respectively $\pi ^{\mu \nu }\equiv x^{\mu \nu }$\ and $\pi
_{\mu \nu }\equiv x_{\mu \nu }$\ in Eqs.(\ref{A-3}) of Appendix B), it is
obvious that in order to preserve the correct covariance properties of the
previous equations the operator $\frac{d}{ds}$\ appearing on the rhs of Eqs.(%
\ref{cov-1}) and (\ref{cov-2}) now must coincide with the covariant $s-$%
derivative acting on the counter- and covariant components of a second-order
$4-$tensor respectively. Therefore upon identifying $\frac{\partial }{%
\partial g_{\mu \nu }}\mathcal{S}\left( g,\widehat{g},r,s;P\right) \equiv
\pi ^{\mu \nu }$ the second equation in the GR-Hamilton equations (\ref%
{canonical evolution equations -2}) is found to be implied by the classical
Hamilton-Jacobi equation. The construction of the corresponding PDE for $%
\frac{dg_{\mu \nu }}{ds}$ is straightforward and analogous to that given in
Ref.\cite{cqg-3}, thus implying the equivalence between the GR-Hamilton
equations and the GR-Hamilton-Jacobi equation. The consequence is therefore
that the classical Hamiltonian structure $\left\{ x_{R},H_{R}\right\} $
remains preserved also in the case of a non-stationary background metric
tensor (\ref{Eq-2a}).

\subsection{4B - GR - Hamilton-Jacobi quantization}

Based on the validity of the classical GR-Hamilton equations as well the
corresponding classical GR-Hamilton-Jacobi equation, it is now formally
straightforward to carry out the analogous extension for covariant quantum
gravity. The conclusion follows at once adopting the quantization approach
developed in Ref.\cite{cqg-5}, \textit{i.e.}, achieved by means of the
so-called\textbf{\ }\emph{Hamilton-Jacobi }$g-$\emph{quantization. }In
detail, this is realized through the\emph{\ }mapping%
\begin{eqnarray}
&&\left. g_{\mu \nu }\rightarrow g_{\mu \nu }^{(q)}\equiv g_{\mu \nu
},\right.   \label{MAP-1} \\
&&\left. \pi _{\mu \nu }\equiv \frac{\partial \mathcal{S}(g,\widehat{g}%
,r,s;P)}{\partial g^{\mu \nu }}\rightarrow \pi _{\mu \nu }^{(q)}\equiv
-i\hbar \frac{\partial }{\partial g^{\mu \nu }},\right.   \label{MAP-1B} \\
&&\left. p\equiv -\frac{\partial \mathcal{S}(g,\widehat{g},r,s;P)}{\partial s%
}\rightarrow p^{(q)}\equiv -i\hbar \frac{d}{ds},\right.   \label{MPA-2} \\
&&\left. H_{R}\left( g,\frac{\partial \mathcal{S}(g,\widehat{g},r,s;P)}{%
\partial g},\widehat{g}(s),r,s\right) \rightarrow H_{R}^{(q)},\right.
\label{MAP-3}
\end{eqnarray}%
with $g_{\mu \nu }^{(q)},\pi _{\mu \nu }^{(q)},p^{(q)}$ and $H_{R}^{(q)}$
denoting the corresponding quantum fields/operators. Accordingly, $\pi _{\mu
\nu }^{(q)},$ $p^{(q)}$ denote the quantum canonical momenta conjugate to $%
g_{\mu \nu }^{(q)}\equiv g_{\mu \nu }$ and $s$ respectively, while%
\begin{eqnarray}
H_{R}^{(q)} &\equiv &T_{R}^{(q)}(\pi ,\widehat{g})+V, \\
T_{R}^{(q)}(\pi ,\widehat{g}) &=&\frac{1}{2\alpha L}\left( -i\hbar \frac{%
\partial }{\partial g^{\mu \nu }}\right) \left( -i\hbar \frac{\partial }{%
\partial g_{^{\mu \nu }}}\right) ,
\end{eqnarray}%
are the quantum Hamiltonian operator (with $V$\ being the effective
potential prescribed according to the second equation of Eq.(\ref{App-1})
given in Appendix A) and the quantum effective kinetic energy operator. The
mapping realized by Eqs.(\ref{MAP-1})-(\ref{MAP-3}) implies the simultaneous
validity of the two fundamental commutator relations
\begin{equation}
\left[ \pi ^{(q)\alpha \beta },g_{\mu \nu }\right] =-i\hslash \delta _{\mu
}^{\alpha }\delta _{\nu }^{\beta },  \label{FIRST COMMUTATOR}
\end{equation}%
\begin{equation}
\left[ p^{(q)},s\right] =-i\hslash ,
\end{equation}%
together with%
\begin{equation}
\left[ g^{\alpha \beta },g_{\mu \nu }\right] =\left[ \pi ^{(q)\alpha \beta
},\pi _{\mu \nu }^{(q)}\right] =0.
\end{equation}%
Here we notice that since both $\pi ^{(q)\alpha \beta }$\ and $g_{\mu \nu
}^{(q)}$\ are symmetric, Eq.(\ref{FIRST COMMUTATOR}) holds for arbitrary
permutations of the indexes. As a consequence, based on the classical
Hamilton-Jacobi equation (\ref{classical Hamilton-Jacobi}) also the mapping
\begin{equation}
\frac{d\mathcal{S}}{ds}+H_{R}=0\Rightarrow \left\{
p^{(q)}+H_{R}^{(q)}\right\} \psi \left( s\right) =0
\end{equation}%
necessarily applies. This warrants the validity of the quantum-wave equation%
\emph{\ }%
\begin{equation}
i\hslash \frac{d}{ds}\psi (s)=H_{R}^{(q)}\psi (s),  \label{QG-WAVW EQUATION}
\end{equation}%
to be denoted as \emph{CQG-wave equation}, with $\frac{d}{ds}$\ denoting
again the total covariant $s-$derivative in Eulerian form defined by Eq.(\ref%
{cov-scalar-1}). According to the notations of Ref.\cite{cqg-4}, and
omitting possible implicit contributions, here $\psi (g,s)\equiv \psi (g,%
\widehat{g},r,s),$\ with $r=r(s),$ denotes, in principle for arbitrary $s$\
belonging to the time axis $I\equiv
\mathbb{R}
,$\ the $4-$scalar quantum wave function associated with a graviton
particle. Furthermore, $g=\left\{ g_{\mu \nu }\right\} $\ is the quantum
generalized-coordinate field which spans the $10-$dimensional real vector
space $U_{g}\subseteq
\mathbb{R}
^{10}$\ of the same wave-function, \textit{i.e.}, the set on which the
associated quantum probability density function $\rho (g,s)=\left\vert \psi
(g,s)\right\vert ^{2}$ (\emph{quantum PDF}) is prescribed.

One notices that, as shown in Ref.\cite{cqg-4}, the CQG-wave equation (\ref%
{QG-WAVW EQUATION}) can be represented in terms of an equivalent set of
quantum hydrodynamic equations \cite{cqg-4,cqg-5}. This requires the
adoption of the Madelung representation%
\begin{equation}
\psi (g,\widehat{g},r,s)=\sqrt{\rho (g,\widehat{g},r,s)}\exp \left\{ \frac{i%
}{\hslash }\mathcal{S}^{(q)}(g,\widehat{g},r,s)\right\} ,  \label{Madelung}
\end{equation}%
where the quantum fluid fields $\left\{ \rho ,S^{(q)}\right\} \equiv \left\{
\rho (g,\widehat{g},r,s),\mathcal{S}^{(q)}(g,\widehat{g},r,s)\right\} $
identify respectively the $4-$scalar quantum PDF and quantum phase-function.
Elementary algebra then shows that based on Eq.(\ref{QG-WAVW EQUATION}) the
same quantum fluid fields must satisfy the set of GR-quantum hydrodynamic
equations (CQG-QHE) realized respectively by continuity and quantum
Hamilton-Jacobi equations. Written again in Eulerian form these are given by%
\begin{eqnarray}
\frac{d\rho }{ds}+\frac{\partial }{\partial g_{\mu \nu }}\left( \rho V_{\mu
\nu }\right) &=&0,  \label{cont-1} \\
\frac{d\mathcal{S}^{(q)}}{ds}+H^{(q)} &=&0.  \label{quantum H-J}
\end{eqnarray}%
Here in both equations, according to the notation (\ref{cov-scalar-1}), $%
\frac{d}{ds}\equiv D_{s}$ \cite{cqg-4}. Thus, in analogy with the classical
phase-function\textbf{\ }$\mathcal{S}(g,\widehat{g},r,s)$\textbf{\ }(see
Eqs.(\ref{AA-CON-2})-(\ref{AA-CON-3})) the quantum phase-function $\mathcal{S%
}^{(q)}$\ is taken of the form\textbf{\ }%
\begin{equation}
\mathcal{S}^{(q)}(g,\widehat{g},r,s)\equiv \mathcal{S}^{(q)}(g,\widehat{g}%
,r,s;P).  \label{B-CON-1}
\end{equation}%
In particular this means that $\mathcal{S}^{(q)}$ should\ depend smoothly
on: 1) the initial tensor field $P\equiv \left\{ P^{\mu \nu }\right\} $\ to
be considered independent of $(g,\widehat{g},r,s)$\ and constant in the
sense $D_{s}P\equiv 0;$\ 2) the variables $(g,\widehat{g},r,s).$\textbf{\ }%
In addition, by assumption $\mathcal{S}^{(q)}$\ is required to satisfy the
constraint and regularity conditions determined respectively by:
\begin{equation}
\left\{
\begin{array}{c}
Q_{\mu \nu }=\frac{\partial \mathcal{S}^{(q)}(g,\widehat{g},r,s;P)}{\partial
P^{\mu \nu }}, \\
\left\vert \frac{\partial \mathcal{S}^{(q)2}(g,\widehat{g},r,s;P)}{\partial
g_{\mu \nu }\partial P^{\mu \nu }}\right\vert \neq 0.%
\end{array}%
\right.  \label{B-CON-2}
\end{equation}%
Here again\textbf{\ }$\left\vert \frac{\partial \mathcal{S}^{(q)2}(g,%
\widehat{g},r,s;P)}{\partial g_{\mu \nu }\partial P^{\mu \nu }}\right\vert $%
\textbf{\ }denotes the determinant of the matrix $\left\{ \frac{\partial
\mathcal{S}^{(q)2}(g,\widehat{g},r,s;P)}{\partial g_{\mu \nu }\partial
P^{\mu \nu }}\right\} $\ while both $P^{\mu \nu }$and $Q_{\mu \nu }$\ are
assumed to be constant phase functions, \textit{i.e.}, such that $%
D_{s}P^{\mu \nu }\equiv 0$ and $D_{s}Q_{\mu \nu }\equiv 0.$

Furthermore,\textbf{\ }$V_{\mu \nu }\equiv V_{\mu \nu }(g,s)$ and $%
H^{(q)}\equiv H^{(q)}(g,s)$ denote respectively the\ quantum $4-$tensor
velocity field identified with%
\begin{equation}
V_{\mu \nu }=\frac{1}{\alpha L}\frac{\partial \mathcal{S}^{(q)}}{\partial
g^{\mu \nu }},  \label{V_mu-nu}
\end{equation}%
and the effective quantum Hamiltonian density%
\begin{equation}
H^{(q)}=\frac{1}{2\alpha L}\frac{\partial \mathcal{S}^{(q)}}{\partial g^{\mu
\nu }}\frac{\partial \mathcal{S}^{(q)}}{\partial g_{\mu \nu }}+V_{QM}+V,
\label{effective Hamiltoniian density}
\end{equation}%
with\ $V\equiv V(g,s)$ being the effective potential defined according to
Eq.(\ref{App-1}) and $V_{QM}\equiv V_{QM}(g,s)$\ being the Bohm effective
quantum potential \cite{bohm1,bohm2,bohm3} given by%
\begin{equation}
V_{QM}\equiv \frac{\hslash ^{2}}{8\alpha L}\frac{\partial \ln \rho }{%
\partial g^{\mu \nu }}\frac{\partial \ln \rho }{\partial g_{\mu \nu }}-\frac{%
\hslash ^{2}}{4\alpha L}\frac{\partial ^{2}\rho }{\rho \partial g_{\mu \nu
}\partial g^{\mu \nu }},  \label{BOHM potential}
\end{equation}%
with $\rho \equiv \rho (g,\widehat{g},r,s)\equiv $ $\left\vert \psi (g,%
\widehat{g},r,s)\right\vert ^{2}$ being the $4-$scalar quantum PDF. Eq.(\ref%
{QG-WAVW EQUATION}) is therefore manifestly covariant also in such a case.
As such it retains its form under the action of arbitrary local point
transformations which preserve the differential manifold of space-time. As
such the same equation is appropriate for the treatment of problems of
quantum gravity and quantum cosmology also in such extended framework.

We conclude that the CQG-wave equation (\ref{QG-WAVW EQUATION}) and the
equivalent set of CQG-QHE are both manifestly covariant also in validity of
a non-stationary background metric tensor of the type (\ref{Eq-2a}).
Therefore they both hold also in such a case.

\section{5 - Hamiltonian representation of the CQG-quantum hydrodynamics
equations}

In this section we intend to uncover a novel feature of CQG-theory not
previously pointed out.

This concerns the Hamiltonian structure, which in analogy to nonrelativistic
quantum mechanics \cite{FoP-2016,FoP-2016b}, is associated with the quantum
wave equation (\textit{i.e.}, the CQG-wave equation (\ref{QG-WAVW EQUATION}%
)) and the corresponding quantum Hamilton-Jacobi equation (\ref{quantum H-J}%
).\textbf{\ }More precisely, we intend to show that such an equation is
actually equivalent to a corresponding set of manifestly-covariant quantum
Hamilton equations, thus establishing "\textit{de facto}" a Hamiltonian
structure analogous to that holding for the\ classical GR-Hamilton
equations. Such a structure not only lies at the basis of CQG-theory
developed in Refs.\cite{cqg-3,cqg-4,cqg-5,cqg-6} but - as shown in Section 4
- remains also preserved under the extended setting considered in the
present paper. Given the equivalence between the classical GR equations and
the corresponding GR-Hamilton-Jacobi equation established in Section 4 (see
also the analogous one first pointed out in Ref.\cite{cqg-3} in the case of
stationary background) such a result is not surprising. Nevertheless, it is
worth stressing its unique and peculiar feature which distinguishes
CQG-theory from other previous non-manifestly covariant quantum theories of
gravity, like the Wheeler-DeWitt equation \cite{wheeler-1}.

In particular, the goal of this section is to display the quantum Hamilton
equations and corresponding Hamiltonian structure represented by a set $%
\left\{ x,H^{(q)}\right\} $ which are associated respectively with an
appropriate $4-$tensor canonical state $x\equiv (g_{\mu \nu },\Pi ^{\mu \nu
})$ and the$\ 4-$scalar effective quantum Hamiltonian density $H^{(q)}.$ As
shown below, here the second-order canonical $4-$tensor momentum $\Pi ^{\mu
\nu }$ will be identified with%
\begin{equation}
\Pi ^{\mu \nu }=\frac{\partial \mathcal{S}^{(q)}(g,\widehat{g},r,s;P)}{%
\partial g_{\mu \nu }},  \label{B-EQQ-1}
\end{equation}%
$\mathcal{S}^{(q)}(g,\widehat{g},r,s;P)$ being the quantum phase function of
CQG-theory and $H^{(q)}$ the$\ $effective quantum Hamiltonian density
prescribed according to Eq.(\ref{effective Hamiltoniian density}). In the
current notation the latter is written as%
\begin{equation}
H^{(q)}(g,\Pi ,\widehat{g}(r,s),r,s)=\frac{1}{2\alpha L}\Pi ^{\mu \nu }\Pi
_{\mu \nu }+V_{QM}+V.  \label{B-EQQ-2}
\end{equation}%
Then the following proposition holds.

\bigskip

\textbf{THM. 1 - Quantum Hamilton equations}

\emph{In validity of the CQG-wave equation (\ref{QG-WAVW EQUATION}) and the
corresponding quantum Hamilton-Jacobi equation (\ref{quantum H-J}), upon
denoting }$H(g,\Pi ,r,s)\equiv H^{(q)}(g,\Pi ,\widehat{g},r,s)$\emph{, the
canonical state }$x\equiv (g_{\mu \nu },\Pi ^{\mu \nu })$ \emph{satisfies
identically the set of manifestly-covariant equations}%
\begin{equation}
\left\{
\begin{array}{c}
\frac{d}{ds}g^{\mu \nu }=\frac{\partial }{\partial \Pi _{\mu \nu }}H(g,\Pi
,r,s), \\
\frac{d}{ds}\Pi _{\mu \nu }=-\frac{\partial }{\partial g^{\mu \nu }}H(g,\Pi
,r,s),%
\end{array}%
\right.  \label{B-EQQ-3}
\end{equation}%
\emph{subject to the initial conditions}%
\begin{equation}
\left\{
\begin{array}{c}
g_{\mu \nu }(s_{1})\equiv g_{\mu \nu }^{(o)}(r(s_{1}),s_{1}), \\
\Pi ^{\mu \nu }(s_{1})\equiv \Pi _{\left( o\right) }^{\mu \nu
}(r(s_{1}),s_{1}).%
\end{array}%
\right.  \label{B-EQQ-4}
\end{equation}%
\emph{Eqs.(\ref{B-EQQ-3})-(\ref{B-EQQ-4}) are referred to here as Quantum
Hamilton equations.}

\emph{Proof - }To prove the thesis one notices preliminarily that in
validity of Eqs.(\ref{B-EQQ-1}) and (\ref{B-EQQ-2}), Eqs.(\ref{B-EQQ-3})-(%
\ref{B-EQQ-4}) are realized by means of the set of equations%
\begin{eqnarray}
\frac{d}{ds}g^{\mu \nu } &=&\frac{\Pi ^{\mu \nu }}{\alpha L},
\label{QH-1-fin} \\
\frac{d}{ds}\Pi _{\mu \nu } &=&-\frac{\partial (V_{QM}+V)}{\partial g^{\mu
\nu }},  \label{QH-2-fin}
\end{eqnarray}%
to be solved subject to initial conditions of the form%
\begin{equation}
\left\{
\begin{array}{c}
g^{\mu \nu }(s_{o})=g_{(o)}^{\mu \nu }(r(s_{o}),s_{o}), \\
\Pi _{\mu \nu }(s_{o})=\Pi _{\mu \nu }^{\left( o\right) }(r(s_{o}),s_{o}).%
\end{array}%
\right.   \label{QH-3-fin}
\end{equation}%
The proof of the canonical equations (\ref{B-EQQ-3})-(\ref{B-EQQ-4}) is
actually analogous to that reached in Section 4. Thus, in particular,
letting $H(g,\Pi ,r,s)\equiv H^{(q)}(g,\Pi ,\widehat{g},r,s)$, the second
equation is obtained in two steps: first, by partial differentiation of the
quantum Hamilton-Jacobi equation (\ref{quantum H-J})\ with respect to $%
g^{\mu \nu }$, while letting again $\widehat{g}_{\mu \nu }(s)$ as constant,
namely%
\begin{eqnarray}
&&\left. \frac{\partial }{\partial g^{\mu \nu }}H\left( g,\Pi ,r,s\right)
+\right.   \notag \\
&&\left. \frac{d}{ds}\left[ \frac{\partial \mathcal{S}^{(q)}(g,\widehat{g}%
,r,s;P)}{\partial g^{\mu \nu }}\right] =0,\right.
\end{eqnarray}%
and second noting that $\frac{\partial }{\partial s}\left[ \frac{\partial
\mathcal{S}^{(q)}(g,\widehat{g},r,s;P)}{\partial g^{\mu \nu }}\right] \equiv
\frac{d}{ds}\Pi _{\mu \nu }$. In a similar way, by evaluating the partial
derivative with respect to $\frac{\partial \mathcal{S}^{(q)}(g,\widehat{g}%
,r,s;P)}{\partial g^{\mu \nu }}\equiv \Pi _{\mu \nu }$ and keeping $\widehat{%
g}_{\mu \nu }(s)$, $g^{\mu \nu }$ and $r^{\mu }$ as constants, gives%
\begin{eqnarray}
&&\left. \frac{\partial }{\partial \Pi _{\mu \nu }}H(g,\Pi ,r,s)+\right.
\notag \\
&&\left. \frac{\partial }{\partial \Pi _{\mu \nu }}\frac{\partial \mathcal{S}%
^{(q)}(g,\widehat{g},r,s;P)}{\partial s}=0.\right.
\end{eqnarray}%
Then, the proof of Eq.(\ref{B-EQQ-1}) follows by invoking the identity%
\begin{gather}
\frac{\partial }{\partial \frac{\partial \mathcal{S}^{(q)}(g,s)}{\partial
g^{\mu \nu }}}\left. \frac{\partial \mathcal{S}^{(q)}(g,\widehat{g},r,s;P)}{%
\partial s}\right\vert _{g^{\mu \nu }}=  \notag \\
\frac{\partial }{\partial \frac{\partial \mathcal{S}^{(q)}(g,s)}{\partial
g^{\mu \nu }}}\frac{\partial \mathcal{S}^{(q)}(g,\widehat{g},r,s;P)}{%
\partial s}-  \notag \\
-\frac{d}{ds}g^{\beta \gamma }\frac{\partial }{\partial \frac{\partial
\mathcal{S}^{(q)}(g,s)}{\partial g^{\mu \nu }}}\frac{\partial \mathcal{S}%
^{(q)}(g,\widehat{g},r,s;P)}{\partial g^{\beta \gamma }},  \label{prev}
\end{gather}%
where $\frac{\partial }{\partial \frac{\partial \mathcal{S}^{(q)}(g,\widehat{%
g},r,s;\beta )}{\partial g^{\mu \nu }}}\frac{\partial \mathcal{S}^{(q)}(g,%
\widehat{g},r,s;P)}{\partial g^{\beta \gamma }}=\delta _{\beta \gamma }^{\mu
\nu }$ and $\delta _{\beta \gamma }^{\mu \nu }\equiv \delta _{\beta }^{\mu
}\delta _{\gamma }^{\nu }.$ Notice that, since the first term on the rhs of
Eq.(\ref{prev}) vanishes identically, here $\frac{\partial \mathcal{S}%
^{(q)}(g,\widehat{g},r,s;P)}{\partial s}$ must be considered as independent
of $\Pi _{\mu \nu }$, because they represent different canonical momenta.
Finally, one notices that the dependence of the quantum phase-function $%
\mathcal{S}^{(q)}$\ in terms of $P\equiv \left\{ P^{\mu \nu }\right\} $\
remains still indeterminate. This\ means that it can still be prescribed in
such a way to satisfy identically the constraint equation%
\begin{equation}
\frac{\partial \mathcal{S}^{(q)}(g,\widehat{g},r,s;P)}{\partial P^{\mu \nu }}%
\equiv Q_{(o)\mu \nu }(r(s_{o})),
\end{equation}%
with\ $Q_{(o)\mu \nu }(r(s_{o}))$\ being the (still arbitrary) initial
constant phase function prescribed in analogy to Eqs.(\ref{QH-3-fin}). As a
consequence the $4-$scalar quantum hydrodynamic equation (\ref{quantum H-J}%
), in analogy with the GR-Hamilton-Jacobi equation discussed above in
Section 4 (see Eq.(\ref{classical Hamilton-Jacobi})), can indeed be
interpreted in a proper sense as a Hamilton-Jacobi equation, \textit{i.e.},
as generating a corresponding set of Hamilton equations. \textbf{Q.E.D.}

\section{6 - Quantum modified Einstein field equations}

Concerning the quantum Hamilton equations (\ref{B-EQQ-3}) the fundamental
question to be answered is whether they actually admit a particular
realization which is analogous to the classical Einstein equations (\ref%
{EINSTEIN FIELD EQS}), for a suitable choice of the initial conditions (\ref%
{B-EQQ-4}) and in close analogy with the classical GR-Hamilton equations (%
\ref{canonical evolution equations -2}) for which such a property was first
pointed out in Ref.\cite{cqg-3}. Being based on the quantum Hamilton
equations (\ref{B-EQQ-3}), the tensor components of such an equation will be
referred to as \emph{quantum-modified Einstein field equations}.

Given the formal analogy of the two sets of Hamiltonian equations, \textit{%
i.e.}, (\ref{canonical evolution equations -2}) and (\ref{B-EQQ-3}), both
holding in validity of the extended functional setting (\ref{Eq-2a}), the
following result holds.

\bigskip

\textbf{THM. 2 - Quantum-modified Einstein field equations}

\emph{Let us assume validity of the initial-value problem represented by the
quantum Hamilton equations (\ref{B-EQQ-3}) and the initial conditions (\ref%
{B-EQQ-4}). For this purpose let us impose that the initial conditions are
prescribed requiring}%
\begin{equation}
\left\{
\begin{array}{c}
g_{\mu \nu }(r(s_{o}),s_{o})=\widehat{g}_{\mu \nu }(r(s_{o}),s_{o}), \\
\Pi ^{\mu \nu }(r(s_{o}),s_{o})\equiv 0,%
\end{array}%
\right.  \label{initial conditions}
\end{equation}%
\emph{("extremal" initial conditions) and that at the initial proper-time }$%
s_{o},$\emph{\ }$\widehat{g}_{\mu \nu }(r(s),s)$\emph{\ is solution of the
PDE}%
\begin{equation}
\left. \frac{\partial }{\partial g^{\mu \nu }}\left[ V\left( g,\widehat{g}%
,r,s\right) +V_{QM}(g,s)\right] \right\vert _{\substack{ g=\widehat{g}  \\ %
s=s_{o}}}=0  \label{initial PDE}
\end{equation}%
\emph{(constraint equation). Then the following propositions hold:}

\emph{P2}$_{1}$\emph{) The quantum Hamilton equations in the general case of
non-vacuum configuration reduce to the single extremal tensor equation}%
\begin{equation}
\left. \frac{\partial }{\partial g^{\mu \nu }}\left[ V\left( g,\widehat{g}%
,r,s\right) +V_{QM}(g,s)\right] \right\vert _{g=\widehat{g}}=0.
\label{extremal equation}
\end{equation}

\emph{P2}$_{2}$\emph{) In the case of vacuum, namely letting}%
\begin{equation}
V\left( g,\widehat{g},r,s\right) =V_{o}\left( g,\widehat{g},r,s\right) ,
\label{zxc}
\end{equation}%
\emph{and setting }$\Lambda =\Lambda _{bare}$ \emph{in Eq.(\ref{App-2}), the
extremal tensor equation (\ref{extremal equation}) becomes}%
\begin{equation}
\widehat{R}_{\mu \nu }-\frac{1}{2}\left[ \widehat{R}-2\Lambda _{bare}\right]
\widehat{g}_{\mu \nu }(r,s)=B_{\mu \nu }(r,s),
\label{quantum-modified Einstein eq}
\end{equation}%
\emph{which identifies the quantum-modified Einstein equation, where }$%
B_{\mu \nu }$\emph{, prescribed in terms of the Bohm interaction potential }$%
V_{QM}$\emph{\ as}%
\begin{equation}
B_{\mu \nu }(r,s)\equiv -\frac{1}{\alpha L}\left. \frac{\partial }{\partial
g^{\mu \nu }}V_{QM}(g,s)\right\vert _{g=\widehat{g}\left( r,s\right) },
\label{Bohm source term}
\end{equation}%
\emph{is referred to as\ Bohm source tensor field.}

\emph{Proof - }To reach the proof of P2$_{1}$ we first obtain an equivalent
explicit representation for the quantum Hamilton equation (\ref{B-EQQ-3})
holding in case of validity of the extended representation (\ref{Eq-2a}).
This follows thanks to Eqs.(\ref{A-3bis}) and the prescription of the
covariant $s-$ derivative given by Eq.(\ref{covariant s-derivative}). Thus
one finds that the same equations become:%
\begin{equation}
\left\{
\begin{array}{c}
D_{s}g^{\mu \nu }-g_{\alpha }^{\mu }\frac{\partial }{\partial s}(\widehat{g}%
^{\nu \alpha })=\frac{\Pi ^{\mu \nu }}{\alpha L}, \\
D_{s}\Pi _{\mu \nu }-\Pi _{\mu }^{\alpha }\frac{\partial }{\partial s}(%
\widehat{g}_{\nu \alpha })= \\
-\frac{\partial }{\partial g^{\mu \nu }}\left[ V\left( g,\widehat{g}%
,r,s\right) +V_{QM}(g,s)\right] ,%
\end{array}%
\right.
\end{equation}%
where%
\begin{equation}
\left\{
\begin{array}{c}
g_{\alpha }^{\mu }\frac{\partial }{\partial s}(\widehat{g}^{\nu \alpha
})=g_{\alpha }^{\mu }D_{s}(\widehat{g}^{\nu \alpha }), \\
\Pi _{\mu }^{\alpha }\frac{\partial }{\partial s}(\widehat{g}_{\nu \alpha
})=\Pi _{\mu }^{\alpha }D_{s}(\widehat{g}_{\nu \alpha }).%
\end{array}%
\right.
\end{equation}%
Hence, straightforward algebra delivers%
\begin{equation}
\left\{
\begin{array}{c}
\widehat{g}^{\mu \alpha }D_{s}g_{\alpha }^{\nu }=\frac{\Pi ^{\mu \nu }}{%
\alpha L}, \\
\widehat{g}_{\nu \alpha }D_{s}\Pi _{\mu }^{\alpha }= \\
-\frac{\partial }{\partial g^{\mu \nu }}\left[ V\left( g,\widehat{g}%
,r,s\right) +V_{QM}(g,s)\right] ,%
\end{array}%
\right.
\end{equation}%
implying, in turn, that the following equivalent explicit representation
must hold for the same equations%
\begin{equation}
\left\{
\begin{array}{c}
D_{s}g_{\alpha }^{\nu }=\frac{\Pi _{\alpha }^{\nu }}{\alpha L}, \\
D_{s}\Pi _{\nu }^{\alpha }=-\frac{\partial }{\partial g_{\alpha }^{\nu }}%
\left[ V\left( g,\widehat{g},r,s\right) +V_{QM}(g,s)\right] .%
\end{array}%
\right.  \label{C-EQ-1}
\end{equation}%
As a second step one notices that the requirement that
\begin{equation}
\forall s\in I:g_{\alpha }^{\nu }(r(s),s)=\widehat{g}_{\alpha }^{\nu
}(r(s),s)  \label{proposition}
\end{equation}%
is obviously equivalent to require that Eq.(\ref{extremal equation}) must
hold identically. The implication, however, is that to reach the thesis it
is actually necessary to prove that the initial conditions (\ref{initial
conditions}) are actually equivalent to the validity of proposition (\ref%
{proposition}). That this is indeed the case follows in fact by direct
inspection of Eqs.(\ref{C-EQ-1}), since the constraint equation (\ref%
{initial PDE}) implies that
\begin{equation}
\left. D_{s}\Pi _{\nu }^{\alpha }\right\vert _{s=s_{o}}=0,
\end{equation}%
while the initial conditions (\ref{initial conditions}) warrant that $\left.
\Pi _{\nu }^{\alpha }\right\vert _{s=s_{o}}=0$ too. Hence, for all $s\geq
s_{o}$ it is identically vanishing as $D_{s}g_{\alpha }^{\nu },$ which means
that the proposition (\ref{proposition}) is necessarily true. As a
consequence the validity of\ the extremal tensor equation (\ref{extremal
equation}) remains warranted too. The proof of P2$_{2}$ then follows by
elementary algebra once the condition (\ref{zxc}) is imposed in Eq.(\ref%
{extremal equation}), which yields as well the representation of the Bohm
source tensor\ field $Q_{\mu \nu }(s)$. \textbf{Q.E.D.}

\bigskip

The main implication arising from THM.2 is the appearance of a quantum
contribution to the stress energy tensor, denoted here $\widehat{T}_{\mu \nu
}^{(q)}.$ In the context of CQG-theory this is prescribed in terms of the
Bohm source tensor field $B_{\mu \nu }$, so that
\begin{equation}
B_{\mu \nu }\equiv \kappa \widehat{T}_{\mu \nu }^{(q)},
\label{quantu energy-momentum tensor}
\end{equation}%
with $\kappa $ being again the universal constant (\ref{UNIVERSAL CONSTANT}%
). The tensor field $B_{\mu \nu }$ is\ therefore ascribed to the\ effect of
\emph{Bohm vacuum self-interaction acting on gravitons, }and more precisely,
as shown below (see Eq.(\ref{vacuum energy density})), to corresponding
\emph{quantum vacuum energy fluctuations produced by gravitons.} The
contribution associated with $B_{\mu \nu }$\ defined by Eq.(\ref{quantu
energy-momentum tensor}) in the quantum-modified Einstein field equations (%
\ref{quantum-modified Einstein eq}) realizes a space-time
second-quantization effect, namely a quantum correction to the classical
metric tensor arising from a non-linear quantum interaction of the
gravitational field with itself. According to this picture, the background
metric tensor solution of Eq.(\ref{quantum-modified Einstein eq}) is
affected by quantum corrections, and this distinguishes the
second-quantization framework from the first-quantization one, where instead
the background metric tensor is prescribed to be a purely-classical
space-time field tensor.

One notices that Eq.(\ref{quantum-modified Einstein eq}) coincides
functionally with the\ set (\ref{EINSTEIN FIELD EQS}) of classical Einstein
equations holding for the components of the background field tensor.\ Hence,
Eq.(\ref{quantum-modified Einstein eq}) will be referred to as \emph{%
quantum-modified Einstein field equations}. Indeed, the two equations
coincide once the tensor field $T_{\mu \nu }$ is replaced with the quantum
field $B_{\mu \nu }(s)$. The notable difference arising in Eqs.(\ref%
{quantum-modified Einstein eq}) lies\textbf{\ }in the non-stationary
character of the quantum source term, \textit{i.e.}, its explicit dependence
on proper-time. Therefore in the present context all the tensor fields,
including the background metric tensor $\widehat{g}\equiv \left\{ \widehat{g}%
_{\mu \nu }\right\} ,$ the Ricci tensor $\widehat{R}_{\mu \nu }$\ and the
Ricci $4-$scalar $\widehat{R}\equiv $\ $\widehat{g}^{\alpha \beta }(r,s)%
\widehat{R}_{\alpha \beta }$\ are necessarily to be considered as
non-stationary too.

Nevertheless, an additional difficulty arises due to the intrinsic quantum
origin of the Bohm source tensor field $Q_{\mu \nu }$, for which the
prescription of the quantum probability density $\rho \equiv \rho (g,%
\widehat{g},r,s)$ is needed. Its determination requires in fact the explicit
solution of the CQG-wave equation (\ref{QG-WAVW EQUATION}) in a highly
non-linear second quantization picture (\textit{i.e.}, in which also the
background field tensor must be consistently evaluated). In fact, the
crucial issue is that the latter must be consistently determined by means of
the quantum-modified Einstein field equations themselves.\ To unfold these
tasks and achieve the construction of explicit solutions of the CQG-wave
equation the Generalized Lagrangian Path Approach (GLP-approach) recently
developed in Ref.\cite{cqg-6} will be adopted. For this purpose, as a
starting point, the generalization of the GLP-approach to the extended
functional setting adopted here is performed.

\section{7 - GLP-approach in the extended functional setting}

In this section we show that also the GLP-approach developed in Ref.\cite%
{cqg-6} remains valid for the extended functional setting (\ref{Eq-2a}) and
actually fulfills the property of manifest covariance also in such a case.
Let us start recalling, for this purpose, that the GLP-approach crucially
depends on the notion of Generalized Lagrangian Path (GLP), \textit{i.e.},
the integral curve $\left\{ \delta G_{L}(s)\equiv \delta G_{L\mu \nu
}(r(s),s),\forall s\in I\right\} $ which is determined by the
GLP-initial-value problem%
\begin{equation}
\left\{
\begin{array}{c}
\frac{d}{ds}\delta G_{L\mu \nu }(s)=V_{\mu \nu }(G_{L}(s),\Delta g,r(s),s),
\\
\delta G_{L\mu \nu }(s_{1})=\delta g_{L\mu \nu }(s_{1})-\Delta g_{\mu \nu
}(s_{1}), \\
\delta g_{L\mu \nu }(s_{1})\equiv \delta g_{L\mu \nu }^{(o)}.%
\end{array}%
\right.  \label{GLP-initial-value problem}
\end{equation}%
More precisely, here $\delta g_{L\mu \nu }(s)$\textbf{\ }and $\delta G_{L\mu
\nu }(s)$\ identify respectively, according to Ref.\cite{cqg-6}, the
deterministic Lagrangian Path (LP) and stochastic Generalized Lagrangian
Path (GLP). We remark that in the extended functional setting $\frac{d}{ds}$
identifies the covariant $s-$derivative (\ref{covariant s-derivative}) while
$r(s)$\ denotes again, also in such a context, an arbitrary geodesic
trajectory with $s_{1}\geq s_{o}$\ being in principle an arbitrary initial
proper-time along it. In particular in the following $r(s)$\ and $s_{1}$\
can always be identified respectively with a\ maximal observer's geodetics
and, upon requiring
\begin{equation}
s_{1}=s_{o}=0,
\end{equation}%
with the Big Bang proper-time (\ref{minimum proper time}). Furthermore,
according to the notations of Ref.\cite{cqg-6} and consistent with Eq.(\ref%
{V_mu-nu}), $\delta g_{\mu \nu }^{(o)}$ and $V_{\mu \nu }(G_{L}(s),\Delta
g,r(s),s)$ denote respectively a deterministic initial tensor field and the
quantum tensor velocity field\textbf{\ }%
\begin{equation}
V^{\mu \nu }(G_{L}(s),\Delta g,r(s),s)=\frac{1}{\alpha L}\frac{\partial
\mathcal{S}^{(q)}(G_{L}(s),\Delta g,r(s),s;P)}{\partial \delta g_{L\mu \nu }}%
,
\end{equation}%
where $\Delta g\equiv \left\{ \Delta g_{\mu \nu }\right\} $\ identifies the
stochastic displacement $4-$tensor%
\begin{equation}
\Delta g_{\mu \nu }(s)\equiv \delta g_{L\mu \nu }(s)-\delta G_{L\mu \nu }(s).
\label{DISPLACEMENT 4-TENSOR}
\end{equation}

\subsection{7A - Formal solution of GLP in the extended functional setting}

It is obvious "\textit{a priori}" that the solution of Eqs.(\ref%
{GLP-initial-value problem}) must depend on the functional setting of the
background field tensor, \textit{i.e.}, again on the validity of the
requirement (\ref{Eq-2a}). Nevertheless a formal exact solution of the same
equations can directly be recovered also in such a case. In fact, upon
denoting respectively%
\begin{equation}
\left\{
\begin{array}{c}
\delta g_{L\nu }^{\alpha }(s)=\widehat{g}^{\mu \alpha }(r(s),s)\delta
g_{L\mu \nu }(r(s),s), \\
\delta G_{L\nu }^{\alpha }(s)=\widehat{g}^{\mu \alpha }(r(s),s)\delta
G_{L\mu \nu }(r(s),s), \\
\Delta g_{\nu }^{\alpha }(s)=\widehat{g}^{\mu \alpha }(r(s),s)\Delta g_{\mu
\nu }(s), \\
V_{\nu }^{\alpha }(G_{L}(s),\Delta g,s)=\widehat{g}^{\mu \alpha }(r,s)V_{\mu
\nu }(G_{L}(s),\Delta g,s),%
\end{array}%
\right.  \label{EQQ-1}
\end{equation}%
one can show that the following result applies.

\bigskip

\textbf{THM. 3 - Construction of a formal representation of GLP}

\emph{Regarding the GLP-initial-value problem (\ref{GLP-initial-value
problem}) the following propositions apply:}

\emph{P3}$_{1}$\emph{) Upon integration, Eq.(\ref{GLP-initial-value problem}%
) actually delivers formal exact solutions both for }$\delta g_{L\nu
}^{\alpha }(s)$\emph{\ and }$\delta G_{L\nu }^{\alpha }(s)$\emph{. These are
realized respectively by the initial-value problems}%
\begin{equation}
\left\{
\begin{array}{c}
\delta g_{L\nu }^{\alpha }(s)=\delta g_{L\nu }^{\alpha
}(s_{o})+\int_{s_{1}}^{s}ds^{\prime }V_{\nu }^{\alpha }(G_{L}(s^{\prime
}),\Delta g,r(s^{\prime }),s^{\prime }), \\
\delta g_{L\nu }^{\alpha }(s_{o})\equiv \delta g_{L\nu }^{\alpha (o)},%
\end{array}%
\right.  \label{EQQ-2A}
\end{equation}%
\emph{and}%
\begin{equation}
\left\{
\begin{array}{c}
\delta G_{L\nu }^{\alpha }(s)=\delta G_{L\nu }^{\alpha
}(s_{o})+\int_{s_{1}}^{s}ds^{\prime }V_{\nu }^{\alpha }(G_{L}(s^{\prime
}),\Delta g,r(s^{\prime }),s^{\prime }), \\
\delta G_{L\nu }^{\alpha }(s_{o})=\delta g_{L\nu }^{\alpha }(s_{o})-\Delta
g_{\nu }^{\alpha }(s_{o}), \\
\delta g_{L\nu }^{\alpha }(s_{o})\equiv \delta g_{L\nu }^{\alpha (o)},%
\end{array}%
\right.  \label{EQQ-2}
\end{equation}%
\emph{where}%
\begin{equation}
V_{\nu }^{\alpha }(G_{L}(s),\Delta g,r(s),s)=\frac{1}{\alpha L}\frac{%
\partial \mathcal{S}^{(q)}(G_{L}(s),\Delta g,r(s),s;P)}{\partial \delta
g_{L\alpha }^{\nu }(s)}.
\end{equation}%
\emph{These equations imply that by construction the stochastic displacement
}$4-$\emph{tensor }$\Delta g_{\nu }^{\mu }(s)$\emph{\ must be a constant,
\textit{i.e.}, such that for all }$s,s_{o}\in I$\emph{\ }%
\begin{equation}
\Delta g_{\nu }^{\mu }(s)=\Delta g_{\nu }^{\mu }(s_{o})\equiv \Delta g_{\nu
}^{\mu },  \label{CNSTRAINT-displacement}
\end{equation}%
\emph{while }$\delta g_{L\nu }^{\alpha }(s)$\emph{\ and }$\delta G_{L\nu
}^{\alpha }(s)$\emph{\ are related by means of the transformation}%
\begin{equation}
\delta G_{L\nu }^{\alpha }(s)=\delta g_{L\nu }^{\alpha }(s)-\Delta g_{\nu
}^{\alpha }.  \label{MAP}
\end{equation}

\emph{P3}$_{2}$\emph{) Eqs.(\ref{EQQ-2A}) and (\ref{EQQ-2}) can once again
be represented in terms of the covariant components }$\delta G_{L\mu \nu
}(s) $\emph{\ (and similarly }$\delta g_{L\mu \nu }(s)$\emph{) yielding}%
\begin{eqnarray}
&&\left. \delta G_{L\mu \nu }(s)=\widehat{g}_{\mu \alpha }(r(s),s)\delta
G_{L\nu }^{\alpha }(s_{o})+\right.  \notag \\
&&+\widehat{g}_{\mu \alpha }(r(s),s)\int_{s_{1}}^{s}ds^{\prime }V_{\nu
}^{\alpha }(G_{L}(s^{\prime }),\Delta g,r(s^{\prime }),s^{\prime }).
\label{EQQ-3}
\end{eqnarray}

\emph{P3}$_{3}$\emph{) Finally, in validity of Eq.(\ref{Eq-2b}) (stationary
background metric tensor), the same equations recover identically the form
determined previously in Ref.\cite{cqg-6}, namely}%
\begin{eqnarray}
&&\left. \delta g_{L\mu \nu }(s)=\delta g_{L\mu \nu }(s_{o})+\right.  \notag
\\
&&+\int_{s_{1}}^{s}ds^{\prime }V_{\mu \nu }(G_{L}(s^{\prime }),\Delta
g,r(s^{\prime }),s^{\prime }),  \label{EQQ-6}
\end{eqnarray}%
\emph{with }$\Delta g_{\mu \nu }(s)$\emph{\ being such that identically}%
\begin{equation}
\Delta g_{\mu \nu }(s)=\Delta g_{\mu \nu }(s_{o})\equiv \Delta g_{\mu \nu },
\end{equation}%
\emph{and}%
\begin{equation}
\delta G_{L\mu \nu }(s)=\delta g_{L\mu \nu }(s)-\Delta g_{\mu \nu }.
\end{equation}%
\emph{Proof -} The proof of the previous statements follows by
straightforward algebra. For this purpose one notices first that Eq.(\ref%
{GLP-initial-value problem}) can be equivalently written as%
\begin{eqnarray}
&&\left. D_{s}\delta G_{L\mu \nu }(s)=V_{\mu \nu }(G_{L}(s),\Delta
g,s)+\right.  \notag \\
&&\frac{1}{2}\delta G_{Lpq}(s)\widehat{g}^{\mu ^{\prime }p}(r,s)\widehat{g}%
^{\nu ^{\prime }q}(r,s)\frac{\partial }{\partial s}\left( \widehat{g}_{\mu
\mu ^{\prime }}(r,s)\widehat{g}_{\nu \nu ^{\prime }}(r,s)\right) ,
\label{EQQ-3B}
\end{eqnarray}%
where it is obvious also that $\frac{\partial }{\partial s}\left( \widehat{g}%
_{\mu \mu ^{\prime }}(r,s)\widehat{g}_{\nu \nu ^{\prime }}(r,s)\right)
\equiv D_{s}\left( \widehat{g}_{\mu \mu ^{\prime }}(r,s)\widehat{g}_{\nu \nu
^{\prime }}(r,s)\right) $ and furthermore
\begin{eqnarray}
&&\left. \frac{1}{2}\delta G_{Lpq}(s)\widehat{g}^{\mu ^{\prime }p}(r,s)%
\widehat{g}^{\nu ^{\prime }q}(r,s)\frac{\partial }{\partial s}\left(
\widehat{g}_{\mu \mu ^{\prime }}(r,s)\widehat{g}_{\nu \nu ^{\prime
}}(r,s)\right) \right.  \notag \\
&&\left. =\delta G_{L\nu }^{\mu ^{\prime }}(s)D_{s}\left( \widehat{g}_{\mu
\mu ^{\prime }}(r,s)\right) .\right.
\end{eqnarray}%
Denoting $\delta G_{L\nu }^{\mu ^{\prime }}(s)\widehat{g}_{\mu \mu ^{\prime
}}(r,s)=\delta G_{L\mu \nu }(s)$, the Leibnitz differentiation rule requires
the identity
\begin{eqnarray}
&&\left. \delta G_{L\nu }^{\mu ^{\prime }}(s)D_{s}\left( \widehat{g}_{\mu
\mu ^{\prime }}(r,s)\right) =D_{s}\left( \delta G_{L\nu }^{\mu ^{\prime }}(s)%
\widehat{g}_{\mu \mu ^{\prime }}(r,s)\right) \right.  \notag \\
&&-\widehat{g}_{\mu \mu ^{\prime }}(r,s)D_{s}\left( \delta G_{L\nu }^{\mu
^{\prime }}(s)\right)  \label{EQQ-4}
\end{eqnarray}%
to hold. Hence, Eq.(\ref{EQQ-3B}) finally yields%
\begin{eqnarray}
&&\left. D_{s}\delta G_{L\mu \nu }(s)=V_{\mu \nu }(G_{L}(s),\Delta
g,s)+\right.  \notag \\
&&D_{s}\delta G_{L\mu \nu }(s)-\widehat{g}_{\mu \mu ^{\prime
}}(r,s)D_{s}\left( \delta G_{L\nu }^{\mu ^{\prime }}(s)\right) ,
\label{EQQ-5}
\end{eqnarray}%
which also in turn implies
\begin{equation}
V_{\mu \nu }(G_{L}(s),\Delta g,s)-\widehat{g}_{\mu \mu ^{\prime
}}(r,s)D_{s}\left( \delta G_{L\nu }^{\mu ^{\prime }}(s)\right) =0.
\label{EQQ.5b}
\end{equation}%
Therefore, in the previous equation, upon multiplying tensorially term by
term by $\widehat{g}^{\mu \alpha }(r,s)$ and noting that $\widehat{g}^{\mu
^{\prime }\alpha }(r,s)\widehat{g}_{\mu \mu ^{\prime }}(r,s)=$ $\delta _{\mu
}^{\alpha }$ and $\widehat{g}^{\mu \alpha }(r,s)V_{\mu \nu }(G_{L}(s),\Delta
g,s)=V_{\nu }^{\alpha }(G_{L}(s),\Delta g,s)$, the differential equation (%
\ref{EQQ.5b}) implies%
\begin{equation}
D_{s}\left( \delta G_{L\nu }^{\alpha }(s)\right) =V_{\nu }^{\alpha
}(G_{L}(s),\Delta g,s),
\end{equation}%
which, in turn, upon integration delivers the integral equation (\ref{EQQ-2}%
) too (or equivalently Eq.(\ref{EQQ-3})). Finally, the proof of P3$_{3}$
follows by noting that in case of a stationary background metric tensor Eq.(%
\ref{EQQ-3B}) reduces to%
\begin{equation}
D_{s}\delta G_{L\mu \nu }(s)=V_{\mu \nu }(G_{L}(s),\Delta g,s),
\end{equation}%
thus implying in turn Eq.(\ref{EQQ-6}). \textbf{Q.E.D.}

\bigskip

One notices, however, that the GLP initial-value problem (\ref%
{GLP-initial-value problem}) can be equivalently replaced with%
\begin{equation}
\left\{
\begin{array}{c}
\frac{d}{ds}\delta G_{L\mu \nu }(s)=V_{\mu \nu }(G_{L}(s),\Delta g,s), \\
\delta G_{L\mu \nu }(s)=\delta g_{L\mu \nu }(s)-\Delta g_{\mu \nu }, \\
\delta g_{L\mu \nu }(s)=\delta g_{L\mu \nu },%
\end{array}%
\right.  \label{EQQ-7}
\end{equation}%
with $\delta g_{L\mu \nu }$\ prescribing now the initial condition
(associated with the deterministic Lagrangian Path). Eq.(\ref{EQQ-7}) admits
the formal solution%
\begin{equation}
\left\{
\begin{array}{c}
\delta G_{L\nu }^{\alpha }(s)=\delta G_{L\nu }^{\alpha
}(s_{1})+\int_{s_{1}}^{s}ds^{\prime }V_{\nu }^{\alpha }(G_{L}(s^{\prime
}),\Delta g,r(s^{\prime }),s^{\prime }), \\
\delta G_{L\nu }^{\alpha }(s)=\delta g_{L\nu }^{\alpha }-\Delta g_{\nu
}^{\alpha },%
\end{array}%
\right.
\end{equation}%
while correspondingly%
\begin{equation}
\left\{
\begin{array}{c}
\delta g_{L\nu }^{\alpha }(s)=\delta g_{L\nu }^{\alpha
}(s_{1})+\int_{s_{1}}^{s}ds^{\prime }V_{\nu }^{\alpha }(G_{L}(s^{\prime
}),\Delta g,r(s^{\prime }),s^{\prime }), \\
\delta g_{L\nu }^{\alpha }(s)=\delta g_{L\nu }^{\alpha }.%
\end{array}%
\right.
\end{equation}%
As a consequence the stochastic displacement $4-$tensor defined by Eq.(\ref%
{DISPLACEMENT 4-TENSOR}) can also be equivalently represented as%
\begin{equation}
\Delta g_{\nu }^{\mu }(s)\equiv \delta g_{L\nu }^{\mu }-\delta G_{L\nu
}^{\mu }(s),
\end{equation}%
where $\delta g_{L\nu }^{\mu }\equiv g_{L\nu }^{\mu }-\widehat{g}_{\nu
}^{\mu }(r,s)$ is considered prescribed and $\delta G_{L\nu }^{\mu }(s)$ is
a function of the proper-time $s.$ Then, introducing as in Ref.\cite{cqg-6}
the Lagrangian derivative realized by the operator%
\begin{equation}
\frac{D}{Ds}\equiv \left. \frac{d}{ds}\right\vert _{\delta g_{L\mu \nu
}(s)}+V_{\nu }^{\mu }(G_{L}(s),\Delta g,r,s)\frac{\partial }{\partial \delta
g_{L\nu }^{\mu }},
\end{equation}%
and upon denoting $\left. \frac{d}{ds}\right\vert _{\delta g_{L\nu }^{\mu
}(s)}\equiv \frac{d}{ds}$ and invoking also Eq.(\ref{EQQ-7}), one finds that
the differential identity
\begin{eqnarray}
\frac{D}{Ds}\Delta g_{\nu }^{\mu }(s) &=&V_{\nu }^{\mu }(G_{L}(s),\Delta
g,r,s)-  \notag \\
&&V_{\nu }^{\mu }(G_{L}(s),\Delta g,r,s)\equiv 0
\end{eqnarray}%
necessarily holds.

\subsection{7B - Properties of \textbf{polynomial GLP-solutions of the
Hamilton-Jacobi equation}}

Next, in analogy to Ref.\cite{cqg-6} let us consider \emph{polynomial
GLP-solutions} for the quantum phase-function $\mathcal{S}%
^{(q)}(G_{L}(s),\Delta g,r,s),$ namely represented in terms of a polynomial
"harmonic" representation, \textit{i.e.}, determined by means of a
second-degree polynomial of the form%
\begin{eqnarray}
&&\left. \mathcal{S}^{(q)}(G_{L}(s),\Delta g,r,s;P)\equiv \right.  \notag \\
&&\frac{a_{pq}^{\alpha \beta }(s)}{2}\Delta g_{\alpha \beta }\Delta
g^{pq}+b_{\alpha \beta }(s)\Delta g^{\alpha \beta }+c(s).  \label{HARM-1}
\end{eqnarray}%
Here $a_{\mu \nu }^{\alpha \beta }(s),$ $b_{\mu \nu }(s)$ and $c(s)$ denote
respectively suitable real $4-$tensors and a $4-$scalar functions of $s$ to
be determined in terms of the quantum H-J equation (\ref{quantum H-J})
recalled above. In particular, consistent again with Ref.\cite{cqg-6} and
upon denoting $\delta _{pq}^{\alpha \beta }\equiv \delta _{p}^{\alpha
}\delta _{q}^{\beta }$, the tensor coefficients $a_{pq}^{\alpha \beta }(s)$
are taken of the form%
\begin{equation}
a_{pq}^{\alpha \beta }(s)=\frac{1}{2}\left[ a_{(o)}(s)\delta _{pq}^{\alpha
\beta }+a_{(1)}(s)\widehat{g}_{pq}(s)\widehat{g}^{\alpha \beta }(s)\right] ,
\label{SOLUTIPON FOR a-tensotr}
\end{equation}%
with $a_{(o)}(s)$ and $a_{(1)}(s)$ being appropriate $4-$scalar functions.%
\textbf{\ }Since%
\begin{equation}
\left\{
\begin{array}{c}
\delta _{pq}^{\alpha \beta }\Delta g_{\alpha \beta }\Delta g^{pq}=\Delta
g_{\alpha \beta }\Delta g^{\alpha \beta }, \\
\widehat{g}_{pq}(s)\widehat{g}^{\alpha \beta }(s)\Delta g_{\alpha \beta
}\Delta g^{pq}=\Delta g_{\alpha }^{\alpha }\Delta g_{\beta }^{\beta },%
\end{array}%
\right.
\end{equation}%
from Eq.(\ref{HARM-1}) it follows%
\begin{equation}
\alpha _{pq}^{\alpha \beta }(s)\Delta g_{\alpha \beta }\Delta g^{pq}=\frac{1%
}{2}\left[ a_{(o)}(s)\Delta g_{\alpha \beta }\Delta g^{\alpha \beta
}+a_{(1)}(s)\Delta g_{\alpha }^{\alpha }\Delta g_{\beta }^{\beta }\right] ,
\end{equation}%
and therefore
\begin{eqnarray}
&&\left. \mathcal{S}^{(q)}(G_{L}(s),\Delta g,s)=\right.  \notag \\
&&\frac{1}{4}\left[ a_{(o)}(s)\Delta g_{\alpha \beta }\Delta g^{\alpha \beta
}+a_{(1)}(s)\Delta g_{\alpha }^{\alpha }\Delta g_{\beta }^{\beta }\right] +
\notag \\
&&b_{\alpha \beta }(s)\Delta g^{\alpha \beta }+c(s).  \label{HARM-3}
\end{eqnarray}%
Notice furthermore that here for consistency with Eqs.(\ref{B-CON-2}) and
the invariance property of the displacement tensor field $\Delta g_{\nu
}^{\mu }(s)$:%
\begin{equation}
\Delta g_{\nu }^{\mu }(s)=\Delta g_{\nu }^{\mu }(s_{1})\equiv \delta g_{L\nu
}^{\mu }(s_{1})-\delta G_{L\nu }^{\mu }(s_{1}),  \label{Invariance property}
\end{equation}%
the constant tensor field $P_{\nu }^{\mu }$\ can always be identified with%
\textbf{\ }%
\begin{equation}
P_{\nu }^{\mu }=\delta g_{L\nu }^{\mu }(s_{1}).  \label{PRESCRIPTION OF P}
\end{equation}%
In particular, this warrants that Eqs.(\ref{B-CON-2}) can indeed by suitably
fulfilled by a particular solution of the form (\ref{HARM-3}). On the same
grounds the effective quantum Hamiltonian density (\ref{effective
Hamiltoniian density}) can equivalently be represented as
\begin{equation}
H^{(q)}=\frac{1}{2\alpha L}\frac{\partial \mathcal{S}^{(q)}}{\partial \delta
g_{L\nu }^{\mu }}\frac{\partial \mathcal{S}^{(q)}}{\partial \delta g_{L\mu
}^{\nu }}+V_{QM}+V,
\end{equation}%
where%
\begin{equation}
\frac{\partial \mathcal{S}^{(q)}}{\partial \delta g_{\nu }^{\mu }}=p(s)\left[
a_{(o)}(s)\Delta g_{\mu }^{\nu }+a_{(1)}(s)\delta _{\mu }^{\nu }\Delta
g_{\beta }^{\beta }\right] +p(s)b_{\mu }^{\nu }(s),
\end{equation}%
and%
\begin{eqnarray}
&&[a_{(o)}(s)\Delta g_{\mu }^{\nu }+a_{(1)}(s)\delta _{\mu }^{\nu }\Delta
g_{\beta }^{\beta }][a_{(o)}(s)\Delta g_{\nu }^{\mu }+  \notag \\
&&\left. a_{(1)}(s)\delta _{\nu }^{\mu }\Delta g_{\beta }^{\beta
}]=a_{(o)}^{2}(s)\Delta g_{\mu }^{\nu }\Delta g_{\nu }^{\mu }+\right.  \notag
\\
&&\left[ 4a_{(1)}^{2}(s)+2a_{(o)}(s)a_{(1)}(s)\right] \Delta g_{\alpha
}^{\alpha }\Delta g_{\beta }^{\beta }.
\end{eqnarray}

Then one can show that the following result applies.

\bigskip

\textbf{THM. 4 - Polynomial GLP-solutions of the Hamilton-Jacobi equation in
the case of vacuum}

\emph{Regarding the existence of polynomial GLP-solutions of the quantum
Hamilton-Jacobi equation (\ref{quantum H-J}) in the case of vacuum which
hold in validity of the extended functional setting (\ref{Eq-2a}) the
following propositions apply:}

\emph{P4}$_{1}$\emph{) The quantum-modified Einstein field equations (\ref%
{quantum-modified Einstein eq}) are recovered by requiring the identical
validity of the extremal equation}%
\begin{equation}
\left. \frac{\partial }{\partial \Delta g^{\mu \nu }}\left[ V_{o}(g+\Delta
g)+V_{QM}(g,s)\right] \right\vert _{\Delta g=0}=0.
\label{GLP-extremal equation}
\end{equation}

\emph{P4}$_{2}$\emph{) (Polynomial solution) - The solution of the quantum
Hamilton-Jacobi equation (\ref{quantum H-J}) takes the polynomial form (\ref%
{HARM-3}) (harmonic representation).}

\emph{P4}$_{3}$\emph{) (Uniqueness property) - The }$4-$\emph{scalar
coefficients }$a_{(o)}(s)$\emph{\ and }$a_{(1)}(s)$\emph{\ are determined by
means of ODEs which are implied by Eq.(\ref{quantum H-J}).}

\emph{P4}$_{4}$\emph{) (Invariance property) - The same ODEs are identical
with the corresponding equations holding in the case of stationary
background field tensor (\textit{i.e.}, the case (\ref{Eq-2a})) reported
previously in Ref.\cite{cqg-6}. Hence the }$4-$\emph{scalar coefficients }$%
a_{(o)}(s)$\emph{\ and }$a_{(1)}(s)$\emph{\ are independent of the choice of
the functional setting for the background field tensor (\textit{i.e.},
respectively Eqs.(\ref{Eq-2a}) or (\ref{Eq-2b})).}

\emph{Proof - }To reach the thesis we introduce preliminarily a
second-order, \textit{i.e.}, harmonic, expansion for the effective potential
$V_{o}(g+\Delta g).$ Elementary algebra shows that this takes the form%
\begin{eqnarray}
&&\left. V_{o}(g+\Delta g)=V_{o}(g)\right.  \notag \\
&&+\Delta g_{\nu }^{\mu }\left. \frac{\partial }{\partial \Delta g_{\nu
}^{\mu }}V_{o}(g+\Delta g)\right\vert _{\Delta g=0}  \notag \\
&&+\frac{1}{2}\Delta g_{\beta }^{\alpha }\Delta g_{\nu }^{\mu }\left. \frac{%
\partial ^{2}}{\partial \Delta g_{\beta }^{\alpha }\partial \Delta g_{\nu
}^{\mu }}V_{o}(g+\Delta g)\right\vert _{\Delta g=0}.
\end{eqnarray}%
Then, substituting this expression in Eq.(\ref{GLP-extremal equation}),
evaluation of the partial derivative gives%
\begin{eqnarray}
&&\left. \frac{\partial }{\partial \Delta g_{k}^{\lambda }}\left. \left[
V_{o}(g+\Delta g)+V_{QM}(g,s)\right] \right\vert _{\Delta g=0}=\right.
\notag \\
&&\left. \frac{\partial }{\partial \Delta g_{k}^{\lambda }}\left[
V_{o}(g)+V_{QM}(g,s)\right] \right\vert _{\Delta g=0}  \notag \\
&&+\left. \frac{\partial }{\partial \Delta g_{k}^{\lambda }}\left[ \Delta
g_{\nu }^{\mu }\frac{\partial }{\partial \Delta g_{\nu }^{\mu }}%
V_{o}(g+\Delta g)\right] \right\vert _{\Delta g=0}  \notag \\
&&+\left. \frac{\partial }{\partial \Delta g_{k}^{\lambda }}\left[ \frac{1}{2%
}\Delta g_{\beta }^{\alpha }\Delta g_{\nu }^{\mu }\frac{\partial ^{2}}{%
\partial \Delta g_{\beta }^{\alpha }\partial \Delta g_{\nu }^{\mu }}%
V_{o}(g+\Delta g)\right] \right\vert _{\Delta g=0}.
\label{polynomil expansion}
\end{eqnarray}%
Only the linear term in the $\Delta g_{\nu }^{\mu }-$expansion and the Bohm
potential contribution\ $V_{QM}(g,s)$ remain, so that explicit calculation
recovers identically Eq.(\ref{GLP-extremal equation}) which is satisfied
being proportional to the quantum-modified Einstein equation (\ref%
{quantum-modified Einstein eq}) (in agreement with Proposition \emph{P2}$%
_{1} $). This proves \emph{P4}$_{1}$. Then one can show (Proposition \emph{P4%
}$_{2}$) by straightforward algebra that, in analogy with Ref.\cite{cqg-6},
a polynomial solution of the quantum Hamilton-Jacobi equation (\ref{quantum
H-J}) exists also in the case of non-stationary background (see \textit{e.g.}%
, Eq.(\ref{Eq-2b})).

Next let us consider the uniqueness property (Proposition \emph{P4}$_{3}$)%
\emph{. }For this purpose one needs to evaluate the covariant $s-$derivative
of the quadratic terms (\textit{i.e.}, proportional to $\Delta g_{\alpha
}^{\alpha }\Delta g_{\beta }^{\beta }$) in $\mathcal{S}^{(q)}(G_{L}(s),%
\Delta g,s).$ For this purpose one notices that upon invoking Eq.(\ref%
{HARM-3}) it follows identically that
\begin{equation}
\left\{
\begin{array}{c}
D_{s}\left[ \Delta g_{\alpha \beta }\Delta g^{\alpha \beta }\right] =0, \\
D_{s}\left[ \Delta g_{\alpha }^{\alpha }\Delta g_{\beta }^{\beta }\right] =0.%
\end{array}%
\right.
\end{equation}%
This implies the differential identity
\begin{eqnarray}
&&\left. D_{s}\left[ \frac{a_{\alpha p}^{\beta q}(s)}{2}\Delta g_{\beta
}^{\alpha }\Delta g_{q}^{p}\right] =\right.   \notag \\
&&\frac{1}{2}\left[ \Delta g_{\alpha }^{\beta }\Delta g_{\beta }^{\alpha
}D_{s}\left[ a_{(o)}(s)\right] +\Delta g_{\alpha }^{\alpha }\Delta g_{\beta
}^{\beta }D_{s}\left[ a_{(1)}(s)\right] \right] ,
\end{eqnarray}%
where the derivatives of $a_{(o)}(s)$ and\emph{\ }$a_{(1)}(s)$ are
respectively proportional to the two $4-$scalars $\Delta g_{\alpha }^{\beta
}\Delta g_{\beta }^{\alpha }$ and $\Delta g_{\alpha }^{\alpha }\Delta
g_{\beta }^{\beta }$, to be considered here as independent and arbitrary.
Explicit evaluation of the coefficient in the quadratic term appearing in
Eq.(\ref{polynomil expansion}) then shows that two distinct ODEs are
determined for $a_{(o)}(s)$\ and\ $a_{(1)}(s).$\ Direct comparison with the
analogous equations determined in Ref.\cite{cqg-6} then shows that the same
ODEs are independent of the specific functional setting, \textit{i.e.},
either Eq.(\ref{Eq-2b}) or (\ref{Eq-2a}). The remarkable conclusion is%
\textbf{\ }therefore realized by the invariance property (Proposition \emph{%
P4}$_{4}$) of the\textbf{\ }$4-$scalar coefficients\textbf{\ }$a_{(o)}(s)$%
\textbf{\ }and $a_{(1)}(s)$.\emph{\ }\textbf{Q.E.D.}

\subsection{7C - GLP Gaussian particular solutions\ of the quantum PDF}

Let us now show that also the quantum continuity equation (\ref{cont-1})
admits, even in the case of an arbitrary non-stationary background field
tensor (\textit{e.g.}, Eq.(\ref{Eq-2a})), Gaussian-like solutions of the
form:%
\begin{eqnarray}
&&\rho (G_{L}(s),\widehat{g}(s),\Delta g,r(s),s)=  \notag \\
&&\rho (G_{L}(s_{o}),\widehat{g}(s_{o}),\Delta g(s_{o}),r(s_{o}),s_{o})
\notag \\
&&\exp \left\{ -\int\limits_{s_{o}}^{s}ds^{\prime }\frac{\partial V_{\nu
}^{\mu }(G_{L}(s^{\prime }),\Delta g,r(s^{\prime }),s^{\prime })}{\partial
g_{L\nu }^{\mu }(s^{\prime })}\right\} ,  \label{SOLUTION OF CONTINUITY EQ}
\end{eqnarray}%
where, introducing the signature parameter $\theta \equiv \pm ,$%
\begin{eqnarray}
&&\left. \rho (G_{L}(s_{o}),\Delta g(s_{o}),r(s_{o}),s_{o})\equiv \right.
\notag \\
&&\frac{1}{\pi ^{5}r_{th}^{10}}\exp \left\{ -\frac{\left( \Delta
g(s_{o})+\theta \widehat{g}(s_{o})\right) ^{2}}{r_{th}^{2}}\right\}  \notag
\\
&\equiv &\rho _{G}(\Delta g(s_{o})+\theta \widehat{g}(s_{o})),
\label{REALIZATION OF THE INITIAL ODF}
\end{eqnarray}%
identifies an initial \emph{shifted Gaussian PDF}, with $\widehat{g}%
(s)\equiv \widehat{g}(r(s),s)$\ and $\widehat{g}(s_{o})\equiv \widehat{g}%
(r(s_{o}),s_{o})$\ denoting a generally non-stationary background metric
tensor and its initial value at the initial proper-time $s_{o}$\ evaluated
along an observer's geodesic curve. In particular, in validity of the
polynomial decomposition (\ref{HARM-3}) for the quantum phase function $%
\mathcal{S}^{(q)}(G_{L}(s),\Delta g,r,s),$ the $4-$scalar function $\frac{%
\partial V_{\mu \nu }(G_{L}(s^{\prime }),\Delta g,s^{\prime })}{\partial
g_{L\mu \nu }(s^{\prime })}$ is found to be function of proper-time only.
More precisely it takes the form%
\begin{equation}
\frac{\partial V_{\nu }^{\mu }(G_{L}(s^{\prime }),\Delta g,r(s^{\prime
})s^{\prime })}{\partial g_{L\nu }^{\mu }(s^{\prime })}\equiv
16p^{2}(s^{\prime })a(s^{\prime }),
\end{equation}%
where $p(s^{\prime })$ is given by Eq.(\ref{A-4BIS}) (see Appendix C) and
the $4-$scalar function $a(s^{\prime })$\ is prescribed by requiring%
\begin{equation}
a(s^{\prime })=\frac{1}{2}\left[ a_{(o)}(s^{\prime })+a_{(1)}(s^{\prime })%
\right]  \label{A-1quaterBIS}
\end{equation}%
(or equivalently Eq.(\ref{A-1quater}) in Appendix C). In addition, one
notices that here both $r_{th}^{2}$\ and $\left( \Delta g+\theta \widehat{g}%
(s_{o})\right) ^{2}$\ are $4-$scalars and%
\begin{eqnarray}
&&\left. \left( \Delta g(s_{o})+\theta \widehat{g}(s_{o})\right) ^{2}\equiv
\right.  \notag \\
&&\left( \Delta g(s_{o})+\theta \widehat{g}(s_{o})\right) _{\mu \nu }\left(
\Delta g(s_{o})+\theta \widehat{g}(s_{o})\right) ^{\mu \nu },
\label{INVARIANCE PROPERTY}
\end{eqnarray}%
and $r_{th}^{2}$\ is a constant independent of both the $4-$position $r^{\mu
}$ and the proper-time $s$. In particular, one can prove also in this case
that the validity of the invariance property%
\begin{equation}
\left( \Delta g(s_{o})+\theta \widehat{g}(s_{o})\right) ^{2}=\left( \Delta
g(s)+\theta \widehat{g}(s)\right) ^{2}
\end{equation}%
remains preserved for arbitrary $s,s_{o}\in I.$ The proof follows from
elementary algebra by noting first that%
\begin{eqnarray}
\left( \Delta g(s_{o})+\theta \widehat{g}(s_{o})\right) ^{2} &\equiv &\Delta
g_{\nu }^{\mu }(s_{o})\Delta g_{\mu }^{\nu }(s_{o})+4  \notag \\
&&+2\theta \Delta g_{\mu \nu }(s_{o})\widehat{g}^{\mu \nu }(s_{o}).
\end{eqnarray}%
Indeed, thanks to Eq.(\ref{CNSTRAINT-displacement})
\begin{equation}
\Delta g_{\nu }^{\mu }(s_{o})\Delta g_{\mu }^{\nu }(s_{o})=\Delta g_{\nu
}^{\mu }(s)\Delta g_{\mu }^{\nu }(s),
\end{equation}%
while for the same reason
\begin{eqnarray}
D_{s}\left[ \Delta g_{\mu \nu }(s)\widehat{g}^{\mu \nu }(s)\right] &=&
\notag \\
D_{s}\left[ \Delta g_{\nu }^{\alpha }(s)\widehat{g}_{\alpha \mu }(s)\widehat{%
g}^{\mu \nu }(s)\right] &=&D_{s}\left[ \Delta g_{\nu }^{\alpha }(s)\delta
_{\alpha }^{\nu }\right] =0.
\end{eqnarray}%
Hence also the equation
\begin{equation}
\Delta g_{\mu \nu }(s)\widehat{g}^{\mu \nu }(s)\equiv \Delta g_{\mu \nu }%
\widehat{g}^{\mu \nu }(s)=\Delta g_{\mu \nu }(s_{o})\widehat{g}^{\mu \nu
}(s_{o})
\end{equation}%
necessarily holds. As a consequence one finds again as in Ref.\cite{cqg-6} (%
\textit{i.e.}, for the case of stationary background) that%
\begin{equation}
\rho _{G}(\Delta g(s_{o})+\theta \widehat{g}(s_{o}))=\rho _{G}(\Delta
g+\theta \widehat{g}(s)),
\end{equation}%
which proves the statement. The conclusion is therefore that Eq.(\ref%
{SOLUTION OF CONTINUITY EQ}) takes the form%
\begin{eqnarray}
&&\left. \rho (G_{L}(s),\Delta g,s)=\rho _{G}(\Delta g+\theta \widehat{g}%
(s))\right.  \notag \\
&&\exp \left\{ -16\int\limits_{s_{o}}^{s}ds^{\prime }p^{2}(s^{\prime
})a(s^{\prime })\right\} ,  \label{quantum PDF}
\end{eqnarray}%
which indeed realizes a Gaussian particular solution for the quantum PDF.
Hence, the realization of the quantum PDF (\ref{SOLUTION OF CONTINUITY EQ})
is again independent of the choice of the functional setting of the
background field tensor, respectively being prescribed either according to
Eq.(\ref{Eq-2a}) or Eq.(\ref{Eq-2b}).

Although in principle both Gaussian solutions corresponding to $\theta =+$
and $\theta =-$\ are admissible from the mathematical point of view, in the
following we shall consider only the one obtained from Eq.(\ref{SOLUTION OF
CONTINUITY EQ}) by setting $\theta \equiv -$. Hence, the initial shifted
Gaussian PDF dealt with in the rest of the calculations takes the form%
\begin{eqnarray}
&&\left. \rho (G_{L}(s_{o}),\Delta g(s_{o}),r(s_{o}),s_{o})\equiv \right.
\notag \\
&&\frac{1}{\pi ^{5}r_{th}^{10}}\exp \left\{ -\frac{\left( \Delta g(s_{o})-%
\widehat{g}(s_{o})\right) ^{2}}{r_{th}^{2}}\right\}  \notag \\
&\equiv &\rho _{G}(\Delta g(s_{o})-\widehat{g}(s_{o})).
\end{eqnarray}%
This choice has a physical basis. In fact, according to the emergent gravity
picture inherent the GLP formulation of CQG-wave equation, it warrants that
the GLP-quantum/stochastic expectation value of the stochastic displacement $%
4-$tensor $\Delta g_{\mu \nu }$\ recovers the correct signature of the
background metric tensor, namely%
\begin{equation}
\left\langle \Delta g_{\mu \nu }\right\rangle =\int_{U_{g}}d(\Delta g)\rho
_{G}(\Delta g-\widehat{g}(r,s))\Delta g_{\mu \nu }=\widehat{g}_{\mu \nu
}(r,s).  \label{EMERGENT GRAVITY}
\end{equation}%
We refer to Ref.\cite{cqg-6} for an exhaustive discussion of the emergent
gravity phenomenon in CQG-theory and for a detailed mathematical definition
of quantum expectation value, see in particular Proposition 3 and related
comments in the same reference.

\subsection{7D - Semiclassical limit}

An important aspect of CQG-theory and the related GLP description concerns
the investigation of the semiclassical limit of the quantum theory, which
establishes the connection with the classical Hamiltonian structure of GR
reported in Ref.\cite{cqg-2} and provides a test of consistency of the
theory itself. The study of the semiclassical limit is\ conveniently
performed on the set of QHE, namely the quantum Hamilton-Jacobi equation (%
\ref{quantum H-J}) and the continuity equation (\ref{cont-1}) through its
explicit analytical Gaussian solution for the quantum PDF given by Eq.(\ref%
{quantum PDF}).

We consider first the quantum Hamilton-Jacobi equation, for which the
semiclassical limit is prescribed letting $\hslash \rightarrow 0$. By
requiring that in the same limit both $\alpha $ and $L(m_{o})$ reduce to
their classical definition and that the real limit function $\lim_{\hslash
\rightarrow 0}\frac{^{(q)}(s)}{\hslash }=\frac{\mathcal{S}(s)}{\alpha }$
exists for arbitrary $s\in I\equiv
\mathbb{R}
$, with $\mathcal{S}(s)$\ identifying the classical reduced Hamilton
principal function\ (see Ref.\cite{EPJ1}), then one can shown that the
quantum Hamilton-Jacobi equation (\ref{quantum H-J}) reduces to the
analogous classical Hamilton-Jacobi equation (\ref{classical Hamilton-Jacobi}%
). In fact, considering without loss of generality the case of vacuum, the
semiclassical limit of Eq.(\ref{quantum H-J}) delivers%
\begin{equation}
\frac{1}{\alpha }\frac{\partial \mathcal{S}(s)}{\partial s}+\frac{1}{2\alpha
^{2}L}\frac{\partial \mathcal{S}(s)}{\partial g^{\mu \nu }}\frac{\partial
\mathcal{S}(s)}{\partial g_{\mu \nu }}+\lim_{\hslash \rightarrow 0}\frac{%
V_{QM}(s)}{\hslash }=0,  \label{evaluation H-J}
\end{equation}%
where the limit $\lim_{\hslash \rightarrow 0}\frac{V_{QM}(s)}{\hslash }=0$
holds identically. As a consequence the quantum Hamiltonian density\emph{\ }$%
H^{(q)}$ necessarily must reduce to the limit function%
\begin{equation}
H_{R}=\frac{1}{2\alpha L}\frac{\partial \mathcal{S}(s)}{\partial g^{\mu \nu }%
}\frac{\partial \mathcal{S}\left( s\right) }{\partial g_{\mu \nu }}.
\end{equation}%
This coincides in form with the classical normalized Hamiltonian density
given above by Eq.(\ref{classical Hamiltonian density}) in the case of
vacuum, while Eq.(\ref{evaluation H-J}) reduces to the classical
GR-Hamilton-Jacobi equation.

Let us now consider the semiclassical limit for the quantum continuity
equation, which is investigated here by direct analysis of the analytical
Gaussian solution. In this case the behavior of the free-parameter $%
r_{th}^{2}$ must be prescribed when the limit $\lim_{\hslash \rightarrow 0}$
holds. Here we require that%
\begin{equation}
r_{th}^{2}\sim \hslash ^{\gamma },  \label{gam}
\end{equation}%
with the exponent $\gamma >0$ to be later determined upon imposing that in
the same semiclassical limit both the quantum Bohm potential and the quantum
cosmological constant expressed in the GLP representation vanish
identically. Thus, under the previous assumption the semiclassical limit on
the quantum PDF is prescribed equivalently letting $\lim_{r_{th}^{2}%
\rightarrow 0}\rho (G_{L}(s),\Delta g,s)$. This amounts to require the
finite width of the Gaussian function to vanish in the semiclassical limit,
which means that the quantum Gaussian PDF becomes a Dirac-delta function
making the quantum gravitational field to "collapse" and coincide with the
classical background metric tensor at initial proper-time $s_{o}$:%
\begin{equation}
\lim_{r_{th}^{2}\rightarrow 0}\rho (G_{L}(s),\Delta g,s)=\delta \left(
\Delta g(s_{o})-\widehat{g}(s_{o})\right) .
\end{equation}

The analysis of the semiclassical limit of the QHE enables us to stress the
character of the quantum modified Einstein field equations and the
underlying Gaussian solution for the quantum PDF. The latter ones in fact
must not be interpreted as modified classical gravitational field equations.
The quantum modified Einstein equations truly include non-stationary quantum
effects arising from CQG-theory, but at the same time they preserve exactly
the classical form of the Einstein theory. This is made manifest by
inspection of the semiclassical limit, which recovers exactly the classical
equations. Thus, the present theory is not providing some type of "ad hoc"
modifications of classical GR, but instead it is consistently including
quantum effects computed in the framework of a covariant quantum theory of
the same gravitational field.

\section{8 - \textbf{Explicit evaluation of the Bohm effective potential and
source term}}

In this section the Bohm effective potential and the corresponding source
term appearing in the quantum-modified Einstein tensor equation (see Eq.(\ref%
{extremal equation})) are determined. The task is achieved based on the
GLP-approach developed here in the context of the extended functional
setting. In this regard it is important to acknowledge the following unique
features:

\begin{itemize}
\item First, as shown in the present paper, the same equation (\ref{extremal
equation}) has been recovered independently also in the GLP-approach, being
provided in such a context by the extremal tensor equation (\ref%
{GLP-extremal equation}) (see THM.4).

\item Second, based on the construction of an analytic solution for the
quantum PDF (see Eq.(\ref{quantum PDF}) above) and of the corresponding
quantum phase-function (the function $\mathcal{S}^{(q)}(G_{L}(s),\Delta g,s)$
determined via the polynomial representation (\ref{HARM-3})), the
GLP-approach permits one to obtain also an explicit representation of the
Bohm effective quantum potential (\ref{BOHM potential}) and corresponding
source term $B_{\mu \nu }(s)$.

\item Third, in the subsequent calculations all integrations are performed
with respect to the local extremal geodesic trajectory. As a consequence the
initial proper-time $s_{o}$\ is set equal to $s_{o}=0.$
\end{itemize}

In fact, based on the construction of the analytic solution for the quantum
PDF indicated above (see Eq.(\ref{quantum PDF}) in Section 7), an explicit
representation of the Bohm effective quantum potential (\ref{BOHM potential}%
) follows. This is determined by a second-degree polynomial in terms of the
quantum displacement field tensor $\Delta g,$ namely
\begin{equation}
V_{QM}=\frac{\hslash ^{2}}{4\alpha L}\frac{8p^{2}(s)}{r_{th}^{2}}-\frac{%
\hslash ^{2}}{8\alpha L}\frac{4p^{2}(s)}{r_{th}^{4}}\left( \Delta g-\widehat{%
g}(s)\right) ^{2},  \label{bohm POTENTIAL - 2}
\end{equation}%
with $\hbar $ being the reduced Planck constant, while $r_{th}$ is still
arbitrary and must be suitably determined. The rest of the notation follows
from Ref.\cite{cqg-4}, with $\alpha $ being the dimensional constant defined
as $\alpha =m_{o}cL$, while $m_{o}$ and $L$ are the graviton mass and $L$
its Compton length, namely $L=\frac{\hbar }{m_{o}c}$. Thanks to this result
also the Bohm source tensor field $B_{\mu \nu }\equiv B_{\mu \nu }(s),$
prescribed by means of Eq.(\ref{Bohm source term}), can be readily evaluated
yielding the formal representation
\begin{equation}
B_{\mu \nu }=-\frac{\hbar ^{2}}{\left( \alpha L\right) ^{2}}\frac{1}{%
r_{th}^{4}}f(s)\widehat{g}_{\mu \nu }(r(s),s),  \label{Bohm source term-2}
\end{equation}%
with $f(s)$ being a function of proper-time defined with respect to a
maximal length local geodesic trajectory. This is determined by the equation
\begin{equation}
f(s)\equiv p^{3}(s).  \label{f(s)}
\end{equation}%
Notice that here $p(s)$\ is prescribed according to Ref.\cite{cqg-6} (see
also Eq.(\ref{A-4BIS})\ recalled in Appendix C). More precisely, it is a
function of the definite integral $\int\limits_{s_{o}}^{s}ds^{\prime
}a(s^{\prime })$\ with respect to the $4-$scalar function $a(s)$\ (see Eq.(%
\ref{A-1quater})), while also requiring $s_{o}=0$\ (\textit{i.e.}, upon
identifying the local geodetics with a maximal geodesic curve). As a
consequence one has that%
\begin{equation}
p(s_{o}=0)=1,
\end{equation}%
while its precise $s-$dependence follows from the behavior of $a(s)$.
Omitting here unnecessary further details on the matter we shall refer for
this purpose to the related discussion already treated in the cited
reference.

\subsection{8A - Determination of the CQG-cosmological constant $\Lambda
_{CQG}(s).$}

The representation given above (\ref{Bohm source term-2}) for the Bohm
source tensor field $B_{\mu \nu }$ suggests its obvious connection with a
suitably-prescribed notion of cosmological constant. The same tensor field $%
B_{\mu \nu }$ can in fact be equivalently represented as
\begin{equation}
B_{\mu \nu }\equiv -\Lambda _{CQG}(s)\widehat{g}_{\mu \nu }(r(s),s),
\label{BOHM SOURCE-3}
\end{equation}%
with $\Lambda _{CQG}(s)$ denoting the \emph{CGQ-cosmological constant}%
\begin{equation}
\Lambda _{CQG}(s)=\frac{\hbar ^{2}}{\left( \alpha L\right) ^{2}}\frac{1}{%
r_{th}^{4}}f(s)  \label{Lambda-CQG}
\end{equation}%
and the $4-$scalar function $f(s)$\ is prescribed by Eq.(\ref{f(s)}). A
number of important features of $\Lambda _{CQG}(s)$ emerge. First, the
function $\Lambda _{CQG}(s)$ does not depend on the $4-$position $r\equiv
\left\{ r^{\mu }\right\} $ and hence it effectively behaves as a constant in
the quantum-modified Einstein field equations (\ref{quantum-modified
Einstein eq}). A further feature of\textbf{\ }$\Lambda _{CQG}(s)$ concerns
its quantum origin. As a consequence one expects that $\Lambda _{CQG}(s)$
should vanish identically in the semiclassical limit so that
\begin{equation}
\lim_{\hslash \rightarrow 0}\Lambda _{CQG}(s)=0,
\end{equation}%
and, in particular, that for $\hslash \rightarrow 0$
\begin{equation}
\Lambda _{CQG}(s)\sim O(\hslash ).  \label{Infinitesimal of h}
\end{equation}%
One can show that such a requirement permits us to determine consistently
the still undetermined exponential factor $\gamma $ previously introduced in
Eq.(\ref{gam}). Given the analytical solution (\ref{Lambda-CQG}) the
prescription of $\gamma $, in fact, follows once demanding for consistency
with Eq.(\ref{Infinitesimal of h}) that the ratio $\frac{\hbar ^{2}}{%
r_{th}^{4}}$\ appearing in Eq.(\ref{Lambda-CQG}) is such that%
\begin{equation}
\frac{\hbar ^{2}}{r_{th}^{4}}\sim O(\hbar ).
\end{equation}%
Invoking Eq.(\ref{gam}) this implies therefore that $\gamma =1/2$, namely $%
r_{th}\sim \hslash ^{1/4}$.

An equivalent representation of the CQG-cosmological constant can be
achieved in terms of a suitable, strictly positive vacuum energy density
\begin{equation}
\rho _{vac}=\frac{1}{\kappa }\Lambda _{CQG}(s).
\label{vacuum energy density}
\end{equation}%
Eq.(\ref{BOHM SOURCE-3}), in fact, can be represented equivalently via Eq.(%
\ref{quantu energy-momentum tensor}) in terms of the corresponding\textbf{\ }%
quantum contribution to the stress energy tensor\textbf{\ }$\widehat{T}_{\mu
\nu }^{(q)}$. This is obtained letting
\begin{equation}
\widehat{T}_{\mu \nu }^{(q)}\equiv \frac{1}{\kappa }B_{\mu \nu }\equiv -\rho
_{vac}\widehat{g}_{\mu \nu }(r,s).  \label{vacuum energy density-2}
\end{equation}%
Eq.(\ref{vacuum energy density-2}) is therefore formally analogous to that
given in Ref.\cite{Barrow}, the realization of $\rho _{vac}$\ being,
however, quite different. In the present case, in fact, $\rho _{vac}$ must
be identified with the \emph{graviton vacuum energy density,} \textit{i.e.},
the vacuum energy density produced by gravitons and arising due to the Bohm
interaction which acts on the same gravitons.

Finally, a notable feature of $\Lambda _{CQG}(s)$ concerns its proper-time
dependence\emph{\ }occurring through the strictly positive $4-$scalar
function $f(s)$ (see Eq.(\ref{f(s)})). In view of the prescription of the
function $p(s)$\ (see Eq.(\ref{A-1quaterBIS}) in Appendix C) it follows that
its initial value is $p(s_{o})=1$\ also for $s_{o}=0$. Hence the initial
value of $\Lambda _{CQG}(s)$\ occurring at $s=s_{o}=0$\ is
\begin{equation}
\Lambda _{CQG}(s_{o})=\frac{\hbar ^{2}}{\left( \alpha L\right) ^{2}}\frac{1}{%
r_{th}^{4}},  \label{LAMDDA INIZIALE}
\end{equation}%
so that the relationship between $\Lambda _{CQG}(s)$ and $\Lambda
_{CQG}(s_{o})$ is simply%
\begin{equation}
\Lambda _{CQG}(s)=\Lambda _{CQG}(s_{o})p^{3}(s).
\label{relazione tra LAMBDA}
\end{equation}

The issue over the prescription of the proper-time dependence of $\Lambda
_{CQG}(s)$ and the related analysis of qualitative properties is addressed
in the next section.

\section{9 - Proper-time behavior of $\Lambda _{CQG}(s)$ and physical
implications}

In view of Eqs.(\ref{f(s)}) and Eq.(\ref{A-4BIS}) (see Appendix C) the
prescription of the proper-time functions $f(s)$ and $p(s)$ requires in turn
the evaluation of the $4-$scalar function $a(s)$ given by Eq.(\ref{A-1quater}%
). As discussed above the same function is realized by the equation (\ref%
{A-1quaterBIS}), with $a_{(o)}(s)$ and $a_{(1)}(s)$ being the $4-$scalar
coefficients appearing in the quadratic term of Harmonic representation of
the quantum-phase function (\ref{HARM-3}). According to Ref.\cite{cqg-6} the
same functions are determined by an initial-value problem associated with
appropriate first-order ODEs. In the present case, one notices that $\Lambda
$ must be identified with the CQG-cosmological constant $\Lambda _{CQG}(s)$
so that Eqs.(\ref{LAMDDA INIZIALE}) and (\ref{relazione tra LAMBDA}) must be
taken into account. One obtains accordingly for $a_{(o)}(s)$\ and $%
a_{(1)}(s) $\ the two equations:
\begin{equation}
\left\{
\begin{array}{c}
\frac{1}{4}\frac{d}{ds}a_{(o)}(s)=\frac{p^{2}(s)}{8\alpha L}a_{(o)}^{2}(s)+%
\frac{\alpha L}{2}\Lambda (s_{o})p^{2}(s)+ \\
-\frac{\alpha L}{2}\Lambda (s_{o})p^{3}(s)+G_{(o)}, \\
\frac{1}{4}\frac{d}{ds}a_{(1)}(s)=\frac{p^{2}(s)}{8\alpha L}\left(
4a_{(1)}^{2}(s)+2a_{(o)}(s)a_{(1)}(s)\right) + \\
-\frac{\alpha L}{2}\Lambda (s_{o})p^{3}(s)+G_{(1)},%
\end{array}%
\right.  \label{BBB-EQ}
\end{equation}%
where $G_{(o)}$ and $G_{(1)}$ are arbitrary constant gauge functions which
can be conveniently chosen in such a way that the same equations admit a
stationary solution. When cast in dimensionless form upon letting%
\begin{equation}
\left\{
\begin{array}{c}
\theta =\frac{2s}{L}, \\
\overline{a}_{(o)}(\theta )=\frac{a_{(o)}(\theta )}{\alpha }, \\
\overline{a}_{(1)}(\theta )=\frac{a_{(1)}}{\alpha }, \\
\overline{\Lambda }_{o}=\Lambda (s_{o})L^{2},%
\end{array}%
\right.
\end{equation}%
these yield%
\begin{equation}
\left\{
\begin{array}{c}
\frac{d}{d\theta }\overline{a}_{(o)}(\theta )=\frac{p^{2}(\theta )}{4}%
\overline{a}_{(o)}^{2}(\theta )-\frac{\overline{a}_{(o)}^{2}}{4}(\theta
_{o})+\overline{\Lambda }(s_{o})\left[ p^{2}(\theta )-1\right] - \\
-\overline{\Lambda }(s_{o})\left[ p^{3}(\theta )-1\right] , \\
\frac{d}{d\theta }\overline{a}_{(1)}(\theta )=\frac{p^{2}(\theta )}{4}\left(
4\overline{a}_{(1)}^{2}(\theta )+2\overline{a}_{(o)}(\theta )\overline{a}%
_{(1)}(\theta )\right) + \\
-\overline{\Lambda }(s_{o})\left[ p^{3}(\theta )-1\right] -\frac{3}{4}%
\overline{a}_{(1)}^{2}(\theta _{o})+\frac{\overline{a}_{(o)}^{2}}{4}(\theta
_{o}).%
\end{array}%
\right.  \label{DIMENSIONLESS BBB-EQ}
\end{equation}%
Then, by setting the initial conditions%
\begin{equation}
\left\{
\begin{array}{c}
a_{(o)}(\theta _{o})=\widehat{a}_{(o)}, \\
a_{(1)}(\theta _{o})=\widehat{a}_{(1)},%
\end{array}%
\right.  \label{BBB-EQ-1}
\end{equation}%
with $\left( \widehat{a}_{(o)},\widehat{a}_{(1)}\right) $ being initial
constants, and by requiring also
\begin{equation}
\widehat{a}_{(1)}=-\widehat{a}_{(o)},  \label{BBB-EQ-1-BIS}
\end{equation}%
it follows that Eqs.(\ref{DIMENSIONLESS BBB-EQ}) admit the stationary
solution $\overline{a}_{(o)}(\theta )\equiv \widehat{a}_{(o)},$ $\overline{a}%
_{(1)}(\theta )\equiv -\widehat{a}_{(o)}$ and $\overline{p}(\theta )\equiv 1$%
.

We now pose the problem of the investigation of the asymptotic property of
the solutions of Eqs.(\ref{DIMENSIONLESS BBB-EQ})-(\ref{BBB-EQ-1}), \textit{%
i.e.}, for solutions such that
\begin{equation}
\left\{
\begin{array}{c}
\lim_{\theta \rightarrow \infty }\overline{a}(\theta )=0, \\
\lim_{\theta \rightarrow \infty }\overline{a}_{(o)}(\theta )=-\lim_{\theta
\rightarrow \infty }\overline{a}_{(1)}(\theta )=\overline{a}_{(o)\infty },
\\
\lim_{\theta \rightarrow \infty }\frac{d}{d\theta }\overline{a}_{(o)}(\theta
)=\lim_{\theta \rightarrow \infty }\frac{d}{d\theta }\overline{a}%
_{(1)}(\theta )=0, \\
\lim_{\theta \rightarrow \infty }p(\theta )=p_{\infty }, \\
\lim_{\theta \rightarrow \infty }\overline{\Lambda }(s)=\overline{\Lambda }%
_{CQG\infty }.%
\end{array}%
\right.  \label{BBB-EQ-2}
\end{equation}%
In this regard the following result holds.

\bigskip

\textbf{THM. 5 - Asymptotic behavior of the solutions of Eqs.(\ref%
{DIMENSIONLESS BBB-EQ})}

\emph{Assuming that Eqs.(\ref{BBB-EQ-2}) and the limits }$\overline{a}%
_{(o)\infty }$\emph{\ and }$p_{\infty }$\emph{\ exist, with }$\overline{a}%
_{(o)\infty }$\emph{\ and }$p_{\infty }$\emph{\ being non-vanishing, then
the following propositions apply:}

\emph{P5}$_{1}$\emph{) First, the equation}%
\begin{equation}
p_{\infty }^{2}=\frac{1+\frac{1}{4}\frac{\overline{a}_{(1)}^{2}(\theta _{o})%
}{\overline{\Lambda }(s_{o})}-\frac{1}{2}\frac{\overline{a}_{(o)}^{2}(\theta
_{o})}{\overline{\Lambda }(s_{o})}}{1-\frac{1}{4}\frac{\overline{a}%
_{(o)\infty }^{2}}{\overline{\Lambda }(s_{o})}}  \label{P5_1}
\end{equation}%
\emph{holds.}

\emph{P5}$_{2}$\emph{) Second, depending whether }$p_{\infty }^{2}>1$\emph{\
or} $p_{\infty }^{2}<1,$\emph{\ it follows respectively that}%
\begin{equation}
\Lambda _{CQG\infty }>\Lambda _{CQG}(s_{o}),  \label{P5_2}
\end{equation}%
\emph{or}%
\begin{equation}
\Lambda _{CQG\infty }<\Lambda _{CQG}(s_{o}).  \label{P5_2-bis}
\end{equation}%
\emph{In particular, in validity of the initial conditions (\ref%
{BBB-EQ-1-BIS}) it follows that}%
\begin{equation}
\Lambda _{CQG\infty }=\Lambda _{CQG}(s_{o}).  \label{P5_3}
\end{equation}

\emph{Proof -} The proof of proposition \emph{P5}$_{1}$ follows from
elementary algebra. Indeed, let us evaluate the limits for $s\rightarrow
+\infty $ of Eqs.(\ref{DIMENSIONLESS BBB-EQ}). In validity of Eqs.(\ref%
{BBB-EQ-2}) these become respectively%
\begin{equation}
\left\{
\begin{array}{c}
0=\frac{p_{\infty }^{2}}{4}\overline{a}_{(o)\infty }^{2}-\frac{\overline{a}%
_{(o)}^{2}}{4}(\theta _{o})+\overline{\Lambda }(s_{o})\left[ p_{\infty
}^{2}-1\right] - \\
-\overline{\Lambda }(s_{o})\left[ p_{\infty }^{3}-1\right] , \\
0=\frac{p_{\infty }^{2}}{4}\left( 4\overline{a}_{(1)\infty }^{2}+2\overline{a%
}_{(o)\infty }(\theta )\overline{a}_{(1)\infty }(\theta )\right) + \\
-\overline{\Lambda }(s_{o})\left[ p_{\infty }^{3}-1\right] -\frac{3}{4}%
\overline{a}_{(1)}^{2}(\theta _{o})+\frac{1}{4}\overline{a}_{(o)}^{2}(\theta
_{o}).%
\end{array}%
\right.  \label{BBB-EQ-3}
\end{equation}%
Subtracting the second equation from the first one it then follows
\begin{equation}
p_{\infty }^{2}=\frac{\frac{1}{2}\overline{a}_{(1)}^{2}(\theta _{o})-\frac{1%
}{4}\overline{a}_{(o)}^{2}(\theta _{o})-\overline{\Lambda }(s_{o})}{\frac{1}{%
4}\overline{a}_{(o)\infty }^{2}-\overline{\Lambda }(s_{o})},  \label{FIRST}
\end{equation}%
which implies Eq.(\ref{P5_1}). Similarly, the inequalities (\ref{P5_2}), (%
\ref{P5_2-bis}) and Eq.(\ref{P5_3}) are immediate consequences of Eqs.(\ref%
{f(s)}) and (\ref{Lambda-CQG}). \textbf{Q.E.D.}

\bigskip

THM. 5 yields sufficient conditions for the establishment of the asymptotic
behavior of the CQG--cosmological constant. It follows, depending on the
initial conditions, that the asymptotic value of the cosmological constant $%
\Lambda _{CQG\infty }$\ can in principle be either larger or smaller than
the initial value $\Lambda _{CQG}(s_{o})$\ provided that the asymptotic
limit $\overline{a}_{(o)\infty }^{2}\equiv \overline{a}_{(1)\infty }^{2}$\
is suitably well behaved.\textbf{\ }However, the issue remains under which
initial conditions (\ref{B-EQQ-1}) $p_{\infty }^{2}$ can be respectively $>1$
or $<1.$

To answer this question let us consider for definiteness the case of small
amplitude solutions corresponding to initial conditions of the type%
\begin{equation}
\left\{
\begin{array}{c}
a_{(o)}(\theta _{o})=\widehat{a}_{(o)}+\Delta a_{(o)}, \\
a_{(1)}(\theta _{o})=-\widehat{a}_{(o)}+\Delta a_{(1)}, \\
\left\vert \frac{\Delta a_{(o)}}{\widehat{a}_{(o)}}\right\vert ,\left\vert
\frac{\Delta a_{(1)}}{\widehat{a}_{(o)}}\right\vert \ll 1,%
\end{array}%
\right.  \label{initial conditions-SMALL AMPLITUDE}
\end{equation}%
namely of the form
\begin{equation}
\left\{
\begin{array}{c}
p_{\infty }=1+\Delta p_{\infty }, \\
a_{(o)\infty }=\widehat{a}_{(o)}+\Delta a_{(o)\infty }, \\
a_{(1)\infty }=-\widehat{a}_{(o)}-\Delta a_{(o)\infty }, \\
\left\vert \Delta p_{\infty }\right\vert ,\left\vert \frac{\Delta
a_{(o)\infty }}{\widehat{a}_{(o)}}\right\vert \ll 1.%
\end{array}%
\right.
\end{equation}%
Then elementary algebra shows from Eqs.(\ref{BBB-EQ}) that
\begin{equation}
\Delta p_{\infty }\cong -\frac{2\widehat{a}_{(o)}\Delta a_{(1)}}{3\overline{%
\Lambda }(\theta _{o})\left( 1-\frac{5}{24}\frac{\widehat{a}_{(o)}^{2}}{%
\overline{\Lambda }(\theta _{o})}\right) }.
\end{equation}%
This implies, therefore, that depending on the initial conditions (see in
particular Eqs.(\ref{initial conditions-SMALL AMPLITUDE})) $\Delta p_{\infty
}$ in principle can indeed be either positive or negative and hence $%
p_{\infty }$ respectively $>1$ or $<1.$

\subsection{9A - Physical implications}

Given the results established so far, physical implications and qualitative
properties of the resulting quantum-modified Einstein field equations can be
established. The main feature arising from the discussion presented above is
that it provides a generally non-stationary realization for $\Lambda
_{CQG}(s),$\ leading in turn to a corresponding non-stationary background
space-time of the form (\ref{Eq-2b}). In the case of vacuum considered here
this means that the latter may be identified, for example, with a \emph{%
non-stationary de Sitter space-time} $\left\{ \mathbf{Q}^{4},\widehat{g}%
(r,s)\right\} $, \textit{i.e.}, an expanding universe having an explicitly
proper-time dependent cosmological constant
\begin{equation}
\Lambda (s)=\Lambda _{bare}+\Lambda _{CQG}(s),  \label{f-1}
\end{equation}%
as the only source of curvature, with the explicit proper-time dependence
contained in $\Lambda _{CQG}(s)$ being prescribed according to CQG-theory
via second-quantization effects. The equation which determines the
cosmological constant (\ref{f-1}), however, still contains arbitrary free
parameters (see also Eq.(\ref{Lambda-CQG})), \textit{i.e.}, besides the
(possible) classical contribution $\Lambda _{bare}$, also $r_{th}$. One
notices,\ in particular, that\ $\Lambda _{bare}$ remains in principle
completely undetermined at this stage. Indeed no account has been given here
for a classical physical mechanism that can possibly justify a non-vanishing
contribution of this kind. For this reason,\ ruling out possible classical
modifications of Einstein field equation, its contribution can be ignored in
the present framework, thus yielding the identification%
\begin{equation}
\Lambda (s)\equiv \Lambda _{CQG}(s).  \label{F-2}
\end{equation}%
As a consequence, upon introducing the function $B(s)\equiv \left( 1-\frac{%
r^{2}}{A(s)^{2}}\right) $, with $A(s)$ identifying the de Sitter
characteristic length, the Riemann distance in the de Sitter space-time $%
\left\{ \mathbf{Q}^{4},\widehat{g}(r,s)\right\} $ when expressed in
spherical coordinates $(ct,r,\vartheta ,\varphi )$ takes the form:%
\begin{eqnarray}
ds^{2} &=&B(s)c^{2}dt^{2}-B(s)^{-1}dr^{2}+r^{2}d\Omega ^{2}  \notag \\
&\equiv &\widehat{g}_{\mu \nu }(r,s)dr^{\mu }dr^{\nu }.  \label{sol-1}
\end{eqnarray}%
Therefore the\ corresponding space-time background metric tensor becomes%
\begin{equation}
\widehat{g}_{\mu \nu }(r,s)=diag\left\{ B(s),B(s)^{-1},r^{2},r^{2}\sin
^{2}\vartheta \right\} .  \label{sol-2}
\end{equation}%
Here the parameter $A\equiv A(s)$\ is related to $\Lambda (s)$\ by means of
the prescription%
\begin{equation}
A(s)=\sqrt{\frac{3}{\Lambda (s)}}.
\end{equation}%
Hence this means that in turn $ds^{2}$\ necessarily must depend on the
maximal geodesic curve on which the tangent infinitesimal displacement $%
dr^{\mu }\equiv dr^{\mu }(s)$\ is evaluated.\ One notices, in particular,
that the same parameter must be suitably associated with the radius of the
de Sitter event horizon, \ \textit{i.e.}, the region of space-time which can
be reached only by particles which, after starting from the Big Bang event,
have traveled at the speed of light. As such they are necessarily realized
by photons whose world-lines are null geodesic trajectories. As a
consequence, for these trajectories it must be that $s=s_{o}\equiv 0$\ so
that both $A(s_{o}=0)$ and $\Lambda (s_{o}=0)$\ are necessarily identified
with pure constants. In particular $\Lambda (s_{o}=0)$\ can be identified
with the radius of the de Sitter event horizon, \textit{i.e.}, prescribed so
that $B(s_{o}=0)=0$.

Regarding the physical identification of the initial value $A(s_{o}=0)\equiv
\sqrt{\frac{3}{\Lambda (s_{o}=0)}}$,\ this would require in turn knowledge
of the precise value of $\Lambda (s_{o}=0)$. However, a possible
order-of-magnitude estimate can be achieved assuming that the initial value
of the cosmological constant was comparable to its current experimental
value, namely
\begin{equation}
\Lambda _{obs}\cong 1.2\times 10^{-52}m^{-2}.  \label{EXPERIMENTAL LAMBDA}
\end{equation}%
Thus, upon letting for example that%
\begin{equation}
\Lambda (s_{o}=0)\sim \Lambda _{obs},  \label{order of magnitude}
\end{equation}%
this yields for $A(s_{o}=0)$\ the estimate%
\begin{equation}
A(s_{o}=0)\sim \sqrt{\frac{3}{\Lambda _{obs}}}\sim 10^{10}ly.
\end{equation}%
As a consequence the initial value $A(s_{o}=0)$ is comparable, in order of
magnitude, with the theoretical estimate of the current radius of the
universe,\ namely%
\begin{equation}
\lambda _{th}\cong 1.38\times 10^{10}ly.
\end{equation}%
It must be stressed, however, that the precise estimate of $A(s_{o}=0)$ is
also subject to the validity of Eq.(\ref{F-2}), namely it can be modified if
additional (classical/quantum) contributions to the cosmological constant
are included in the theory. Regarding, instead, the still undetermined
quantum 4-scalar and dimensionless parameter $r_{th}$ a numerical estimate
can be obtained as follows. First, thanks to the prescriptions for $\alpha $
and $L$ recalled above (see Ref. \cite{cqg-4}), the ratio $\frac{\hbar ^{2}}{%
\left( \alpha L\right) ^{2}}$ becomes $\frac{\hbar ^{2}}{\left( \alpha
L\right) ^{2}}=\frac{1}{L^{2}}$, with $L$ denoting again the graviton
Compton length defined above. Next, in validity of Eq.(\ref{F-2}), let us
now require for definiteness that $\Lambda (s)$\ coincides in order of
magnitude with $\Lambda _{obs}$. In this case adopting for the graviton mass
the theoretical estimate given in Ref.\cite{cqg-4} one finds that $%
r_{th}^{2}\cong 0.326$. As a consequence the Gaussian quantum PDF (\ref%
{quantum PDF}) remains prescribed, with its half-way amplitude (namely $%
r_{th}^{2}$) being necessarily of $O\left( 1\right) $, \textit{i.e.}, safely
in the quantum regime (in fact validity of the semiclassical regime would
require instead $r_{th}^{2}\rightarrow 0$). In other words, the same PDF has
a finite "thermal spread", so that it exhibits an intrinsic quantum
character. Furthermore, the same result warrants also validity of the
equation\ (\ref{F-2}), a choice which is consistent with the graviton mass
estimate given in Ref.\cite{cqg-4}.

Equation (\ref{F-2}) is the main result of the paper. The physical
implications of the CQG-prediction of the cosmological constant are
potentially wide-range. The main one, besides the identification of the
non-stationary de Sitter space-time, concerns the physical interpretation of
the quantum origin of the cosmological constant. Unlike earlier conjectures
that the quantum contribution to the cosmological constant should be
ascribed to quantum-vacuum energy density arising from all possible quantum
fields \cite{vilenkin2001}, CQG-theory shows that $\Lambda _{QM}\equiv
\Lambda _{CQG}(s)$\ is actually produced by the Bohm interaction only due to
a quantum vacuum populated by gravitons only, \textit{i.e.}, without
requiring any additional quantum or classical field. More precisely, as
shown by Eq.(\ref{Bohm source term}), the cosmological constant behaves
generally as a non-stationary (with respect to the invariant proper-time $s$%
) field generated by the gravitational field itself through quantum self
interaction. However, it follows clearly that it is not the vacuum energy
density per se which is responsible for a non-vanishing cosmological
constant, but rather the fluctuations of its probability density. From the
mathematical point of view this is expressed by the fact that the quantum
cosmological constant term entering Eqs.(\ref{quantum-modified Einstein eq})
is generated specifically due to the gradient of the Bohm potential, which
here has the physical meaning of a\ vacuum gravitational quantum
interaction. In addition, it can be concluded that this same mechanism of
generation of $\Lambda _{QM}\ $is also consistent with the existence of
quantum massive gravitons, as it follows from the results previously
reported in Ref.\cite{cqg-4} where the inclusion of a cosmological constant
in the quantum wave equation was shown to generate a discrete
invariant-energy spectrum for the same massive gravitons characteristic of
manifestly-covariant quantum gravity theory.

Further interesting implications concern the comparison with the large-scale
structure of the universe and in particular evidences \#1-\#5 (see
subsection 1A in the Introduction). We stress that for this purpose a
systematic, \textit{i.e.}, detailed numerical, analysis of the solution (\ref%
{sol-1}) or equivalent (\ref{sol-2}) (both to be considered in validity of
Eq.(\ref{F-2})) is required, being left to future work. Nevertheless its
consistency with the large-scale structure of the universe can still be
formally established as discussed above. Let us briefly outline some of the
interesting conclusions that emerge in this way.

Consider, in particular, evidence \#1 about the flatness property of the de
Sitter space-time. Indeed, this is characterized by a Ricci curvature $4-$%
scalar $R=R(s)$\ such that $R(s)=4\Lambda (s)$. If one assumes the ordering
estimate $\Lambda (s_{o}=0)\sim \Lambda (s)$\ it follows, consistent with
such a property, that $R(s)\ll 1.$\ Similarly, evidence \#2 (the lack of
large-scale correlations among "distant" regions of universe) is direct
consequence of spherical symmetry property of the de Sitter metric tensor.
The third evidence about the isotropic and uniform character of the
expansion/acceleration of the universe at large distances is again
consistent with the same symmetry property and the strict positivity of the
quantum cosmological constant determined here. Furthermore, regarding the
"Big Bang hypothesis" (see evidence \#4), as discussed in Section 2, this is
already built-in in CQG-theory itself. Finally, concerning the issue of the
inflationary transient phase of the early universe (evidence \#5 ), we
conclude that its possible existence - based on the inequality estimates
given above - cannot be ruled out.

Finally, a comment must be made on the possible implications for cosmology
and actual physical relevance related to the determination of the
CQG-cosmological constant and specifically to the prediction of its possible
proper-time dependence achieved here. This refers to the issue whether the
theory presented here may be adequate in a quantitative sense to explain the
large-scale phenomenology\ of the universe. The same issue is particularly
relevant for the possible theoretical prediction and suggested explanation
of the observed values of the expansion rate and acceleration of the
universe.\ It seems wise to state that at this stage definite conclusions
are still premature. Indeed, before drawing definite conclusions, a
systematic analysis of the proper-time dependence predicted here for the
cosmological constant as well a deeper analysis of the same phenomenology
are required.\textbf{\ }Nevertheless, the fact is that the prediction of the
proper-time behavior determined by the CQG-cosmological constant appears%
\textbf{\ - }at least in qualitative sense - compatible or in agreement both
with the possible existence of an inflationary stage in the early universe
and the tentative suggested explanation of the phenomena of expansion and
acceleration of the universe based on CQG-theory. According to CQG-theory,
in fact, such phenomena should arise due to the quantum self-generation of
the cosmological constant, namely the vacuum quantum Bohm interaction
occurring among massive gravitons. As a consequence no additional external
sources or classical/quantum interactions are actually needed to determine
the proper-time dependence and observed value of the cosmological constant.

This conclusion departs from previous literature in at least three respects.
The first one is the physical origin of the cosmological constant. In
previous literature in fact, by far the most popular conjecture is usually
regarded to be the vacuum energy density ($\rho _{A}$) associated with dark
matter/energy as the possible physical cause able to explain both the
expansion and acceleration of the universe as well as the cosmological
constant \cite{EV3-bbb,Perlmutter,Peebles}.\textbf{\ }The second one, is
that former theoretical approaches are phenomenological in character and do
not provide a self-consistent theory for constructing either the
quantum-modified Einstein field equations or the energy density $\rho _{A}$\
itself. Third and final, there is no obvious connection between the same
quantum-modified Einstein field equations (which are manifestly covariant)
and previous quantum theories of gravity (which typically are not so).

\section{10 - Concluding remarks}

In this paper key issues have been address which are related to the
determination of the (quantum) cosmological constant in the context of
manifestly-covariant quantum gravity (CQG-theory). These have included in
particular:

\begin{enumerate}
\item The definition of the observer's proper-time ($s$), consistent with
the treatment adopted in CQG-theory of gravitons as classical
point-particles and with the Big Bang hypothesis. This is prescribed as the
arc length of a suitable non-null geodesic world-line associated with the
background metric tensor\ $\widehat{g}$,\textbf{\ }which represents a
virtual trajectory, namely one of the infinite possible physically
admissible worldlines, associated with a massive graviton. To this end the
same curve is identified in a cosmological framework with an observer's
maximal geodetics, \textit{i.e.}, a geodesic curve having the maximal arc
length and with origin point $r^{\mu }(s_{o}),$ the\ point of creation of
the same particle, coinciding with (or suitably close to) the Big Bang
event. By construction for the initial $4-$position $r^{\mu }(s_{o})$\ is
therefore such that $r^{\mu }(s_{o})\equiv r^{\mu }(s_{o}=0)$.\textbf{\ }

\item The establishment of\ the Hamiltonian structure of CQG-theory. This is
represented by a set of continuous canonical equations (referred to here as
quantum Hamilton equations) whose validity is implied by the quantum-wave
equation through its corresponding quantum Hamilton-Jacobi equation. As
shown in THM. 1, the same Hamiltonian structure remains preserved also in
validity of the said extended setting (\textit{i.e.}, for non-stationary
background metric tensor).

\item The discovery of\emph{\ }quantum-modified Einstein field equation. In
fact, the quantum Hamilton equations have been shown to admit a particular
realization in terms of a set of PDEs which is analogous to the classical
Einstein equations (\ref{EINSTEIN FIELD EQS}) but in which quantum source
terms are taken into account. Remarkably also such an equation remains
preserved under the same extended functional setting (see THM.2).

\item The establishment of the corresponding formulation of the generalized
Lagrangian path (GLP) approach.\ The issues indicated above have been cast
in the framework provided by the said, earlier formulated, GLP-approach. The
key feature of the GLP-approach unveiled here (THM.3 and THM.4) concerns its
validity also in the context of the extended functional setting and the
determination of explicit vacuum solutions of the quantum hydrodynamic
equations associated with the CQG-wave equation, with particular reference
to quantum solutions characterized by Gaussian quantum PDFs.

\item The prescription of the quantum cosmological constant, its estimate
achieved in the framework of CQG-theory and its dynamical behavior.\textbf{\
}In fact it has been shown that the cosmological constant $\Lambda \equiv
\Lambda _{CQG}(s)$ is non-stationary, \textit{i.e.},\ dependent on the
observer's proper-time $s$. The determination of the proper-time dependence
of the quantum cosmological constant has been based on the GLP-approach
which permits the construction of dynamically-consistent analytic solutions
for the quantum wave-function. As a result the relevant asymptotic
properties (for $s\rightarrow \infty $) of the $s-$dependent quantum
cosmological constant have been established (THM.5).

\item The implications and\ possible interpretation of the large-scale
phenomenology of the universe by means of an extended formulation of
CQG-theory in which the background space-time itself is non-stationary. For
this purpose the associated background metric field tensor $\widehat{g}%
\equiv \left\{ \widehat{g}_{\mu \nu }\right\} $ has been couched in an
extended functional setting in which the same tensor field is considered of
the form $\widehat{g}(r,s)\equiv \left\{ \widehat{g}_{\mu \nu }(r,s)\right\}
,$\ namely again explicitly dependent on the same proper-time $s.$
\end{enumerate}

The conclusions are relevant at least for two main reasons.

The first one refers to a peculiar emergent-gravity feature, previously
referred to (see Ref.\cite{cqg-6}) as "\emph{first-type emergent-gravity
paradigm}", according to which the Einstein field equations themselves
should be implied by quantum theory of SF-GR. As a consequence this means
that in such a context also the precise form of the background space-time,
\textit{i.e.}, the background field tensor itself should be determined in
terms of a \emph{suitable particular solution of the quantum-wave equation
appropriate for the same quantum theory}. Such a feature, in our view, can
be regarded as a true test of consistency for arbitrary quantum theories of
gravity. Indeed the ultimate goal of any theory of this type\ should be the
prediction of the background metric tensor of the universe and its
corresponding tensor field equation which in the context of SF-GR coincides
with the Einstein field equations. In this paper such a property has been
shown to hold for CQG-theory based on the quantum-modified Einstein field
equations indicated above. The remarkable feature is, in fact, that as shown
by THM.2 these are achieved without performing any limiting approximation,
such as the semiclassical continuum limit which is obtained letting $\hslash
\rightarrow 0$ in the quantum-wave equation.

However, in this connection, two additional side-consequences follow.\textbf{%
\ }The first one is a fundamental physical restriction. Indeed, due to the
manifest-covariance property of the Einstein field equations a stringent
condition arises also on the class of possible, \textit{i.e.},
physically-admissible, quantum field theories of gravity. In fact, it is
obvious that the same ones should necessarily be restricted to
manifestly-covariant ones, namely realized, as CQG-theory, by means of a $4-$%
tensor field theory endowed with tensor properties prescribed with respect
to the same background field tensor $\widehat{g}$.\textbf{\ }The second
side-implication concerns the physical interpretation and role of
CQG-theory. In fact it cannot merely be viewed as one of the possible
background space-time theories, \textit{i.e.}, in which the background field
tensor is arbitrarily prescribed as a particular solution of the classical
Einstein field equations. On the contrary it must be intended as a truly
self-consistent quantum theory of gravity in which the background field
tensor $\widehat{g}$\ is a solution of the corresponding quantum-modified
Einstein field equations, namely in which the cosmological constant is, in
turn, uniquely prescribed by means of CQG-theory\ itself.

The further aspect worth to be mentioned concerns the explicit prescription
and properties of the background space-time arising in the context of
CQG-theory.\ This is identified here with the non-stationary de Sitter
space-time $\left\{ \mathbf{Q}^{4},\widehat{g}(r,s)\right\} ,$\ being $%
\widehat{g}(r,s)$\ the corresponding metric field-tensor characterized by
the non-stationary cosmological constant $\Lambda \equiv \Lambda _{CQG}(s)$%
.\ In particular, remarkable features which emerge in this connection are
that: A) The prediction of the initial value $\Lambda _{CQG}(s_{o}=0)$\
obtained here is consistent with the graviton mass estimate established in a
previous paper in the case (see Ref.\cite{cqg-4}). B) The unique
second-quantization character of the cosmological constant, which arises due
to the Bohm interaction and more precisely due to the gradient of the Bohm
effective quantum potential.\textbf{\ }C) The universal property of the
cosmological constant, i.e., the fact that its existence is independent of
the possible presence of additional external fields and further classical or
quantum interactions.

These conclusions provide also further theoretical insight on the axiomatic
foundations as well as physical implications of CQG-theory for large-scale
phenomena of the universe. In the present investigation crucially-important
second-quantization aspects of the theory have been studied. These concern
the possible explanation, in such a context, of the physical mechanism
responsible for the occurrence of the cosmological constant, as well the
possible existence of an inflationary phenomenon in the early universe, and,
in turn, also a suggested physical explanation for the closely-related
observed values of the expansion rate and acceleration of the universe.

Nevertheless, CQG-theory is still built upon a first-quantization approach
(realized by the so-called $g$-quantization) which fulfills the quantum
unitarity principle and, consequently, the conservation of quantum
probability associated with the quantum wave function. As such, no
trans-Planckian effects, nor possible information losses arising at event
horizons in black-hole space-times are taken into account in the current
formulation of CQG-theory. Nevertheless, the inclusion of additional
second-quantization effects is in principle possible, with particular
reference to quantum modifications of the background space-time at the
Planck length, such as the inclusion of localized quantum particle sources.%
\textbf{\ }In view of these considerations, CQG-theory may be expected to
provide fertile grounds for new conceptual developments and a variety of
applications in quantum gravity and quantum cosmology.

\textit{Acknowledgments -} This work is dedicated to the dearest memory of
Flavia, wife of M.T. and motherly friend of C.C., recently passed away.
Investigation developed within the research project of the Albert Einstein
Center for Gravitation and Astrophysics, Czech Science Foundation No.
14-37086G (M.T.). The authors acknowledge institutional support by the
Silesian University in Opava, Czech Republic. The authors are grateful to
the International Center for Theoretical Physics (ICTP\ Trieste, Italy) for
the hospitality during the initial preparation of the manuscript.

\section{Appendix A - Classical kinetic and normalized effective potential
densities}

The effective kinetic and the normalized effective potential density $T_{R}$
and $V$ appearing in the classical Hamiltonian density $H_{R}$ (\ref%
{classical Hamiltonian density})\ take the form (see Refs.\cite{cqg-3,cqg-4}%
)
\begin{equation}
\left\{
\begin{array}{c}
T_{R}\equiv \frac{1}{2\alpha Lf(h)}\pi _{\mu \nu }\pi ^{\mu \nu }, \\
V\left( g,\widehat{g},r,s\right) \equiv \sigma V_{o}\left( g,\widehat{g}%
,r,s\right) +\sigma V_{F}\left( g,\widehat{g},r,s\right) ,%
\end{array}%
\right.  \label{App-1}
\end{equation}%
with $h$ being the variational weight-factor%
\begin{equation}
h(g,\widehat{g}(r,s))=2-\frac{1}{4}g^{\alpha \beta }g^{\mu \nu }\widehat{g}%
_{\alpha \mu }(r,s)\widehat{g}_{\beta \nu }(r,s),
\end{equation}%
while $L$ and $\alpha $ are constants, \textit{i.e.}, suitable $4-$scalars
both identified according to the treatment given in Ref.\cite{cqg-4}. In
addition, $V$ and $V_{F}$ represent respectively the vacuum and external
field contributions (see definitions in Ref.\cite{cqg-3}),
\begin{equation}
\begin{array}{c}
V_{o}\equiv h\alpha L\left[ g^{\mu \nu }\widehat{R}_{\mu \nu }-2\Lambda %
\right] , \\
V_{F}\left( g,\widehat{g},r\right) \equiv hL_{F}\left( g,\widehat{g}%
,r\right) ,%
\end{array}
\label{App-2}
\end{equation}%
where $\widehat{R}_{\mu \nu }\equiv R_{\mu \nu }(\widehat{g})$ and $\Lambda $
identify respectively the background Ricci tensor and the cosmological
constant, $L_{F}$ being associated with a non-vanishing stress-energy
tensor, while $f(h)$\ and $\sigma $\ denote suitable multiplicative gauge%
\emph{\ }functions identified with $f(h)=1$ and $\sigma =-1.$

\section{Appendix B - Covariant partial derivative}

In the case of non-stationary background metric tensor (see Eq.(\ref{Eq-2a}%
)) the operator $\left. \frac{d}{ds}\right\vert _{r}$ appearing in Eq.(\ref%
{covariant s-derivative}) must be prescribed in such a way to satisfy the
covariance property (\ref{LPT}) also for the covariant and countervariant
components of the tensor field%
\begin{equation}
\left\{
\begin{array}{c}
H_{\mu \nu }\equiv \left. \frac{d}{ds}\right\vert _{r}x_{\mu \nu }, \\
H^{\mu \nu }\equiv \left. \frac{d}{ds}\right\vert _{r}x^{\mu \nu },%
\end{array}%
\right.  \label{A-0}
\end{equation}%
with $x_{\mu \nu }$ denoting the canonical $4-$tensors $x_{\mu \nu }=g_{\mu
\nu },\pi _{\mu \nu }$ for which by construction%
\begin{equation}
\left\{
\begin{array}{c}
x^{pq}\widehat{g}_{\mu ^{\prime }p}\widehat{g}_{\nu ^{\prime }q}=x_{\mu
^{\prime }\nu ^{\prime }}, \\
x_{pq}\widehat{g}^{\mu ^{\prime }p}\widehat{g}^{\nu ^{\prime }q}=x^{\mu
^{\prime }\nu ^{\prime }}.%
\end{array}%
\right.  \label{A-1a}
\end{equation}%
For definiteness, denoting in the following $\widehat{g}_{\mu \alpha }=%
\widehat{g}_{\mu \alpha }(r,s)$ and $\widehat{g}^{\mu \beta }=\widehat{g}%
^{\mu \beta }(r,s),$ the identities $\left. \frac{d}{ds}\right\vert _{s}%
\widehat{g}_{\mu \alpha }\equiv \left. \frac{d}{ds}\right\vert _{s}\widehat{g%
}^{\mu \beta }\equiv 0$ hold. Furthermore, noting that the non-stationary
background metric tensor must satisfy as well the orthogonality condition%
\begin{equation}
\widehat{g}_{\mu \alpha }\widehat{g}^{\mu \beta }=\delta _{\alpha }^{\beta },
\label{A-1}
\end{equation}%
it follows that also the identity%
\begin{equation}
\widehat{g}^{\mu \beta }\frac{\partial }{\partial s}\widehat{g}_{\mu \alpha
}+\widehat{g}_{\mu \alpha }\frac{\partial }{\partial s}\widehat{g}^{\mu
\beta }=0  \label{A-2}
\end{equation}%
must hold. Then one can prove that, thanks to Eqs.(\ref{A-1a}),(\ref{A-1})
and (\ref{A-2}), the following prescriptions for the covariant partial
derivatives $\left. \frac{d}{ds}\right\vert _{r}$ hold:%
\begin{equation}
\left\{
\begin{array}{c}
\left. \frac{d}{ds}\right\vert _{r}x^{\alpha \beta }=\frac{\partial }{%
\partial s}x^{\alpha \beta }-\frac{1}{2}x^{pq}\widehat{g}_{\mu ^{\prime }p}%
\widehat{g}_{\nu ^{\prime }q}\frac{\partial }{\partial s}(\widehat{g}%
^{\alpha \mu ^{\prime }}\widehat{g}^{\beta \nu ^{\prime }}), \\
\left. \frac{d}{ds}\right\vert _{r}x_{\alpha \beta }=\frac{\partial }{%
\partial s}x_{\alpha \beta }-\frac{1}{2}x_{pq}\widehat{g}^{\mu ^{\prime }p}%
\widehat{g}^{\nu ^{\prime }q}\frac{\partial }{\partial s}(\widehat{g}%
_{\alpha \mu ^{\prime }}\widehat{g}_{\beta \nu ^{\prime }}).%
\end{array}%
\right.  \label{A-3}
\end{equation}%
In view of the orthogonality conditions (\ref{A-1}) these can be
equivalently written as%
\begin{equation}
\left\{
\begin{array}{c}
\left. \frac{d}{ds}\right\vert _{r}x^{\alpha \beta }=\frac{\partial }{%
\partial s}x^{\alpha \beta }-\frac{1}{2}x_{\mu ^{\prime }\nu ^{\prime }}%
\frac{\partial }{\partial s}(\widehat{g}^{\alpha \mu ^{\prime }}\widehat{g}%
^{\beta \nu ^{\prime }}), \\
\left. \frac{d}{ds}\right\vert _{r}x_{\alpha \beta }=\frac{\partial }{%
\partial s}x_{\alpha \beta }-\frac{1}{2}x^{\mu ^{\prime }\nu ^{\prime }}%
\frac{\partial }{\partial s}(\widehat{g}_{\alpha \mu ^{\prime }}\widehat{g}%
_{\beta \nu ^{\prime }}).%
\end{array}%
\right.  \label{A-3bis}
\end{equation}%
To prove Eqs.(\ref{A-3}) one notices in fact that%
\begin{eqnarray}
&&\left. \widehat{g}_{\mu \alpha }\widehat{g}_{\nu \beta }\left. \frac{d}{ds}%
\right\vert _{r}x^{\alpha \beta }=\widehat{g}_{\mu \alpha }\widehat{g}_{\nu
\beta }\frac{\partial }{\partial s}x^{\alpha \beta }\right.  \notag \\
&&-\widehat{g}_{\mu \alpha }\widehat{g}_{\nu \beta }\frac{1}{2}x^{pq}%
\widehat{g}_{\mu ^{\prime }p}\widehat{g}_{\nu ^{\prime }q}\frac{\partial }{%
\partial s}(\widehat{g}^{\alpha \mu ^{\prime }}\widehat{g}^{\beta \nu
^{\prime }}),
\end{eqnarray}%
where%
\begin{equation}
\widehat{g}_{\mu \alpha }\widehat{g}_{\nu \beta }\frac{\partial }{\partial s}%
x^{\alpha \beta }=\frac{\partial }{\partial s}x_{\mu \nu }-x^{\alpha \beta }%
\frac{\partial }{\partial s}(\widehat{g}_{\mu \alpha }\widehat{g}_{\nu \beta
}),
\end{equation}%
and%
\begin{eqnarray}
&&\left. \widehat{g}_{\mu \alpha }\widehat{g}_{\nu \beta }\frac{1}{2}x^{pq}%
\widehat{g}_{\mu ^{\prime }p}\widehat{g}_{\nu ^{\prime }q}\frac{\partial }{%
\partial s}(\widehat{g}^{\alpha \mu ^{\prime }}\widehat{g}^{\beta \nu
^{\prime }})=\right.  \notag \\
&&\left. -\widehat{g}^{\alpha \mu ^{\prime }}\widehat{g}^{\beta \nu ^{\prime
}}\frac{1}{2}x^{pq}\widehat{g}_{\mu ^{\prime }p}\widehat{g}_{\nu ^{\prime }q}%
\frac{\partial }{\partial s}(\widehat{g}_{\mu \alpha }\widehat{g}_{\nu \beta
})=\right.  \notag \\
&&\left. -\frac{1}{2}x^{\alpha \beta }\frac{\partial }{\partial s}(\widehat{g%
}_{\mu \alpha }\widehat{g}_{\nu \beta }).\right.
\end{eqnarray}%
As a consequence, elementary algebra yields%
\begin{eqnarray}
\widehat{g}_{\mu \alpha }\widehat{g}_{\nu \beta }\left. \frac{d}{ds}%
\right\vert _{r}x^{\alpha \beta } &=&\frac{\partial }{\partial s}x_{\mu \nu
}-\frac{1}{2}x^{\alpha \beta }\frac{\partial }{\partial s}(\widehat{g}_{\mu
\alpha }\widehat{g}_{\nu \beta })\equiv  \notag \\
&\equiv &\left. \frac{d}{ds}\right\vert _{r}x_{\mu \nu },
\end{eqnarray}%
and similarly%
\begin{equation}
\widehat{g}^{\mu \alpha }\widehat{g}^{\nu \beta }\left. \frac{d}{ds}%
\right\vert _{r}x_{\alpha \beta }=\left. \frac{d}{ds}\right\vert _{r}x^{\mu
\nu }.
\end{equation}%
Furthermore, as a consequence of the prescriptions (\ref{A-3}) (or
equivalent of Eqs.(\ref{A-3bis})), it is immediate to prove the distributive
property holds%
\begin{equation}
D_{s}\left( x^{\alpha \beta }x_{\alpha \beta }\right) =x^{\alpha \beta }%
\frac{d}{ds}x_{\alpha \beta }+x_{\alpha \beta }\frac{d}{ds}x^{\alpha \beta },
\end{equation}%
with $D_{s}$\ and $\frac{d}{ds}$\ being respectively the covariant $s-$%
derivatives (\ref{cov-scalar-1}) and (\ref{covariant s-derivative}). This
proves the validity of the covariance property for the $4-$tensor field
defined by Eqs.(\ref{A-0}) and hence of the prescriptions (\ref{A-3}) for
the covariant partial derivatives $\left. \frac{d}{ds}\right\vert _{r}$.

\section{Appendix C - Determination of the $4-$scalar factor $p(s)$}

In analogy to Ref.\cite{cqg-6} let us introduce for the quantum
phase-function $\mathcal{S}^{(q)}(G_{L}(s),\Delta g,s)$ the "harmonic"
polynomial decomposition realized in terms of a second-degree polynomial of
the form (\ref{HARM-1}). Then, in analogy to Ref.\cite{cqg-6} the
determination of the $4-$scalar factor $p(s)$ is provided by the following
propositions (with proofs analogous to those given in Appendix B of Ref.\cite%
{cqg-6}).

\bigskip

\textbf{Proposition C1 - Determination of the tensor field }$\frac{\partial
\Delta g_{\beta }^{\alpha }}{\partial \delta g_{L\nu }^{\mu }(s^{\prime })}$

\emph{Given validity of the polynomial representation (\ref{HARM-1}), the
tensor field }$\frac{\partial \Delta g_{\beta }^{\alpha }}{\partial \delta
g_{L\nu }^{\mu }(s^{\prime })}$\emph{\ takes the form}%
\begin{equation}
\frac{\partial \Delta g_{\beta }^{\alpha }}{\partial \delta g_{L\nu }^{\mu
}(s^{\prime })}=-\frac{\partial \Delta g_{\beta }^{\alpha }}{\partial
G_{L\nu }^{\mu }(s^{\prime })},  \label{A-0A}
\end{equation}%
\emph{with}%
\begin{equation}
\frac{\partial \Delta g_{\beta }^{\alpha }}{\partial \delta g_{L\nu }^{\mu
}(s^{\prime })}=\delta _{\mu }^{\alpha }\delta _{\beta }^{\nu }p(s),
\label{A-1A}
\end{equation}%
\emph{and }$p(s)$\emph{\ being the }$4-$\emph{scalar function determined by
the integral equation}%
\begin{equation}
p(s)=\frac{1}{1+\int_{s_{o}}^{s}ds^{\prime }\frac{1}{\alpha L}a(s^{\prime
})p(s^{\prime })}.  \label{A-1ter}
\end{equation}%
\emph{Here }$a(s)$ \emph{is the }$4-$\emph{scalar function }%
\begin{equation}
a(s)\equiv \frac{1}{4}a_{q\beta }^{p\alpha }(s)\delta _{p}^{q}\emph{\ }%
\delta _{\alpha }^{\beta },  \label{A-1quater}
\end{equation}%
\emph{with }$a_{q\beta }^{p\alpha }(s)$ \emph{being the tensor introduced in
the polynomial decomposition of the phase function }$\mathcal{S}^{(q)}$\emph{%
\ given above by Eq.(\ref{HARM-1}).}

\textbf{Proposition C2 - Determination of the }$4-$\textbf{scalar function }$%
p(s)$

\emph{In validity of Eq.(\ref{A-1ter}) it follows that}%
\begin{equation}
\left\vert p(s)\right\vert =\frac{1}{\left( 1+\frac{2}{\alpha L}%
\int\limits_{s_{o}}^{s}ds^{\prime }a(s^{\prime })\right) ^{1/2}},
\label{A-4}
\end{equation}%
\emph{which upon requiring }$p(s_{o})=1$\emph{\ delivers}
\begin{equation}
p(s)=\frac{1}{\left( 1+\frac{2}{\alpha L}\int\limits_{s_{o}}^{s}ds^{\prime
}a(s^{\prime })\right) ^{1/2}}.  \label{A-4BIS}
\end{equation}

\bigskip


\begin{thebibliography}{999}
\bibitem{cqg-1} C. Cremaschini and M. Tessarotto,\textit{\ Synchronous
Lagrangian variational principles in General Relativity. }Eur. Phys. J. Plus
\textbf{130}, 123 (2015).

\bibitem{cqg-2} C. Cremaschini and M. Tessarotto,\textit{\ Manifest
covariant Hamiltonian theory of General Relativity. }Applied Physics
Research \textbf{8}, 2 (2016).

\bibitem{cqg-3} C. Cremaschini and M. Tessarotto, \textit{Hamiltonian
approach to GR -- Part 1: covariant theory of classical gravity. }Eur. Phys.
J. C \textbf{77}, 329 (2017).

\bibitem{cqg-4} C. Cremaschini and M. Tessarotto, \textit{Hamiltonian
approach to GR -- Part 2: covariant theory of quantum gravity. }Eur. Phys.
J. C \textbf{77}, 330 (2017).

\bibitem{cqg-5} C. Cremaschini and M. Tessarotto, \textit{Quantum-wave
equation and Heisenberg inequalities of covariant quantum gravity. }Entropy
\textbf{19}(7), 339 (2017).

\bibitem{cqg-6} M. Tessarotto and C. Cremaschini, \textit{Generalized
Lagrangian path approach to manifestly-covariant quantum gravity theory. }%
Entropy \textbf{20}(3), 205 (2018).

\bibitem{eli0} A. Escofet and E. Elizalde, \textit{Gauss-Bonnet modified
gravity models with bouncing behavior.} Modern Physics Letters A \textbf{31}%
, 1650108 (2016).

\bibitem{eli1} E. Elizalde, S.V. Odintsov, L. Sebastiani and R. Myrzakulov,
\textit{Beyond-one-loop quantum gravity action yielding both inflation and
late-time acceleration.} Nuclear Physics B \textbf{921}, 411 (2017).

\bibitem{Messiah} A. Messiah, \textit{Quantum Mechanics}, Dover Pubs, New
York, U.S.A. (1999).

\bibitem{ein1} A. Einstein, \textit{The Meaning of Relativity}, Princeton
University Press, Princeton, N.J., U.S.A. (2004).

\bibitem{LL} L. D. Landau and E.M. Lifschitz, \textit{Field Theory,
Theoretical Physics Vol.2}, Addison-Wesley, N.Y., U.S.A. (1957).

\bibitem{gravi} C.W. Misner, K.S. Thorne and J.A. Wheeler, \textit{%
Gravitation}, W.H. Freeman, 1st edition (1973).

\bibitem{noi4} M. Tessarotto and C. Cremaschini, \textit{Theory of Nonlocal
Point Transformations in General Relativity}, Adv. Math. Phys. \textbf{2016}%
, 9619326 (2016). DOI: http://dx.doi.org/10.1155/2016/9619326.

\bibitem{Jordi-Narciso} Gaset Jordi and Rom\'{a}n-Roy Narciso, \textit{%
Multisymplectic unified formalism for Einstein-Hilbert gravity.} Journal of
Mathematical Physics \textbf{59}, 032502 (2018).

\bibitem{FoP-2016} M. Tessarotto and C. Cremaschini, \textit{Generalized
Lagrangian-path representation of non-relativistic quantum mechanics. }%
Found. Phys. \textbf{46(8)}, 1022 (2016).

\bibitem{FoP-2016b} M. Tessarotto, M. Mond and D. Batic, \textit{Hamiltonian
Structure of the Schr\"{o}dinger Classical Dynamical System.} Found. Phys.
\textbf{46(9)}, 1127 (2016).

\bibitem{flat4} M.J. Mortonson, \textit{Testing flatness of the universe
with probes of cosmic distances and growth.} Phys. Rev. D \textbf{80},
123504 (2009).

\bibitem{flat3} P. M. Okouma, Y. Fantaye and B.A. Bassett, \textit{How flat
is our Universe really?} Phys. Lett. B \textbf{719}, 1-4 (2013).

\bibitem{flat2} T. Paw\l owski, R. Pierini and E. Wilson-Ewing, \textit{Loop
quantum cosmology of a radiation-dominated flat FLRW universe.} Phys. Rev. D
\textbf{90}, 123538 (2014).

\bibitem{flat1} M. Fathi, S. Jalalzadeh and P.V. Moniz, \textit{Classical
universe emerging from quantum cosmology without horizon and flatness
problems.} Eur. Phys. J. C \textbf{76}, 527 (2016).

\bibitem{Winberg2000} S. Weinberg, \textit{The Cosmological Constant Problem,%
} Rev. Mod. Phys. \textbf{16}, 1-23 (1989).

\bibitem{Carroll2004} S. Carroll, \textit{Spacetime and Geometry,} Addison
Wesley, San Francisco, CA. 171-174 (2004).

\bibitem{EV1-a} T. Paw\l owski and A. Ashtekar, \textit{Positive
cosmological constant in loop quantum cosmology.} Phys. Rev. D \textbf{85},
064001 (2012).

\bibitem{EV1-aa} J. Holland and S. Hollands, \textit{A small cosmological
constant due to non-perturbative quantum effects.} Class. Quant. Grav.
\textbf{31}, 125006 (2014).

\bibitem{EV1-aaa} M. Szyd\l owski, \textit{Cosmological model with decaying
vacuum energy from quantum mechanics.} Phys. Rev. D \textbf{91}, 123538
(2015).

\bibitem{EV1-aaaa} I. Oda, \textit{Quantum aspects of nonlocal approach to
the cosmological constant problem.} Phys. Rev. D \textbf{96}, 024027 (2017).

\bibitem{EV1-b} \L . Szulc, \textit{An open FRW model in loop quantum
cosmology.} Class. Quant. Grav. \textbf{24}, 6191 (2007).

\bibitem{EV1-c} D. Brizuela, G.A. Mena Marugan and T. Paw\l owski, \textit{%
FAST TRACK COMMUNICATION: Big Bounce and inhomogeneities}. Class. Quant.
Grav. \textbf{27}, 052001 (2010).

\bibitem{EV1-d} H.M. Sadjadi, \textit{On solutions of loop quantum cosmology.%
} Eur. Phys. J. C \textbf{73}, 2571 (2013).

\bibitem{EV1-e} Yi-Fu Cai and E. Wilson-Ewing, \textit{Non-singular bounce
scenarios in loop quantum cosmology and the effective field description.}
Journal of Cosmology and Astroparticle Physics \textbf{03}, 026 (2014).

\bibitem{EV1-f} I, Agullo, \textit{Loop quantum cosmology, non-Gaussianity,
and CMB power asymmetry. }Phys. Rev. D \textbf{92}, 064038 (2015).

\bibitem{EV1-g} V.K. Oikonomou, \textit{Inflation and bounce from classical
and loop quantum cosmology imperfect fluids.} Int. J. Mod. Phys. D \textbf{26%
}, 1750110 (2017).

\bibitem{EV1-h} E. Alesci, G. Botta, F. Cianfrani and S. Liberati, \textit{%
Cosmological singularity resolution from quantum gravity: the
emergent-bouncing universe.} Phys. Rev. D \textbf{96}, 046008 (2017).

\bibitem{EV1-j} Bao-Fei Li, P. Singh and A. Wang, \textit{Towards
cosmological dynamics from loop quantum gravity.} Phys. Rev. D \textbf{97},
084029 (2018).

\bibitem{EV2-a} C.W. Misner, \textit{The Isotropy of the Universe. }%
Astrophysical Journal \textbf{151}, 431 (1968).

\bibitem{EV2-b} G. Gamow, \textit{Observational Properties of the
Homogeneous and Isotropic Expanding Universe.} Phys. Rev. Lett. \textbf{20},
1310 (1968).

\bibitem{EV2-c} C.B. Collins and S.W. Hawking, \textit{Why is the Universe
Isotropic?} Astrophysical Journal \textbf{180}, 317 (1973).

\bibitem{EV2-d} S.W. Hawking and J.C. Luttrell, \textit{The isotropy of the
universe.} Phys. Lett. B \textbf{143}, 83 (1984).

\bibitem{EV2-e} P. Anninos, R.A. Matzner, T. Rothman, M.P. Ryan Jr., \textit{%
How does inflation isotropize the Universe?} Phys. Rev. D \textbf{43}, 3821
(1991).

\bibitem{EV2-f} J.D. Barrow and H. Kodama, \textit{The isotropy of compact
universes. }Class. Quant. Grav. \textbf{18}, 1753 (2001).

\bibitem{EV2-g} S. R\"{a}s\"{a}nen, \textit{Relation between the isotropy of
the CMB and the geometry of the universe.} Phys. Rev. D \textbf{79}, 123522
(2009).

\bibitem{EV2-h} D. Saadeh, S.M. Feeney, A. Pontzen, H.V. Peiris, J.D.
McEwen, \textit{How Isotropic is the Universe?} Phys. Rev. Lett. \textbf{117}%
, 131302 (2016).

\bibitem{EV2-i} J. \v{R}\'{\i}pa and A. Shafieloo, \textit{Testing the
Isotropic Universe Using the Gamma-Ray Burst Data of Fermi/GBM. }The
Astrophysical Journal \textbf{851}, 15 (2017).

\bibitem{EV2-j} Y. Wang and F.Y. Wang, \textit{Testing the isotropy of the
Universe with Type Ia supernovae in a model-independent way.} Monthly
Notices of the Royal Astronomical Society \textbf{474}, 3516 (2018).

\bibitem{EV3-b} N. Pinto-Neto and E.S. Santini, \textit{The accelerated
expansion of the Universe as a quantum cosmological effect.} Phys. Lett. A
\textbf{315}, 36 (2003).

\bibitem{EV3-bb} B.P. Schmidt, \textit{Nobel Lecture: Accelerating expansion
of the Universe through observations of distant supernovae.} Reviews of
Modern Physics \textbf{84}, 1151 (2012).

\bibitem{EV3-a} B.S. Haridasu, V.V. Lukovi\'{c}, R. D'Agostino and N.
Vittorio, \textit{Strong evidence for an accelerating Universe.} Astronomy
\& Astrophysics \textbf{600}, L1 (2017).

\bibitem{Susskind2005} Leonard Susskind, \textit{The Cosmic Landscape:
String Theory and the Illusion of Intelligent Design}, Little, Brown and
Company, New York (2005).

\bibitem{EV4-b} H. Nussbaumer, \textit{The Discovery of the Expanding
Universe and 80 Years of Big Bang}. Int. J. Mod. Phys. D \textbf{20}, 87
(2011).

\bibitem{EV4-a} E. Rebhan, \textit{Cosmic inflation and big bang interpreted
as explosions.} Phys. Rev. D \textbf{86}, 123012 (2012).

\bibitem{EV4-d} R. Penrose, \textit{Singularities in big-bang cosmology.}
Royal Astronomical Society, Quarterly Journal (ISSN 0035-8738) \textbf{29},
61 (1988).

\bibitem{EV4-dd} M.V. Battisti and G. Montani, \textit{The Big-Bang
singularity in the framework of a Generalized Uncertainty Principle. }Phys.
Lett. B \textbf{656}, 96 (2007).

\bibitem{EV4-c} T.A. Koslowski, F. Mercati and D. Sloan, \textit{Through the
big bang: Continuing Einstein's equations beyond a cosmological singularity.}
Phys. Lett. B \textbf{778}, 339 (2018).

\bibitem{inflation-5} A. Ashtekar and A. Barrau, \textit{Some Conceptual
Issues in Loop Quantum Cosmology.} Class. Quant. Grav. \textbf{32}, 234001
(2015).

\bibitem{inflation-4} M. Benetti, S.J. Landau and J.S. Alcaniz, \textit{%
Constraining quantum collapse inflationary models with CMB data. }Journal of
Cosmology and Astroparticle Physics \textbf{12}, 035 (2016).

\bibitem{inflation-3} G. Le\'{o}n, \textit{Eternal inflation and the quantum
birth of cosmic structure.} Eur. Phys. J. C \textbf{77}, 705 (2017).

\bibitem{inflation-2} R. Brandenberger,\ \textit{Initial conditions for
inflation --- A short review.} Int. J. Mod. Phys. D \textbf{26}, 1740002-126
(2017).

\bibitem{inflation-1} M. Szyd\l owski and A. Stachowski, \textit{Simple
cosmological model with inflation and late times acceleration.} Eur. Phys.
J. C \textbf{78}, 249 (2018).

\bibitem{inflation} A.H. Guth,\textit{\ Inflationary universe: A possible
solution to the horizon and flatness problems.} Phys. Rev. \textbf{D 23},
347--356 (1981).

\bibitem{Planck} P.A.R. Ade, N Aghanim, C Armitage-Caplan, M Arnaud, et al.,
\textit{Planck 2015 results. XIII. Cosmological parameters}. Astronomy \&
Astrophysics \textbf{594}, A13 (2016).

\bibitem{Einstein1917} A. Einstein, \textit{Kosmologische Betrachtungen zur
allgemeinen Relativit\"{a}tstheorie} (\textit{Cosmological Considerations in
the General Theory of Relativity}), Koniglich Preu\ss ische Akademie der
Wissenschaften, Sitzungsberichte (Berlin): 142 (1917).

\bibitem{ivanov2015} A.N. Ivanov and M. Wellenzohn, \textit{Standard
electroweak interactions in gravitational theory with chameleon field and
torsion.} Phys. Rev. D \textbf{91}, 085025 (2015).

\bibitem{ivanov2016} A.N. Ivanov and M. Wellenzohn, \textit{Einstein-Cartan
Gravity with Torsion Field Serving as an Origin for the Cosmological
Constant or Dark Energy Density. }The Astrophysical Journal \textbf{829}, 47
(2016).

\bibitem{azri2012} H. Azri and A. Bounames, \textit{Geometrical origin of
the cosmological constant.} General Relativity and Gravitation \textbf{44},
2547 (2012).

\bibitem{Lu-2012} J. Lu, L. Ma, M. Liu and Y. Wu, \textit{Time variable
cosmological constant of holographic origin with interaction in Brans--Dicke
theory}, Int. J. Mod. Phys. D \textbf{21}, 1250005 (2012).

\bibitem{dym-1} I. Dymnikova, \textit{The cosmological term as a source of
mass.} Class. Quant. Grav. \textbf{19}, 725 (2002).

\bibitem{dym-2} I. Dymnikova, \textit{From vacuum nonsingular black hole to
variable cosmological constant.} Grav. Cosmol. Suppl. \textbf{8N1}, 131
(2002).

\bibitem{dym-3} K.A. Bronnikov, A. Dobosz and I.G. Dymnikova, \textit{%
Nonsingular vacuum cosmologies with a variable cosmological term.} Class.
Quant. Grav. \textbf{20}, 3797 (2003).

\bibitem{maje2003} V. Majern\'{\i}k, \textit{Letter: A Cosmological Constant
Interpreted as the Field Energy of a Quaternionic Field.} General Relativity
and Gravitation \textbf{35}, 1833 (2003).

\bibitem{azri2017} H. Azri and A. Bounames, \textit{Cosmological
Consequences of a Variable Cosmological Constant Model}. Int. J. Mod. Phys.
D \textbf{26}, 1750060 (2017).

\bibitem{vacuum} S. Rugh and H. Zinkernagel, \textit{The quantum vacuum and
the cosmological constant problem. }Studies in History and Philosophy of
Modern Physics \textbf{33}, 663 (2001).

\bibitem{Wang0} Q. Wang, Z. Zhu and W. Unruh, \textit{How the huge energy of
quantum vacuum gravitates to drive the slow accelerating expansion of the
Universe. }Phys. Rev. D \textbf{95}, 103504 (2017).

\bibitem{Wang} S. S. Cree, T.M. Davis, T.C. Ralph, Q. Wang, Z.Zhu and
W.G.Unruh, \textit{Can the fluctuations of the quantum vacuum solve the
cosmological constant problem?} arXiv:1805.12293v1 [gr-qc] (2018).

\bibitem{vilenkin2001} J. Garriga and A. Vilenkin, \textit{Solutions to the
Cosmological Constant Problems.} Phys. Rev. D \textbf{64}, 023517 (2001).

\bibitem{Critte} R. Crittenden, E. Majerotto and F. Piazza, \textit{%
Measuring Deviations from a Cosmological Constant: A Field-Space
Parametrization.} Phys. Rev. Lett. \textbf{98}, 251301 (2007).

\bibitem{Barrow} J.D. Barrow and D.J. Shaw, \textit{New solution of the
cosmological constant problems. }Phys. Rev. Lett.\textbf{106, }101302 (2011).

\bibitem{Cicciarella2017} F. Cicciarella and M. Pieroni, \textit{%
Universality for quintessence.} Journal of Cosmology and Astroparticle
Physics \textbf{08}, 010 (2017).

\bibitem{Asenjo2017} F.A. Asenjo and S.A. Hojman, \textit{Class of Exact
Solutions for a Cosmological Model of Unified Gravitational and Quintessence
Fields.} Found. Phys. \textbf{47}, 887 (2017).

\bibitem{Rovelli2011} E. Bianchi, T. Krajewski, C. Rovelli and F. Vidotto,
\textit{Cosmological constant in spinfoam cosmology. }Phys. Rev. D \textbf{83%
}, 104015 (2011).

\bibitem{Nicolini} R. Garattini and P. Nicolini, \textit{Noncommutative
approach to the cosmological constant problem.} Phys. Rev. D \textbf{83},
064021 (2011).

\bibitem{ray1} A.F. Ali and S. Das, \textit{Cosmology from quantum potential.%
} Phys. Lett. B \textbf{741}, 276 (2014).

\bibitem{ray2} S. Das, \textit{Quantum Raychaudhuri equation.} Phys. Rev. D
\textbf{89}, 084068 (2014).

\bibitem{ray3} E.I. Lashin, \textit{On the correctness of cosmology from
quantum potential.} Mod. Phys. Lett. A \textbf{31}, 07 (2016).

\bibitem{EV3-bbb} A.G. Riess, \textit{Nobel Lecture: My path to the
accelerating Universe.} Reviews of Modern Physics \textbf{84}, 1165 (2012).

\bibitem{Perlmutter} S. Perlmutter, et al., \textit{Measurements of Omega
and Lambda from 42 High-Redshift Supernovae.} Astron. J.\textbf{\ 517}, 565
(1999).

\bibitem{Peebles} P. J. E. Peebles and B. Ratra, \textit{The cosmological
constant and dark energy.} Rev. Mod. Phys.\textbf{\ 75}, 559 (2003).

\bibitem{Han} M. Han, \textit{Einstein equation from covariant loop quantum
gravity in semiclassical continuum limit}. Phys. Rev. D \textbf{96}, 024047
(2017).

\bibitem{ADM} R. Arnowitt, S. Deser and C.W. Misner, \textit{Gravitation: An
introduction to current research}, L. Witten ed., Wiley, N.Y., U.S.A. (1962).

\bibitem{zzz2} Z.B. Etienne, Y.T. Liu and S.L. Shapiro,\textit{\
Relativistic magnetohydrodynamics in dynamical spacetimes: A new AMR
implementation. }Phys. Rev. D \textbf{82}, 084031 (2010).

\bibitem{alcu} M. Alcubierre, \textit{Introduction to 3+1 numerical
relativity}, Oxford University Press, Oxford, U.K. (2008).

\bibitem{Vaca5} T. Gheorghiu, O. Vacaru and S. Vacaru, \textit{Off-diagonal
deformations of kerr black holes in Einstein and modified massive gravity
and higher dimensions.} Eur. Phys. J. C \textbf{74}, 3152 (2014).

\bibitem{Vaca6} V. Ruchin, O. Vacaru and S. Vacaru, \textit{On relativistic
generalization of Perelman's W-entropy and thermodynamic description of
gravitational fields and cosmology.} Eur. Phys. J. C \textbf{77}, 184 (2017).

\bibitem{EPJ1} C. Cremaschini and M. Tessarotto, \textit{Exact solution of
the EM radiation-reaction problem for classical finite-size and Lorentzian
charged particles. }Eur. Phys. J. Plus \textbf{126}, 42 (2011).

\bibitem{EPJ2} C. Cremaschini and M. Tessarotto, \textit{Hamiltonian
formulation for the classical EM radiation-reaction problem: Application to
the kinetic theory for relativistic collisionless plasmas. }Eur. Phys. J.
Plus \textbf{126}, 63 (2011).

\bibitem{Wald} R. Wald, \textit{General Relativity}, Chicago, University of
Chicago Press (1984).

\bibitem{Rovelli} C. Rovelli, \textit{Space and Time in Loop Quantum Gravit}%
y, in "Beyond Spacetime: The Philosophical Foundations of Quantum Gravity",
edited by Baptiste Le Biha, Keizo Matsubara and Christian Wuthrich (2018).

\bibitem{bohm1} D. Bohm, B.J. Hiley and P.N. Kaloyerou, \textit{An
ontological basis for the quantum theory.} Physics Reports \textbf{144},
321--375 (1987).

\bibitem{bohm2} G. Gr\"{o}ssing, \textit{On the thermodynamic origin of the
quantum potential.} Physica A: Statistical Mechanics and its Applications
\textbf{388}, 811-823 (2009).

\bibitem{bohm3} G. Dennis, M.A. de Gosson, B.J. Hiley, \textit{Bohm's
quantum potential as an internal energy.} Physics Letters A \textbf{379},
1224-1227 (2015).

\bibitem{wheeler-1} B.S. DeWitt, \textit{Quantum Theory of Gravity. I. The
Canonical Theory.} Phys. Rev. \textbf{160}, 1113 (1967).
\end{thebibliography}
\end{document}